\documentclass[11pt,a4paper]{article}
\usepackage{jheppub}

\usepackage[english]{babel}

\usepackage{float}
\usepackage{amsmath}
\usepackage{amssymb}
\usepackage{tikz, ifthen}
\usetikzlibrary{decorations.markings} 
\usepackage{xcolor} 
\usepackage{dsfont}
\usepackage{graphicx}
\usepackage{subcaption} 
\usepackage{hyperref}
\graphicspath{{figures/}}

\title{Hamiltonian Lattice Formulation of Compact Maxwell-Chern-Simons Theory}

\author[a]{Changnan Peng,}
\author[b]{Maria Cristina Diamantini,}
\author[c]{Lena Funcke,}
\author[d]{Syed Muhammad Ali Hassan,}
\author[d,e]{Karl Jansen,}
\author[e]{Stefan Kühn,}
\author[a,f,g]{Di Luo,}
\author[d,e]{Pranay Naredi}

\affiliation[a]{Department of Physics, Massachusetts Institute of Technology, Cambridge, Massachusetts 02139, USA}
\affiliation[b]{NiPS Laboratory, INFN and Dipartimento di Fisica e Geologia, University of Perugia, Via A. Pascoli, I-06100 Perugia, Italy}
\affiliation[c]{Transdisciplinary Research Area ``Building Blocks of Matter and Fundamental Interactions'' (TRA Matter) and Helmholtz Institute for Radiation and Nuclear Physics (HISKP), University of Bonn, Nussallee 14-16, 53115 Bonn, Germany}
\affiliation[d]{Computation-Based Science and Technology Research Center,
The Cyprus Institute, 20 Kavafi Street, 2121 Nicosia, Cyprus}
\affiliation[e]{Deutsches Elektronen-Synchrotron DESY, Platanenallee 6, 15738 Zeuthen, Germany}
\affiliation[f]{The NSF AI Institute for Artificial Intelligence and Fundamental Interactions}
\affiliation[g]{Department of Physics, Harvard University, Cambridge, MA 02138, USA}

\emailAdd{cnpeng@mit.edu}
\emailAdd{cristina.diamantini@pg.infn.it}
\emailAdd{lfuncke@uni-bonn.de}
\emailAdd{diluo@mit.edu}

\abstract{In this paper, a Hamiltonian lattice formulation for 2+1D compact Maxwell-Chern-Simons theory is derived.  We analytically solve this theory and demonstrate that the mass gap in the continuum limit matches the well-known continuum formula. Our formulation preserves topological features such as the quantization of the Chern-Simons level, the degeneracy of energy eigenstates, the non-trivial properties of Wilson loops, and the mutual and self statistics of anyons. This work lays the groundwork for future Hamiltonian-based simulations of Maxwell-Chern-Simons theory on classical and quantum computers.}

\keywords{Field Theories in Lower Dimensions, Lattice Gauge Field Theories, Maxwell-Chern-Simons Theory}

\begin{document}

\maketitle

\section{Introduction}

Chern-Simons theory is a topological quantum field theory with numerous applications in condensed matter and high-energy physics, including the study of anomalies, fermion/boson dualities, and the fractional quantum Hall effect~\cite{Chern:1974simons}. Recently, a Lagrangian lattice formulation of pure Chern-Simons theory and its canonical quantization has been proposed~\cite{Sulejmanpasic2023,Sulejmanpasic2024}. However, pure lattice Chern-Simons theory faces a doubling problem, similar to the fermion doubling problem. The presence of a Maxwell term with Chern-Simons theory solves the doubling problem, in a way similar to the Wilson term that eliminates the fermion doubling problem~\cite{Berruto:2000fb}. The resulting theory is lattice Maxwell-Chern-Simons theory, which we will consider in this paper. 
In the continuum, the Maxwell-Chern-Simons theory has been first introduced in Ref.~\cite{Deser:1981wh}, where it was shown that in 2+1D, when a Chern-Simons term is added to the Maxwell Lagrangian, the photon becomes massive through a topological mechanism, without spontaneous symmetry breaking via the Higgs mechanism. This mechanism of topological mass generation is relevant for topological phases of matter (for a review see Ref.~\cite{wen1}), which are gapped systems with a new type of quantum order that is not due to symmetry breaking~\cite{wen}. 
These phases of matter can have various exotic quantum features, like excitations with non-trivial statistics, or ground-state degeneracy on manifolds with non-trivial topology.

The Lagrangian formulation of compact Maxwell-Chern-Simons theory for 2+1D Euclidean spacetime lattices has been first derived in Ref.~\cite{Diamantini:1993iu} using the self-dual approximation introduced in Ref.~\cite{roman}, and later studied in other works (see, e.g., Refs.~\cite{Dunne2000,Colatto:2002pe,Oxman:2002wa}). 
The Hamiltonian formulation has so far been derived only for non-compact gauge fields~\cite{Luscher:1989kk}, but the corresponding formulation for compact gauge fields remains absent. Such a formulation would be particularly important for future Hamiltonian-based simulations of the theory on classical and quantum computers. The Hamiltonian formulation provides a promising approach to simulate sign-problem afflicted regimes in lattice field theory, including topological terms, chemical potentials, and out-of-equilibrium dynamics. 
Recently, there has been growing interest in the development of Hamiltonian-based simulation methods, including tensor network states~\cite{Banuls:2018jag, Banuls2019}, non-Gaussian states~\cite{SHI2018245, PhysRevD.98.034505, julian_gaussian_st2020,julian_vmc_2023}, machine-learning-based approaches~\cite{luo2021gauge_equ,luo2022gauge,luo2021gauge_inv,chen2022simulating}, and quantum computing~\cite{Funcke:2023jbq, DiMeglio:2023nsa}. 
In particular, topological terms have been numerically investigated in 1+1D using tensor network states~\cite{Byrnes:2002gj, Buyens:2017crb, Funcke:2019zna, Nakayama:2021iyp, Funcke:2023lli} and quantum algorithms~\cite{Thompson_2022,Angelides:2023noe}, as well as in 3+1D using exact diagonalization~\cite{Kan:2021nyu}. Topological terms in 2+1D have not yet been simulated, due to the lack of a suitable Hamiltonian lattice formulation.

In this paper, we derive a Hamiltonian lattice formulation of compact Maxwell-Chern-Simons theory using the Villain approximation. We analytically solve this theory and show that the mass gap in the continuum limit matches the well-known continuum formula. Moreover, topological features such as the quantization of the Chern-Simons level, the degeneracy of the energy eigenstates, and the non-trivial properties of Wilson loops are reproduced. When fermions are introduced into the theory, numerical methods become necessary. Thus, our Hamiltonian formulation of Maxwell-Chern-Simons theory lays the groundwork for future Hamiltonian-based simulations of the theory on classical and quantum computers.

The paper is organized as follows. In Sec.~\ref{sec:review}, we review the Lagrangian lattice formulation of compact Abelian Chern-Simons theory as proposed in Ref.~\cite{Sulejmanpasic2023}. In Sec.~\ref{sec:review_maxwell}, we review the compact Maxwell lattice Hamiltonian and emphasize there are two versions of Hamiltonian with instantons allowed or suppressed. In Sec.~\ref{sec:Hamiltonian}, we derive the Hamiltonian lattice formulation of compact Maxwell-Chern-Simons theory. Here, we discuss the quantization of the Chern-Simons level and the degeneracy of the states. We also discuss some physical properties of the lattice Maxwell-Chern-Simons theory, including the perimeter law of Wilson loops and their non-trivial linking, as well as the mutual and self statistics of anyons. In Sec.~\ref{sec:mode}, we derive the solution to the constant mode wave functions. In Sec.~\ref{sec:solution}, we analytically solve the quadratic Hamiltonian that we derived in Sec.~\ref{sec:Hamiltonian}, and we plot the resulting band structure. In Sec.~\ref{sec:conclusion}, we summarize and discuss our results.

\section{Review of Compact Chern-Simons Lattice  Action}
\label{sec:review}

In the following, we review the Lagrangian lattice formulation of compact Abelian Chern-Simons theory, based on Ref.~\cite{Sulejmanpasic2023}.
In the continuum, the Abelian Chern-Simons action is defined as
\begin{align} \label{eq:S_CS_continuum}
    S_{CS}(A) = -\frac{i k}{4\pi} \int d^3 x\, \epsilon^{\mu\nu\rho} A_\mu \partial_\nu A_\rho,
\end{align} 
where $A$ is the U(1) gauge field configuration and $k$ is the quantized Chern-Simons level~\cite{Dunne:1989it}.

For the lattice discretization of this action, we consider a 2+1D cubic lattice with periodic boundary conditions in all directions. The U(1) gauge field $A$ lives on the links of the lattice. A naive lattice discretization of the continuous Chern-Simons action in Eq.~\eqref{eq:S_CS_continuum} is
\begin{align} \label{eq:S_CS_lattice}
    S_{CS}(A) =  -\frac{i k}{4\pi} a^2 d \tau\sum_{x\in \text{sites}} \epsilon^{\mu\nu\rho} A_{x;\mu} \Delta_\nu A_{x+\hat{\mu};\rho},
\end{align}
where $a$ and $d\tau$ are the lattice spacings in the space and time directions, respectively, and $\hat{mu}$ is a unit vector in the corresponding direction. Here, the finite (forward) difference operator of the $A$ field on the lattice is defined as
\begin{align}
    \Delta_0 A_{x;\mu} &= \frac{A_{x+\hat{0};\mu} - A_{x;\mu}}{d\tau}, \label{eq:diff0}\\
    \Delta_i A_{x;\mu} &= \frac{A_{x+\hat{i};\mu} - A_{x;\mu}}{a},\label{eq:diffi}
\end{align}
where $\Delta_0$ represents the time difference and $\Delta_{i\in\{1,2\}}$ represents the spacial difference.

We note that after integration by parts, Eq.~\eqref{eq:S_CS_lattice} can be equivalently written in terms of the backward lattice difference operator:
\begin{align} \label{eq:S_CS_lattice_backward}
    S_{CS}(A) &=  -\frac{i k}{4\pi} a^2 d \tau \sum_{x\in \text{sites}}\epsilon^{\mu\nu\rho} A_{x;\mu} \hat{\Delta}_\nu A_{x-\hat{\rho};\rho},
\end{align}
where
\begin{align}
\hat{\Delta}_0 A_{x;\mu} &= \frac{A_{x;\mu} - A_{x-\hat{0};\mu}}{d\tau}, \\
    \hat{\Delta}_i A_{x;\mu} &= \frac{A_{x;\mu} - A_{x-\hat{i};\mu}}{a},
\end{align}
Thus, we can try to write the action symmetrically with both the forward and backward difference operators:
\begin{align} \label{eq:S_CS_lattice_forward_and_backward}
    S_{CS}(A) =  -\frac{i k}{4\pi} a^2 d \tau \sum_{x\in \text{sites}} \epsilon^{\mu\nu\rho} \frac{A_{x;\mu} \Delta_\nu A_{x+\hat{\mu};\rho} + A_{x;\mu} \hat{\Delta}_\nu A_{x-\hat{\rho};\rho}}{2}.
\end{align}
As shown in Refs.~\cite{froelich,muller}, this symmetric form of the action respects Osterwalder-Schrader reflection positivity.
Next, we introduce a pictorial illustration of the lattice action. For this, we first explicitly write out the epsilon tensor in Eq.~\eqref{eq:S_CS_lattice}, which results in 
\begin{align}
\label{eq:CSexp}
    S_{CS}(A) =  -\frac{i k}{4\pi} a^2 d \tau \sum_{x\in \text{sites}} 
    & \left[ A_{x;0} \left(\Delta_1 A_{x+\hat{0};2} - \Delta_2 A_{x+\hat{0};1}\right) \right. \nonumber \\
    &- A_{x;1} \left( \Delta_0 A_{x+\hat{1};2} - \Delta_2 A_{x+\hat{1};0}\right) \nonumber \\
    &+ \left. A_{x;2} \left(\Delta_0 A_{x+\hat{2};1} - \Delta_1 A_{x+\hat{2};0}\right)
    \right].
\end{align}
This action can be visualized as
\begin{align}
\label{eq:CSdia}
    S_{CS}(A) =  -\frac{i k}{4\pi} a^2 d \tau \sum_{\text{cubes}} & \left[  \ 
    \begin{tikzpicture}[baseline=-0.5ex]
        \draw[thick,->] (-0.5,-0.5,0.5) -- (-0.5,0.5,0.5);
        \draw[thick,dashed] (0.5,0.5,-0.5) -- (-0.5,0.5,-0.5);
        \draw[thick,dashed] (0.5,0.5,0.5) -- (0.5,0.5,-0.5);
        \draw[thick,dashed] (-0.5,0.5,0.5) -- (0.5,0.5,0.5);
        \draw[thick,dashed] (-0.5,0.5,-0.5) -- (-0.5,0.5,0.5);
        \node[right] at (-0.5,0,0.5) {$A_0$};
    \end{tikzpicture}
    \times \frac{1}{a}
    \left(
    \begin{tikzpicture}[baseline=-0.5ex]
        \draw[thick,dashed] (-0.5,-0.5,0.5) -- (-0.5,0.5,0.5);
        \draw[thick,->] (0.5,0.5,-0.5) -- (-0.5,0.5,-0.5);
        \draw[thick,->] (0.5,0.5,0.5) -- (0.5,0.5,-0.5);
        \draw[thick,->] (-0.5,0.5,0.5) -- (0.5,0.5,0.5);
        \draw[thick,->] (-0.5,0.5,-0.5) -- (-0.5,0.5,0.5);
        \node[below] at (0,0.5,0.5) {$A_1$};
        \node[above] at (0,0.5,-0.5) {$-A_1$};
        \node[left] at (-0.5,0.6,0) {$-A_2$};
        \node[right] at (0.5,0.4,0) {$A_2$};
    \end{tikzpicture}
    \right) \right. \nonumber \\
    & \quad - 
    \begin{tikzpicture}[baseline=-0.5ex]
        \draw[thick,->] (-0.5,-0.5,0.5) -- (0.5,-0.5,0.5);
        \draw[thick,dashed] (0.5,0.5,-0.5) -- (0.5,0.5,0.5);
        \draw[thick,dashed] (0.5,0.5,0.5) -- (0.5,-0.5,0.5);
        \draw[thick,dashed] (0.5,-0.5,0.5) -- (0.5,-0.5,-0.5);
        \draw[thick,dashed] (0.5,-0.5,-0.5) -- (0.5,0.5,-0.5);
        \node[above] at (0,-0.5,0.5) {$A_1$};
    \end{tikzpicture}
    \times \frac{1}{d\tau}
    \left(
    \begin{tikzpicture}[baseline=-0.5ex]
        \draw[thick,dashed] (-0.5,-0.5,0.5) -- (0.5,-0.5,0.5);
        \draw[thick,->] (0.5,0.5,-0.5) -- (0.5,-0.5,-0.5);
        \draw[thick,->] (0.5,0.5,0.5) -- (0.5,0.5,-0.5);
        \draw[thick,->] (0.5,-0.5,0.5) -- (0.5,0.5,0.5);
        \draw[thick,->] (0.5,-0.5,-0.5) -- (0.5,-0.5,0.5);
        \node[left] at (0.5,0,0.5) {$\frac{d\tau}{a}A_0$};
        \node[right] at (0.5,0,-0.5) {$-\frac{d\tau}{a}A_0$};
        \node[above] at (0.3,0.5,0) {$A_2$};
        \node[below] at (0.7,-0.5,0) {$-A_2$};
    \end{tikzpicture}
    \right) \nonumber \\
    & \quad + \left. 
    \begin{tikzpicture}[baseline=-0.5ex]
        \draw[thick,->] (-0.5,-0.5,0.5) -- (-0.5,-0.5,-0.5);
        \draw[thick,dashed] (0.5,0.5,-0.5) -- (-0.5,0.5,-0.5);
        \draw[thick,dashed] (-0.5,0.5,-0.5) -- (-0.5,-0.5,-0.5);
        \draw[thick,dashed] (-0.5,-0.5,-0.5) -- (0.5,-0.5,-0.5);
        \draw[thick,dashed] (0.5,-0.5,-0.5) -- (0.5,0.5,-0.5);
        \node[right] at (-0.5,-0.5,0) {$A_2$};
    \end{tikzpicture}
    \times \frac{1}{d\tau}
    \left(
    \begin{tikzpicture}[baseline=-0.5ex]
        \draw[thick,dashed] (-0.5,-0.5,0.5) -- (-0.5,-0.5,-0.5);
        \draw[thick,->] (0.5,-0.5,-0.5) -- (-0.5,-0.5,-0.5);
        \draw[thick,->] (0.5,0.5,-0.5) -- (0.5,-0.5,-0.5);
        \draw[thick,->] (-0.5,0.5,-0.5) -- (0.5,0.5,-0.5);
        \draw[thick,->] (-0.5,-0.5,-0.5) -- (-0.5,0.5,-0.5);
        \node[left] at (-0.5,0,-0.5) {$\frac{d\tau}{a}A_0$};
        \node[right] at (0.5,0,-0.5) {$-\frac{d\tau}{a}A_0$};
        \node[above] at (0,0.5,-0.5) {$A_1$};
        \node[below] at (0,-0.5,-0.5) {$-A_1$};
    \end{tikzpicture}
    \right)
    \right].
\end{align}
Here, the solid lines correspond to the actual terms in the action, while the dashed lines are for visual guidance only; the dashed lines denote the positions of the other gauge fields in the products. Moreover, $A_0$, $A_1$, and $A_2$ denote the gauge fields on the links in the $0$, $1$, and $2$ directions, which correspond to the directions upwards, right, and into the plane, respectively.  Each solid ``plaquette'' in Eq.~\eqref{eq:CSdia} corresponds to the two terms in the round brackets of Eq.~\eqref{eq:CSexp}, which---according to the definition of the finite difference operators in Eqs.~\eqref{eq:diff0} and \eqref{eq:diffi}---can be expressed in terms of four gauge fields multiplied by a prefactor of $1/d\tau$ or $1/a$, respectively. 

This naive definition of the lattice action in Eq.~\eqref{eq:CSdia} is invariant under local gauge transformations, $A\to A+d\lambda$. However, it is not invariant under the discrete shift of the compact gauge variable, $A_0\to A_0 + \frac{2\pi}{d\tau}$ and $A_i\to A_i + \frac{2\pi}{a}$, $i=1,2$. This discrete shift invariance is necessary because we aim to formulate lattice Chern-Simons theory with a compact gauge group, i.e., U(1) rather than $\mathbb R$. The compactness of the gauge group implies the existence of quantized magnetic fluxes, $
\int_{\Sigma} dA \in 2\pi \mathbb Z\,,$
where $\Sigma$ is a closed surface. This indicates the presence of a monopole configuration if the surface is contractible. Consequently, lattice discretizations of compact U(1) gauge theory generically comprise dynamical lattice-scale monopole configurations. However, monopoles are known to violate gauge invariance in the presence of a Chern-Simons term, as shown in the continuum limit~\cite{Pisarski:1986gr,Affleck:1989qf}. While this has been considered an obstacle for formulating a gauge-invariant Chern-Simons theory on the lattice, this issue can be circumvented using the modified Villain approach~\cite{Villain:1974ir,Gross:1990ub,Sulejmanpasic:2019ytl,JingYuanChen2019,Gorantla:2021svj,JingYuanChen2022,JingYuanChen2023}. In the conventional Villain approach~\cite{Villain:1974ir}, the standard algebra-valued gauge fields $A\in \mathbb R$ are accompanied by discrete plaquette variables $n\in \mathbb Z$, which encode the quantized magnetic flux. These variables $n$ can be interpreted as discrete gauge fields for the 
$\mathbb Z$ sub-symmetry of the non-compact $\mathbb R$
gauge symmetry, which is precisely the discrete shift symmetry mentioned above, $A_0\to A_0 + \frac{2\pi}{d\tau}$ and $A_i\to A_i + \frac{2\pi}{a}$, $i=1,2$. Upon gauging these discrete shifts, one effectively studies compact $\text{U}(1) = \mathbb R /2\pi\mathbb Z$ gauge theory instead of non-compact $\mathbb R$ gauge theory. 
This so-called Villain formulation of compact U(1) gauge theory contains instantons (i.e. monopoles in 2+1D), which can be elimated by a Lagrange multiplier $\varphi$ that constrains the discrete gauge field to be flat, called the modified Villain approach~\cite{Sulejmanpasic:2019ytl,JingYuanChen2019}. The lattice action and Hamiltonian formulations of the double Chern-Simons theory have been studied using the modified Villain approach~\cite{JingYuanChen2019,JingYuanChen2022,JingYuanChen2023}.

Following the modified Villain approach, the authors of Ref.~\cite{Sulejmanpasic2023} have built a gauge-invariant Chern-Simons lattice action by introducing integer degrees of freedom on each plaquette, as well as angular variables on each lattice site. In the following, we will use a different sign convention of the Chern-Simons action and therefore obtain a slightly different Chern-Simons lattice action (for the original Chern-Simons lattice action introduced in Ref.~\cite{Sulejmanpasic2023}, see their Eq.~(17)):

\begin{align}
    S_{CS}(A, n, \varphi) =  -\frac{i k}{4\pi} a^2 d \tau \sum_{\text{cubes}} & \left[ \ 
    \begin{tikzpicture}[baseline=-0.5ex]
        \draw[thick,->] (-0.5,-0.5,0.5) -- (-0.5,0.5,0.5);
        \draw[thick,dashed] (0.5,0.5,-0.5) -- (-0.5,0.5,-0.5);
        \draw[thick,dashed] (0.5,0.5,0.5) -- (0.5,0.5,-0.5);
        \draw[thick,dashed] (-0.5,0.5,0.5) -- (0.5,0.5,0.5);
        \draw[thick,dashed] (-0.5,0.5,-0.5) -- (-0.5,0.5,0.5);
        \node[right] at (-0.5,0,0.5) {$A_0$};
    \end{tikzpicture}
    \times \frac{1}{a}
    \left(
    \begin{tikzpicture}[baseline=-0.5ex]
        \draw[thick,dashed] (-0.5,-0.5,0.5) -- (-0.5,0.5,0.5);
        \draw[thick,->] (0.5,0.5,-0.5) -- (-0.5,0.5,-0.5);
        \draw[thick,->] (0.5,0.5,0.5) -- (0.5,0.5,-0.5);
        \draw[thick,->] (-0.5,0.5,0.5) -- (0.5,0.5,0.5);
        \draw[thick,->] (-0.5,0.5,-0.5) -- (-0.5,0.5,0.5);
        \node[below] at (0,0.5,0.5) {$A_1$};
        \node[above] at (0,0.5,-0.5) {$-A_1$};
        \node[left] at (-0.5,0.6,0) {$-A_2$};
        \node[right] at (0.5,0.4,0) {$A_2$};
    \end{tikzpicture}
    \right) \right. \nonumber \\
    & \quad -  
    \begin{tikzpicture}[baseline=-0.5ex]
        \draw[thick,->] (-0.5,-0.5,0.5) -- (0.5,-0.5,0.5);
        \draw[thick,dashed] (0.5,0.5,-0.5) -- (0.5,0.5,0.5);
        \draw[thick,dashed] (0.5,0.5,0.5) -- (0.5,-0.5,0.5);
        \draw[thick,dashed] (0.5,-0.5,0.5) -- (0.5,-0.5,-0.5);
        \draw[thick,dashed] (0.5,-0.5,-0.5) -- (0.5,0.5,-0.5);
        \node[above] at (0,-0.5,0.5) {$A_1$};
    \end{tikzpicture}
    \times \frac{1}{d\tau}
    \left(
    \begin{tikzpicture}[baseline=-0.5ex]
        \draw[thick,dashed] (-0.5,-0.5,0.5) -- (0.5,-0.5,0.5);
        \draw[thick,->] (0.5,0.5,-0.5) -- (0.5,-0.5,-0.5);
        \draw[thick,->] (0.5,0.5,0.5) -- (0.5,0.5,-0.5);
        \draw[thick,->] (0.5,-0.5,0.5) -- (0.5,0.5,0.5);
        \draw[thick,->] (0.5,-0.5,-0.5) -- (0.5,-0.5,0.5);
        \node[left] at (0.5,0,0.5) {$\frac{d\tau}{a}A_0$};
        \node[right] at (0.5,0,-0.5) {$-\frac{d\tau}{a}A_0$};
        \node[above] at (0.3,0.5,0) {$A_2$};
        \node[below] at (0.7,-0.5,0) {$-A_2$};
    \end{tikzpicture}
    \right) \nonumber \\
    & \quad + 
    \begin{tikzpicture}[baseline=-0.5ex]
        \draw[thick,->] (-0.5,-0.5,0.5) -- (-0.5,-0.5,-0.5);
        \draw[thick,dashed] (0.5,0.5,-0.5) -- (-0.5,0.5,-0.5);
        \draw[thick,dashed] (-0.5,0.5,-0.5) -- (-0.5,-0.5,-0.5);
        \draw[thick,dashed] (-0.5,-0.5,-0.5) -- (0.5,-0.5,-0.5);
        \draw[thick,dashed] (0.5,-0.5,-0.5) -- (0.5,0.5,-0.5);
        \node[right] at (-0.5,-0.5,0) {$A_2$};
    \end{tikzpicture}
    \times \frac{1}{d\tau}
    \left(
    \begin{tikzpicture}[baseline=-0.5ex]
        \draw[thick,dashed] (-0.5,-0.5,0.5) -- (-0.5,-0.5,-0.5);
        \draw[thick,->] (0.5,-0.5,-0.5) -- (-0.5,-0.5,-0.5);
        \draw[thick,->] (0.5,0.5,-0.5) -- (0.5,-0.5,-0.5);
        \draw[thick,->] (-0.5,0.5,-0.5) -- (0.5,0.5,-0.5);
        \draw[thick,->] (-0.5,-0.5,-0.5) -- (-0.5,0.5,-0.5);
        \node[left] at (-0.5,0,-0.5) {$\frac{d\tau}{a}A_0$};
        \node[right] at (0.5,0,-0.5) {$-\frac{d\tau}{a}A_0$};
        \node[above] at (0,0.5,-0.5) {$A_1$};
        \node[below] at (0,-0.5,-0.5) {$-A_1$};
    \end{tikzpicture}
    \right) \nonumber \\
    & \quad + \frac{2\pi}{a^2} \ 
    \begin{tikzpicture}[baseline=-0.5ex]
        \draw[thick,->] (-0.5,-0.5,0.5) -- (-0.5,0.5,0.5);
        \draw[thick,dashed] (0.5,0.5,-0.5) -- (-0.5,0.5,-0.5);
        \draw[thick,dashed] (0.5,0.5,0.5) -- (0.5,0.5,-0.5);
        \draw[thick,dashed] (-0.5,0.5,0.5) -- (0.5,0.5,0.5);
        \draw[thick,dashed] (-0.5,0.5,-0.5) -- (-0.5,0.5,0.5);
        \node[right] at (-0.5,0,0.5) {$A_0$};
    \end{tikzpicture}
    \times 
    \begin{tikzpicture}[baseline=-0.5ex]
        \draw[thick,dashed] (-0.5,-0.5,0.5) -- (-0.5,0.5,0.5);
        \draw[thick,dashed] (0.5,0.5,-0.5) -- (-0.5,0.5,-0.5);
        \draw[thick,dashed] (0.5,0.5,0.5) -- (0.5,0.5,-0.5);
        \draw[thick,dashed] (-0.5,0.5,0.5) -- (0.5,0.5,0.5);
        \draw[thick,dashed] (-0.5,0.5,-0.5) -- (-0.5,0.5,0.5);
        \draw[thick] (-0.25,0.5,-0.25) -- (-0.25,0.5,0.25);
        \draw[thick] (-0.25,0.5,0.25) -- (0.25,0.5,0.25);
        \draw[thick] (0.25,0.5,0.25) -- (0.25,0.5,-0.25);
        \draw[thick] (0.25,0.5,-0.25) -- (-0.25,0.5,-0.25);
        \node[below] at (0,0.3,0) {$n_{1,2}$};
    \end{tikzpicture} 
    + \frac{2\pi}{a^2} \ 
    \begin{tikzpicture}[baseline=-0.5ex]
        \draw[thick,->] (0.5,-0.5,-0.5) -- (0.5,0.5,-0.5);
        \draw[thick,dashed] (0.5,-0.5,-0.5) -- (-0.5,-0.5,-0.5);
        \draw[thick,dashed] (0.5,-0.5,0.5) -- (0.5,-0.5,-0.5);
        \draw[thick,dashed] (-0.5,-0.5,0.5) -- (0.5,-0.5,0.5);
        \draw[thick,dashed] (-0.5,-0.5,-0.5) -- (-0.5,-0.5,0.5);
        \node[left] at (0.5,0,-0.5) {$A_0$};
    \end{tikzpicture}
    \times 
    \begin{tikzpicture}[baseline=-0.5ex]
        \draw[thick,dashed] (0.5,-0.5,-0.5) -- (0.5,0.5,-0.5);
        \draw[thick,dashed] (0.5,-0.5,-0.5) -- (-0.5,-0.5,-0.5);
        \draw[thick,dashed] (0.5,-0.5,0.5) -- (0.5,-0.5,-0.5);
        \draw[thick,dashed] (-0.5,-0.5,0.5) -- (0.5,-0.5,0.5);
        \draw[thick,dashed] (-0.5,-0.5,-0.5) -- (-0.5,-0.5,0.5);
        \draw[thick] (-0.25,-0.5,-0.25) -- (-0.25,-0.5,0.25);
        \draw[thick] (-0.25,-0.5,0.25) -- (0.25,-0.5,0.25);
        \draw[thick] (0.25,-0.5,0.25) -- (0.25,-0.5,-0.25);
        \draw[thick] (0.25,-0.5,-0.25) -- (-0.25,-0.5,-0.25);
        \node[above] at (0,-0.3,0) {$n_{1,2}$};
    \end{tikzpicture} 
    \nonumber \\
    & \quad - \frac{2\pi}{a d\tau} \ 
    \begin{tikzpicture}[baseline=-0.5ex]
        \draw[thick,->] (-0.5,-0.5,0.5) -- (0.5,-0.5,0.5);
        \draw[thick,dashed] (0.5,0.5,-0.5) -- (0.5,0.5,0.5);
        \draw[thick,dashed] (0.5,0.5,0.5) -- (0.5,-0.5,0.5);
        \draw[thick,dashed] (0.5,-0.5,0.5) -- (0.5,-0.5,-0.5);
        \draw[thick,dashed] (0.5,-0.5,-0.5) -- (0.5,0.5,-0.5);
        \node[above] at (0,-0.5,0.5) {$A_1$};
    \end{tikzpicture}
    \times 
    \begin{tikzpicture}[baseline=-0.5ex]
        \draw[thick,dashed] (-0.5,-0.5,0.5) -- (0.5,-0.5,0.5);
        \draw[thick,dashed] (0.5,0.5,-0.5) -- (0.5,-0.5,-0.5);
        \draw[thick,dashed] (0.5,0.5,0.5) -- (0.5,0.5,-0.5);
        \draw[thick,dashed] (0.5,-0.5,0.5) -- (0.5,0.5,0.5);
        \draw[thick,dashed] (0.5,-0.5,-0.5) -- (0.5,-0.5,0.5);
        \draw[thick] (0.5,-0.25,-0.25) -- (0.5,-0.25,0.25);
        \draw[thick] (0.5,-0.25,0.25) -- (0.5,0.25,0.25);
        \draw[thick] (0.5,0.25,0.25) -- (0.5,0.25,-0.25);
        \draw[thick] (0.5,0.25,-0.25) -- (0.5,-0.25,-0.25);
        \node[left] at (0.3,0,0) {$n_{0,2}$};
    \end{tikzpicture} 
    - \frac{2\pi}{a d\tau} \ 
    \begin{tikzpicture}[baseline=-0.5ex]
        \draw[thick,->] (-0.5,0.5,-0.5) -- (0.5,0.5,-0.5);
        \draw[thick,dashed] (-0.5,0.5,-0.5) -- (-0.5,0.5,0.5);
        \draw[thick,dashed] (-0.5,0.5,0.5) -- (-0.5,-0.5,0.5);
        \draw[thick,dashed] (-0.5,-0.5,0.5) -- (-0.5,-0.5,-0.5);
        \draw[thick,dashed] (-0.5,-0.5,-0.5) -- (-0.5,0.5,-0.5);
        \node[below] at (0,0.5,-0.5) {$A_1$};
    \end{tikzpicture}
    \times 
    \begin{tikzpicture}[baseline=-0.5ex]
        \draw[thick,dashed] (-0.5,0.5,-0.5) -- (0.5,0.5,-0.5);
        \draw[thick,dashed] (-0.5,0.5,-0.5) -- (-0.5,-0.5,-0.5);
        \draw[thick,dashed] (-0.5,0.5,0.5) -- (-0.5,0.5,-0.5);
        \draw[thick,dashed] (-0.5,-0.5,0.5) -- (-0.5,0.5,0.5);
        \draw[thick,dashed] (-0.5,-0.5,-0.5) -- (-0.5,-0.5,0.5);
        \draw[thick] (-0.5,-0.25,-0.25) -- (-0.5,-0.25,0.25);
        \draw[thick] (-0.5,-0.25,0.25) -- (-0.5,0.25,0.25);
        \draw[thick] (-0.5,0.25,0.25) -- (-0.5,0.25,-0.25);
        \draw[thick] (-0.5,0.25,-0.25) -- (-0.5,-0.25,-0.25);
        \node[right] at (-0.3,0,0) {$n_{0,2}$};
    \end{tikzpicture} 
    \nonumber \\
    & \quad + \left.
    \frac{2\pi}{a d\tau} \ 
    \begin{tikzpicture}[baseline=-0.5ex]
        \draw[thick,->] (-0.5,-0.5,0.5) -- (-0.5,-0.5,-0.5);
        \draw[thick,dashed] (0.5,0.5,-0.5) -- (-0.5,0.5,-0.5);
        \draw[thick,dashed] (-0.5,0.5,-0.5) -- (-0.5,-0.5,-0.5);
        \draw[thick,dashed] (-0.5,-0.5,-0.5) -- (0.5,-0.5,-0.5);
        \draw[thick,dashed] (0.5,-0.5,-0.5) -- (0.5,0.5,-0.5);
        \node[right] at (-0.5,-0.5,0) {$A_2$};
    \end{tikzpicture}
    \times 
    \begin{tikzpicture}[baseline=-0.5ex]
        \draw[thick,dashed] (-0.5,-0.5,0.5) -- (-0.5,-0.5,-0.5);
        \draw[thick,dashed] (0.5,-0.5,-0.5) -- (-0.5,-0.5,-0.5);
        \draw[thick,dashed] (0.5,0.5,-0.5) -- (0.5,-0.5,-0.5);
        \draw[thick,dashed] (-0.5,0.5,-0.5) -- (0.5,0.5,-0.5);
        \draw[thick,dashed] (-0.5,-0.5,-0.5) -- (-0.5,0.5,-0.5);
        \draw[thick] (-0.35,-0.35,-0.5) rectangle (0.35,0.35,-0.5);
        \node at (0,0,-0.5) {$n_{0,1}$};
    \end{tikzpicture} 
    + \frac{2\pi}{a d\tau} \ 
    \begin{tikzpicture}[baseline=-0.5ex]
        \draw[thick,->] (0.5,0.5,0.5) -- (0.5,0.5,-0.5);
        \draw[thick,dashed] (0.5,0.5,0.5) -- (-0.5,0.5,0.5);
        \draw[thick,dashed] (-0.5,0.5,0.5) -- (-0.5,-0.5,0.5);
        \draw[thick,dashed] (-0.5,-0.5,0.5) -- (0.5,-0.5,0.5);
        \draw[thick,dashed] (0.5,-0.5,0.5) -- (0.5,0.5,0.5);
        \node[left] at (0.4,0.5,0) {$A_2$};
    \end{tikzpicture}
    \times 
    \begin{tikzpicture}[baseline=-0.5ex]
        \draw[thick,dashed] (0.5,0.5,0.5) -- (0.5,0.5,-0.5);
        \draw[thick,dashed] (0.5,-0.5,0.5) -- (-0.5,-0.5,0.5);
        \draw[thick,dashed] (0.5,0.5,0.5) -- (0.5,-0.5,0.5);
        \draw[thick,dashed] (-0.5,0.5,0.5) -- (0.5,0.5,0.5);
        \draw[thick,dashed] (-0.5,-0.5,0.5) -- (-0.5,0.5,0.5);
        \draw[thick] (-0.35,-0.35,0.5) rectangle (0.35,0.35,0.5);
        \node at (0,0,0.5) {$n_{0,1}$};
    \end{tikzpicture} 
    \ \right] \nonumber \\
    & \hspace{-1.8cm} +i \sum_{\text{cubes}} 
     \left(
    \ 
    \begin{tikzpicture}[baseline=-0.5ex]
        \draw[thick,dashed] (0.5,0.5,0.5) -- (0.5,0.5,-0.5);
        \draw[thick,dashed] (0.5,-0.5,0.5) -- (0.5,-0.5,-0.5);
        \draw[thick,dashed] (-0.5,0.5,0.5) -- (-0.5,0.5,-0.5);
        \draw[thick,dashed] (-0.5,-0.5,0.5) -- (-0.5,-0.5,-0.5);
        \draw[thick,dashed] (-0.5,-0.5,0.5) rectangle (0.5,0.5,0.5);
        \draw[thick,dashed] (-0.5,-0.5,-0.5) rectangle (0.5,0.5,-0.5);
        \fill (0.5,0.5,-0.5) circle (3pt);
        \node[right] at (0.5,0.5,-0.5) {$\varphi$};
    \end{tikzpicture} 
    - \frac{1}{2} \cdot
    \begin{tikzpicture}[baseline=-0.5ex]
        \draw[thick,->] (-0.5,-0.5,0.5) -- (-0.5,0.5,0.5);
        \draw[thick,->] (-0.5,0.5,0.5) -- (-0.5,0.5,-0.5);
        \draw[thick,->] (-0.5,0.5,-0.5) -- (0.5,0.5,-0.5);
        \draw[thick,dashed] (0.5,0.5,0.5) -- (0.5,0.5,-0.5);
        \draw[thick,dashed] (0.5,-0.5,0.5) -- (0.5,-0.5,-0.5);
        \draw[thick,dashed] (-0.5,0.5,0.5) -- (-0.5,0.5,-0.5);
        \draw[thick,dashed] (-0.5,-0.5,0.5) -- (-0.5,-0.5,-0.5);
        \draw[thick,dashed] (-0.5,-0.5,0.5) rectangle (0.5,0.5,0.5);
        \draw[thick,dashed] (-0.5,-0.5,-0.5) rectangle (0.5,0.5,-0.5);
        \node[left] at (-0.5,0,0.5) {$k d\tau A_0$};
        \node[left] at (-0.6,0.5,0) {$k a A_2$};
        \node[above] at (0,0.5,-0.5) {$k a A_1$};
    \end{tikzpicture} 
    \right)
    \times
    \begin{tikzpicture}[baseline=-0.5ex]
        \draw[thick,red] (-0.35,-0.35,-0.5) rectangle (0.35,0.35,-0.5);
        \draw[thick] (-0.35,-0.35,0.5) rectangle (0.35,0.35,0.5);
        \draw[thick,red] (-0.5,-0.25,-0.25) -- (-0.5,-0.25,0.25);
        \draw[thick,red] (-0.5,-0.25,0.25) -- (-0.5,0.25,0.25);
        \draw[thick,red] (-0.5,0.25,0.25) -- (-0.5,0.25,-0.25);
        \draw[thick,red] (-0.5,0.25,-0.25) -- (-0.5,-0.25,-0.25);
        \draw[thick] (0.5,-0.25,-0.25) -- (0.5,-0.25,0.25);
        \draw[thick] (0.5,-0.25,0.25) -- (0.5,0.25,0.25);
        \draw[thick] (0.5,0.25,0.25) -- (0.5,0.25,-0.25);
        \draw[thick] (0.5,0.25,-0.25) -- (0.5,-0.25,-0.25);
        \draw[thick,red] (-0.25,0.5,-0.25) -- (-0.25,0.5,0.25);
        \draw[thick,red] (-0.25,0.5,0.25) -- (0.25,0.5,0.25);
        \draw[thick,red] (0.25,0.5,0.25) -- (0.25,0.5,-0.25);
        \draw[thick,red] (0.25,0.5,-0.25) -- (-0.25,0.5,-0.25);
        \draw[thick] (-0.25,-0.5,-0.25) -- (-0.25,-0.5,0.25);
        \draw[thick] (-0.25,-0.5,0.25) -- (0.25,-0.5,0.25);
        \draw[thick] (0.25,-0.5,0.25) -- (0.25,-0.5,-0.25);
        \draw[thick] (0.25,-0.5,-0.25) -- (-0.25,-0.5,-0.25);
        \draw[thick,dashed] (0.5,0.5,0.5) -- (0.5,0.5,-0.5);
        \draw[thick,dashed] (0.5,-0.5,0.5) -- (0.5,-0.5,-0.5);
        \draw[thick,dashed] (-0.5,0.5,0.5) -- (-0.5,0.5,-0.5);
        \draw[thick,dashed] (-0.5,-0.5,0.5) -- (-0.5,-0.5,-0.5);
        \draw[thick,dashed] (-0.5,-0.5,0.5) rectangle (0.5,0.5,0.5);
        \draw[thick,dashed] (-0.5,-0.5,-0.5) rectangle (0.5,0.5,-0.5);
        \node at (-0.8,-0.9) {$n_{0,1}$};
        \node at (0.5,-1) {$n_{1,2}$};
        \node at (1,-0.5) {$n_{0,2}$};
        \node at (0.5,0.9) {$-n_{0,1}$};
        \node at (-0.5,1) {$-n_{1,2}$};
        \node at (-1.2,0.5) {$-n_{0,2}$};
    \end{tikzpicture} 
    ,
\label{eq:S_CS_full_diagram}
\end{align}
where $n_{0,1}$ and $n_{0,2}$ are integer variables living on time-like plaquettes, and $n_{1,2}$ are integer variables living on space-like plaquettes. These integer variables are drawn as smaller squares in each plaquette for visual clarity, and the negative $n$ values are indicated by red squares. In the last term in Eq.~\eqref{eq:S_CS_full_diagram}, there is an angular variable $\varphi\in [0,2\pi)$ on each lattice site as the Lagrangian multiplier to impose the zero-instanton constraint for each unit cell. For any configuration where a unit cell contains non-zero total flux (i.e. instanton number, shown as the cube in Eq.~\eqref{eq:S_CS_full_diagram}), summing over the corresponding $\varphi$ from $0$ to $2\pi$ in the partition function makes the partition function vanish. Therefore, the $\varphi$ variables impose hard constraints that eliminate all instantons. Later, we will show that $\varphi$ is not necessary in our lattice Hamiltonian formulation in Eq.~\eqref{eq:Hamiltonian} because we impose those constraints directly by ignoring the instanton configurations in the counting of degrees of freedom (see the detailed derivation in Appendix~\ref{subsec:appendix_Maxwell_Chern_Simons}).

There are two types of gauge transformations associated with the action $S_{CS}(A, n, \varphi)$ (see Eq.~(22) in Ref.~\cite{Sulejmanpasic2023} for the corresponding mathematical formula). For $\forall \lambda\in \mathbb{R}$, $\forall k\in 2\mathbb{Z}$,
\begin{align}
    S_{CS}\left(
    \begin{tikzpicture}[baseline=-0.5ex]
        \draw[thick] (-1,0) -- (1,0);
        \draw[thick] (0,-1) -- (0,1);
        \draw[thick] (0,0,-1) -- (0,0,1);
        \fill (0,0) circle (3pt);
        \node at (-0.7,0.3) {$\varphi - k \lambda$};
        \node at (-1.1,-0.3) {$A_1 - \frac{\lambda}{a}$};
        \node at (-0.8,-0.8) {$A_2 - \frac{\lambda}{a}$};
        \node at (0.8,-0.8) {$A_0 - \frac{\lambda}{d\tau}$};
        \node at (1,-0.3) {$A_1 + \frac{\lambda}{a}$};
        \node at (1.1,0.3) {$A_2 + \frac{\lambda}{a}$};
        \node at (0.8,0.8) {$A_0 + \frac{\lambda}{d\tau}$};
    \end{tikzpicture}
    , *
    \right)
    =
    S_{CS}\left(
    \begin{tikzpicture}[baseline=-0.5ex]
        \draw[thick] (-1,0) -- (1,0);
        \draw[thick] (0,-1) -- (0,1);
        \draw[thick] (0,0,-1) -- (0,0,1);
        \fill (0,0) circle (3pt);
        \node at (-0.3,0.3) {$\varphi$};
        \node at (-1.1,-0.3) {$A_1$};
        \node at (-0.6,-0.6) {$A_2$};
        \node at (0.4,-0.8) {$A_0$};
        \node at (1,-0.3) {$A_1$};
        \node at (0.7,0.3) {$A_2$};
        \node at (0.4,0.8) {$A_0$};
    \end{tikzpicture}
    , *
    \right) 
\label{eq:local_transformation}
\end{align}
\begin{align}
    e^{-S_{CS}}\left(
    \begin{tikzpicture}[baseline=-0.5ex]
        \draw[thick,->] (0,-0.5) -- (0,0.5);
        \draw[thick] (0.15,-0.35,0) rectangle (0.85,0.35,0);
        \draw[thick] (-0.15,-0.35,0) rectangle (-0.85,0.35,0);
        \draw[thick] (0,0.25,0.25) -- (0,0.25,0.75);
        \draw[thick] (0,0.25,0.75) -- (0,-0.25,0.75);
        \draw[thick] (0,-0.25,0.75) -- (0,-0.25,0.25);
        \draw[thick] (0,-0.25,0.25) -- (0,0.25,0.25);
        \draw[thick] (0,0.25,-0.25) -- (0,0.25,-0.75);
        \draw[thick] (0,0.25,-0.75) -- (0,-0.25,-0.75);
        \draw[thick] (0,-0.25,-0.75) -- (0,-0.25,-0.25);
        \draw[thick] (0,-0.25,-0.25) -- (0,0.25,-0.25);
        \node at (-0.5,0.8) {$A_0+\frac{2\pi}{d\tau}$};
        \node at (0.9,0.8) {$n_{0,2}-1$};
        \node at (1.5,0) {$n_{0,1}-1$};
        \node at (-0.9,-0.8) {$n_{0,2}+1$};
        \node at (-1.5,0) {$n_{0,1}+1$};
    \end{tikzpicture}
    , *
    \right) 
    =
    e^{-S_{CS}}\left(
    \begin{tikzpicture}[baseline=-0.5ex]
        \draw[thick,->] (0,-0.5) -- (0,0.5);
        \draw[thick] (0.15,-0.35,0) rectangle (0.85,0.35,0);
        \draw[thick] (-0.15,-0.35,0) rectangle (-0.85,0.35,0);
        \draw[thick] (0,0.25,0.25) -- (0,0.25,0.75);
        \draw[thick] (0,0.25,0.75) -- (0,-0.25,0.75);
        \draw[thick] (0,-0.25,0.75) -- (0,-0.25,0.25);
        \draw[thick] (0,-0.25,0.25) -- (0,0.25,0.25);
        \draw[thick] (0,0.25,-0.25) -- (0,0.25,-0.75);
        \draw[thick] (0,0.25,-0.75) -- (0,-0.25,-0.75);
        \draw[thick] (0,-0.25,-0.75) -- (0,-0.25,-0.25);
        \draw[thick] (0,-0.25,-0.25) -- (0,0.25,-0.25);
        \node at (-0.2,0.8) {$A_0$};
        \node at (0.5,0.8) {$n_{0,2}$};
        \node at (1.3,0) {$n_{0,1}$};
        \node at (-0.5,-0.8) {$n_{0,2}$};
        \node at (-1.3,0) {$n_{0,1}$};
    \end{tikzpicture}
    , *
    \right) 
\label{eq:A_0_transform}
\end{align}
\begin{align}
    e^{-S_{CS}}\left(
    \begin{tikzpicture}[baseline=-0.5ex]
        \draw[thick,->] (0,0) -- (1,0);
        \draw[thick] (0.15,0.15,0) rectangle (0.85,0.85,0);
        \draw[thick] (0.15,-0.15,0) rectangle (0.85,-0.85,0);
        \draw[thick] (0.25,0,0.25) -- (0.25,0,0.75);
        \draw[thick] (0.25,0,0.75) -- (0.75,0,0.75);
        \draw[thick] (0.75,0,0.75) -- (0.75,0,0.25);
        \draw[thick] (0.75,0,0.25) -- (0.25,0,0.25);
        \draw[thick] (0.25,0,-0.25) -- (0.25,0,-0.75);
        \draw[thick] (0.25,0,-0.75) -- (0.75,0,-0.75);
        \draw[thick] (0.75,0,-0.75) -- (0.75,0,-0.25);
        \draw[thick] (0.75,0,-0.25) -- (0.25,0,-0.25);
        \node at (1.7,0) {$A_1+\frac{2\pi}{a}$};
        \node at (1.7,0.4) {$n_{1,2}-1$};
        \node at (1.5,0.8) {$n_{0,1}+1$};
        \node at (-0.7,-0.4) {$n_{1,2}+1$};
        \node at (-0.5,-0.8) {$n_{0,1}-1$};
    \end{tikzpicture}
    , *
    \right) 
    =
    e^{-S_{CS}}\left(
    \begin{tikzpicture}[baseline=-0.5ex]
        \draw[thick,->] (0,0) -- (1,0);
        \draw[thick] (0.15,0.15,0) rectangle (0.85,0.85,0);
        \draw[thick] (0.15,-0.15,0) rectangle (0.85,-0.85,0);
        \draw[thick] (0.25,0,0.25) -- (0.25,0,0.75);
        \draw[thick] (0.25,0,0.75) -- (0.75,0,0.75);
        \draw[thick] (0.75,0,0.75) -- (0.75,0,0.25);
        \draw[thick] (0.75,0,0.25) -- (0.25,0,0.25);
        \draw[thick] (0.25,0,-0.25) -- (0.25,0,-0.75);
        \draw[thick] (0.25,0,-0.75) -- (0.75,0,-0.75);
        \draw[thick] (0.75,0,-0.75) -- (0.75,0,-0.25);
        \draw[thick] (0.75,0,-0.25) -- (0.25,0,-0.25);
        \node at (1.4,0) {$A_1$};
        \node at (1.5,0.4) {$n_{1,2}$};
        \node at (1.3,0.8) {$n_{0,1}$};
        \node at (-0.5,-0.4) {$n_{1,2}$};
        \node at (-0.3,-0.8) {$n_{0,1}$};
    \end{tikzpicture}
    , *
    \right) 
\label{eq:A_1_transform}
\end{align}
\begin{align}
    e^{-S_{CS}}\left(
    \begin{tikzpicture}[baseline=-0.5ex]
        \draw[thick,->] (0,0,0.5) -- (0,0,-0.5);
        \draw[thick] (0,0.25,0.25) -- (0,0.25,-0.25);
        \draw[thick] (0,0.25,-0.25) -- (0,0.75,-0.25);
        \draw[thick] (0,0.75,-0.25) -- (0,0.75,0.25);
        \draw[thick] (0,0.75,0.25) -- (0,0.25,0.25);
        \draw[thick] (0,-0.25,0.25) -- (0,-0.25,-0.25);
        \draw[thick] (0,-0.25,-0.25) -- (0,-0.75,-0.25);
        \draw[thick] (0,-0.75,-0.25) -- (0,-0.75,0.25);
        \draw[thick] (0,-0.75,0.25) -- (0,-0.25,0.25);
        \draw[thick] (0.25,0,0.25) -- (0.25,0,-0.25);
        \draw[thick] (0.25,0,-0.25) -- (0.75,0,-0.25);
        \draw[thick] (0.75,0,-0.25) -- (0.75,0,0.25);
        \draw[thick] (0.75,0,0.25) -- (0.25,0,0.25);
        \draw[thick] (-0.25,0,0.25) -- (-0.25,0,-0.25);
        \draw[thick] (-0.25,0,-0.25) -- (-0.75,0,-0.25);
        \draw[thick] (-0.75,0,-0.25) -- (-0.75,0,0.25);
        \draw[thick] (-0.75,0,0.25) -- (-0.25,0,0.25);
        \node at (0.8,0.5) {$A_2+\frac{2\pi}{a}$};
        \node at (1.5,0) {$n_{1,2}+1$};
        \node at (0.8,-0.5) {$n_{0,2}-1$};
        \node at (-1.5,0) {$n_{1,2}-1$};
        \node at (-0.8,0.5) {$n_{0,2}+1$};
    \end{tikzpicture}
    , *
    \right) 
    =
    e^{-S_{CS}}\left(
    \begin{tikzpicture}[baseline=-0.5ex]
        \draw[thick,->] (0,0,0.5) -- (0,0,-0.5);
        \draw[thick] (0,0.25,0.25) -- (0,0.25,-0.25);
        \draw[thick] (0,0.25,-0.25) -- (0,0.75,-0.25);
        \draw[thick] (0,0.75,-0.25) -- (0,0.75,0.25);
        \draw[thick] (0,0.75,0.25) -- (0,0.25,0.25);
        \draw[thick] (0,-0.25,0.25) -- (0,-0.25,-0.25);
        \draw[thick] (0,-0.25,-0.25) -- (0,-0.75,-0.25);
        \draw[thick] (0,-0.75,-0.25) -- (0,-0.75,0.25);
        \draw[thick] (0,-0.75,0.25) -- (0,-0.25,0.25);
        \draw[thick] (0.25,0,0.25) -- (0.25,0,-0.25);
        \draw[thick] (0.25,0,-0.25) -- (0.75,0,-0.25);
        \draw[thick] (0.75,0,-0.25) -- (0.75,0,0.25);
        \draw[thick] (0.75,0,0.25) -- (0.25,0,0.25);
        \draw[thick] (-0.25,0,0.25) -- (-0.25,0,-0.25);
        \draw[thick] (-0.25,0,-0.25) -- (-0.75,0,-0.25);
        \draw[thick] (-0.75,0,-0.25) -- (-0.75,0,0.25);
        \draw[thick] (-0.75,0,0.25) -- (-0.25,0,0.25);
        \node at (0.5,0.5) {$A_2$};
        \node at (1.2,0) {$n_{1,2}$};
        \node at (0.5,-0.5) {$n_{0,2}$};
        \node at (-1.2,0) {$n_{1,2}$};
        \node at (-0.5,0.5) {$n_{0,2}$};
    \end{tikzpicture}
    , *
    \right) 
\label{eq:A_2_transform}
\end{align}
The $*$ in the parentheses represents all other variables, which are the same on the left and right hand sides in the equations. Equation~\eqref{eq:local_transformation} shows a local gauge transformation, under which $S_{CS}$ itself is invariant. Equations~\eqref{eq:A_0_transform}--\eqref{eq:A_2_transform} are the discrete gauge transformations, under which the partition function $e^{-S_{CS}}$ is invariant. In fact, $S_{CS}$ changes by $i \pi k \mathbb{Z}$ from the last term in the action Eq.~\eqref{eq:S_CS_full_diagram} under discrete gauge transformations. When $k$ is an odd integer, there can be an extra $\pi$ phase in $S_{CS}$, and the partition function changes by an extra minus sign. This extra minus sign can be cancelled by the anomaly inflow from an auxiliary 4D bulk~\cite{Sulejmanpasic2023}, or by introducing additional fermionic degrees of freedom~\cite{Sulejmanpasic2024}. More precisely, the odd-$k$ theory is a theory of fermions, where the Chern-Simons action depends on the choice of the spin structure of the manifold~\cite{Witten:2016spinstructure}. Here, we only consider the pure gauge theory in 2+1D, so we assume that $k$ is an even integer in the beginning. 
What may be surprising is that, in Sec.~\ref{sec:Hamiltonian}, after we add the Maxwell term and derive the compact Maxwell-Chern-Simons theory in Hamiltonian formaluation, the corresponding Hamiltonian works for any integer $k$. We will show that the fermionic nature for odd $k$ is still present and can be probed with our formulation. 

The authors of Ref.~\cite{Sulejmanpasic2023} have shown that the lattice action in Eq.~\eqref{eq:S_CS_full_diagram} reproduces various aspects of the continuum Chern-Simons theory, including the level quantization and the discrete $\mathbb Z$ 1-form symmetry. In a more recent work by the same authors~\cite{Sulejmanpasic2024}, they have tried to canonically quantize the lattice action in Eq.~\eqref{eq:S_CS_full_diagram}. However, without the Maxwell term, the quantization of the pure Chern-Simons action unavoidably has non-trivial commutators between gauge field variables on different links (see Eq.~(23) in Ref.~\cite{Sulejmanpasic2024}). The non-commuting gauge field variables make it hard to find an orthonormal basis of the Hilbert space, and thus lose the capability of performing numerical simulations with their formulation.

\section{Review of Compact Maxwell Lattice Hamiltonian}
\label{sec:review_maxwell}

Before we discuss Maxwell-Chern-Simons theory, let's take a step back and review the pure Maxwell theory. In this section, we revisit the compact Maxwell lattice gauge theory in 2+1D, also known as the 2+1D lattice QED. We note that there is no Chern-Simons term in this section, and the purpose of this section is to see the effect of instanton suppression on the Hamiltonian formalism in our familiar lattice QED setting, which will be useful in the understanding of the later sections.

In 2+1D lattice QED, it is common to consider the following compact Maxwell action~\cite{wilson1974confinement}: 
\begin{align}
    S_{\rm Maxwell}(A) &=  -\frac{\beta_0}{d\tau} \sum_{x\in \text{sites}} 
     \left[ \cos  \left( a d\tau ( \Delta_0 A_{x;1} - \Delta_1 A_{x;0} )\right) + \cos \left( a d\tau ( \Delta_0 A_{x;2} - \Delta_2 A_{x;0} ) \right) \right] \nonumber \\
    &\quad -\frac{\beta d\tau}{a^2} \sum_{x\in \text{sites}}  \left[ \cos \left( a^2  ( \Delta_1 A_{x;2} - \Delta_2 A_{x;1} ) \right)
    \right],
\end{align}
where $\beta_0$ and $\beta$ are the temporal and spatial gauge coupling coefficients, respectively, $d\tau$ and $a$ are the temporal and spatial lattice spacings, respectively, and $A_0$ and $A_i$ with $i=1,2$ are the gauge field in the temporal and spatial directions, respectively.
This lattice action can be visualized as
\begin{align}
S_{\rm Maxwell}(A) = &-\frac{\beta_0}{d\tau} \sum_{\substack{\text{time-like} \\ \text{plaquettes}}} 
\left[ \cos \left(a \cdot \begin{tikzpicture}[baseline=-0.5ex]
\draw[thick,->] (0.5,-0.5) -- (-0.5,-0.5);
\draw[thick,->] (0.5,0.5) -- (0.5,-0.5);
\draw[thick,->] (-0.5,0.5) -- (0.5,0.5);
\draw[thick,->] (-0.5,-0.5) -- (-0.5,0.5);
\node[left] at (-0.5,0) {$\frac{d\tau}{a}A_0$};
\node[right] at (0.5,0) {$-\frac{d\tau}{a}A_0$};
\node[above] at (0,0.5) {$A_1$};
\node[below] at (0,-0.5) {$-A_1$};
\end{tikzpicture} \right) + \cos \left(a \cdot \begin{tikzpicture}[baseline=-0.5ex]
\draw[thick,->] (0,0.5,-0.5) -- (0,-0.5,-0.5);
\draw[thick,->] (0,0.5,0.5) -- (0,0.5,-0.5);
\draw[thick,->] (0,-0.5,0.5) -- (0,0.5,0.5);
\draw[thick,->] (0,-0.5,-0.5) -- (0,-0.5,0.5);
\node[left] at (0,0,0.5) {$\frac{d\tau}{a}A_0$};
\node[right] at (0,0,-0.5) {$-\frac{d\tau}{a}A_0$};
\node[above] at (-0.2,0.5,0) {$A_2$};
\node[below] at (0.2,-0.5,0) {$-A_2$};
\end{tikzpicture} \right) \right] \nonumber \\
&- \frac{\beta d\tau}{a^2} \sum_{\substack{\text{space-like} \\ \text{plaquettes}}} \left[ \cos \left(a \cdot \begin{tikzpicture}[baseline=-0.5ex]
\draw[thick,->] (0.5,0,-0.5) -- (-0.5,0,-0.5);
\draw[thick,->] (0.5,0,0.5) -- (0.5,0,-0.5);
\draw[thick,->] (-0.5,0,0.5) -- (0.5,0,0.5);
\draw[thick,->] (-0.5,0,-0.5) -- (-0.5,0,0.5);
\node[below] at (0,0,0.5) {$A_1$};
\node[above] at (0,0,-0.5) {$-A_1$};
\node[left] at (-0.5,0.1) {$-A_2$};
\node[right] at (0.5,-0.1) {$A_2$};
\end{tikzpicture} \right) \right].
\label{eq:S_maxwell_review}
\end{align}

In order to do the canonical quantization to obtain the Hamiltonian formulation, we need to use the Villain approximation to expand the cosine terms in the action into quadratic forms~\cite{Villain1975}. The Villain approximated action is
\begin{align}
    S_{\rm Maxwell}(A, n) &= \frac{\beta_0}{2 d\tau} \sum_{\substack{x\in\text{sites} \\ i\in\{1,2\}}} (a d\tau (\Delta_0 A_{x;i} - \Delta_i A_{x;0}) + 2\pi n_{x;0,i})^2 \nonumber \\
    &\quad + \frac{\beta d\tau}{2 a^2} \sum_{x\in\text{sites}} (a^2(\Delta_1 A_{x;2} - \Delta_2 A_{x;1}) + 2\pi n_{x;1,2})^2,
\end{align}
where the integer degrees of freedom $n$ live on every plaquettes. 
This Villain approximated lattice action can be visualized as
\begin{align}
S_{\rm Maxwell}(A,n) &= \frac{\beta_0}{2 d\tau} \sum_{\substack{\text{time-like} \\ \text{plaquettes}}} 
\left[  \left(a \cdot \begin{tikzpicture}[baseline=-0.5ex]
\draw[thick,->] (0.5,-0.5) -- (-0.5,-0.5);
\draw[thick,->] (0.5,0.5) -- (0.5,-0.5);
\draw[thick,->] (-0.5,0.5) -- (0.5,0.5);
\draw[thick,->] (-0.5,-0.5) -- (-0.5,0.5);
\node[left] at (-0.5,0) {$\frac{d\tau}{a}A_0$};
\node[right] at (0.5,0) {$-\frac{d\tau}{a}A_0$};
\node[above] at (0,0.5) {$A_1$};
\node[below] at (0,-0.5) {$-A_1$};
\end{tikzpicture} 
+ 2\pi\ 
\begin{tikzpicture}[baseline=-0.5ex]
\draw[thick,dashed] (0.5,-0.5) -- (-0.5,-0.5);
\draw[thick,dashed] (0.5,0.5) -- (0.5,-0.5);
\draw[thick,dashed] (-0.5,0.5) -- (0.5,0.5);
\draw[thick,dashed] (-0.5,-0.5) -- (-0.5,0.5);
\draw[thick] (-0.35,-0.35) rectangle (0.35,0.35);
\node at (0,0) {$n_{0,1}$};
\end{tikzpicture} 
\right)^2 \right. \nonumber \\
&\left. \quad\qquad\qquad\qquad +  \left(a \cdot \begin{tikzpicture}[baseline=-0.5ex]
\draw[thick,->] (0,0.5,-0.5) -- (0,-0.5,-0.5);
\draw[thick,->] (0,0.5,0.5) -- (0,0.5,-0.5);
\draw[thick,->] (0,-0.5,0.5) -- (0,0.5,0.5);
\draw[thick,->] (0,-0.5,-0.5) -- (0,-0.5,0.5);
\node[left] at (0,0,0.5) {$\frac{d\tau}{a}A_0$};
\node[right] at (0,0,-0.5) {$-\frac{d\tau}{a}A_0$};
\node[above] at (-0.2,0.5,0) {$A_2$};
\node[below] at (0.2,-0.5,0) {$-A_2$};
\end{tikzpicture} 
+ 2\pi\ 
\begin{tikzpicture}[baseline=-0.5ex]
\draw[thick,dashed] (0,0.5,-0.5) -- (0,-0.5,-0.5);
\draw[thick,dashed] (0,0.5,0.5) -- (0,0.5,-0.5);
\draw[thick,dashed] (0,-0.5,0.5) -- (0,0.5,0.5);
\draw[thick,dashed] (0,-0.5,-0.5) -- (0,-0.5,0.5);
\draw[thick] (0,-0.25,-0.25) -- (0,-0.25,0.25);
\draw[thick] (0,-0.25,0.25) -- (0,0.25,0.25);
\draw[thick] (0,0.25,0.25) -- (0,0.25,-0.25);
\draw[thick] (0,0.25,-0.25) -- (0,-0.25,-0.25);
\node[right] at (0.15,0,0) {$n_{0,2}$};
\end{tikzpicture} 
\right)^2 \  \right] \nonumber \\
&\quad + \frac{\beta d\tau}{2 a^2} \sum_{\substack{\text{space-like} \\ \text{plaquettes}}} \left[  \left(a \cdot \begin{tikzpicture}[baseline=-0.5ex]
\draw[thick,->] (0.5,0,-0.5) -- (-0.5,0,-0.5);
\draw[thick,->] (0.5,0,0.5) -- (0.5,0,-0.5);
\draw[thick,->] (-0.5,0,0.5) -- (0.5,0,0.5);
\draw[thick,->] (-0.5,0,-0.5) -- (-0.5,0,0.5);
\node[below] at (0,0,0.5) {$A_1$};
\node[above] at (0,0,-0.5) {$-A_1$};
\node[left] at (-0.5,0.1) {$-A_2$};
\node[right] at (0.5,-0.1) {$A_2$};
\end{tikzpicture} 
+ 2\pi\ 
\begin{tikzpicture}[baseline=-0.5ex]
\draw[thick,dashed] (0.5,0,-0.5) -- (-0.5,0,-0.5);
\draw[thick,dashed] (0.5,0,0.5) -- (0.5,0,-0.5);
\draw[thick,dashed] (-0.5,0,0.5) -- (0.5,0,0.5);
\draw[thick,dashed] (-0.5,0,-0.5) -- (-0.5,0,0.5);
\draw[thick] (-0.25,0,-0.25) -- (-0.25,0,0.25);
\draw[thick] (-0.25,0,0.25) -- (0.25,0,0.25);
\draw[thick] (0.25,0,0.25) -- (0.25,0,-0.25);
\draw[thick] (0.25,0,-0.25) -- (-0.25,0,-0.25);
\node[above] at (0,0.15,0) {$n_{1,2}$};
\end{tikzpicture} 
\right)^2 \  \right],
\label{eq:maxwell_villain_S}
\end{align}
which converges to the non-approximated action in Eq.~\eqref{eq:S_maxwell_review} in the $d\tau \to 0$ limit~\cite{Villain1975}. 

From the above Villain approximated lattice action, if we allow the presence of instantons, we can derive the familiar lattice QED Hamiltonian~\cite{Drell:1979cosB}. In the derivation, the presence of instantons allows an integer degree of freedom on every cubic unit cell in the lattice (i.e.\ the instanton number in every cube). Summing over these instanton degrees of freedom in the partition function provides an reversed Villain approximation and allows us to wrap the space-like plaquettes into the familiar cosine term in the Hamiltonian again, while the time-like plaquettes becomes the $E_i^2$ term.  
\begin{equation}
    H_{\rm Maxwell\;with\;instantons} ={e^2 \over 2 a^2} \sum_{\rm links} E_i^2 + {1 \over e^2a^2} \sum_{\rm plaquettes} \left(1 -\cos{a^2 B}\right) \ ,
    \label{qedhamiltonian}
\end{equation}
where $E_i$ is the electric field in the $i$ direction, and $B$ is the magnetic field. 
The coefficients in Eq.~\eqref{eq:maxwell_villain_S} and the coupling constant are related by $\beta_0=\beta=1/{e^2}$. See Appendix~\ref{subsec:appendix_Maxwell_Instanton_Allowed} for a detailed derivation.

Meanwhile, if we totally suppress the instantons, from the same Villain approximated lattice action in Eq.~\eqref{eq:maxwell_villain_S}, we can derive a different version of the lattice Hamiltonian. In our derivation, we enforce the zero-intanton constraint by taking out the integer degrees of freedom which correspond to instanton numbers in cubes (this process is equivalent to adding the Lagrangian multipliers to the action and then integrate them out in the partition function). Without summing over instanton degrees of freedom, the space-like plaquette terms cannot be wrap into cosine terms and stay quadratic.\footnote{
Because the space we study is a closed surface (2D torus due to periodic boundary conditions), there is one leftover integer degree of freedom - the total magnetic flux through the surface. Each integer $n$ of total magnetic flux defines a sector of the theory, with the Hamiltonian $$H_n={e^2 \over 2 a^2} \sum_{\rm links} E_i^2 + {a^2 \over 2 e^2} [(B'+\frac{2\pi n}{a^2})^2+\sum_{\rm plaq.}^{}{}' B^2],$$ where $B'$ is the magnetic field on one special plaquette, and $\sum_{\rm plaq.}'$ means summing all other plaquettes.
The theory is gauge invariant after taking into account all the sectors with $n\in(-\infty,+\infty)$. In~\eqref{qedhamiltonian2} for simplicity we show $H_0$ as a demonstration. The appearance of integer operators in canonical quantization of Villain fields can also be found in other studies~\cite{Sulejmanpasic2024,Fazza:2023integeroperator,Cheng:2023integeroperator,Berkowitz:2024AlekseyCherman}.

As a spoiler, in later sections when we discuss the Maxwell-Chern-Simons theory, the gauge invariant condition from the Chern-Simons term will enforce the total flux $n=0$. Therefore we will not see an $\hat{n}$ operator later in~\eqref{eq:Hamiltonian}.
} 
\begin{equation}
    H_{\rm Maxwell\;without\;instantons} ={e^2 \over 2 a^2} \sum_{\rm links} E_i^2 + {a^2 \over 2 e^2} \sum_{\rm plaquettes} B^2,
    \label{qedhamiltonian2}
\end{equation}
which comes together with two more constraints on its Hilbert space to guarantee the compactness of the theory. We emphasize that this Hamiltonian in Eq.~\eqref{qedhamiltonian2} is a Hamiltonian of a {\bf compact} Maxwell theory, and it is different from Eq.~\eqref{qedhamiltonian} because we enforce the zero-instanton constraint. Since we study a 2+1D cubic lattice with periodic boundary conditions in all directions, the 2D space is a torus, and thus the zero-instanton constraint is nontrivial. We introduce this Hamiltonian and the associated Hilbert space constraints in Appendix~\ref{subsec:appendix_Maxwell_Instanton_Suppressed}. 

To summarize, there are two types of Maxwell theory: instanton-allowed or instanton-suppressed. Sometimes the instanton-suppressed Maxwell theory is referred to as the ``non-compact QED''. However, we would like to emphasize that the instanton-suppressed Maxwell theory is a compact theory due to the additional Hilbert space constraints. Later when we discuss the Maxwell-Chern-Simons theory, we will only talk about the the instanton-suppressed Maxwell theory. This is because accompanied with the Chern-Simons action, instantons carry charges, which is similar to the Witten effect in 3+1D Maxwell theory with the $\theta$ term~\cite{Witten:1979monopole}. Therefore, in order to study the pure gauge theory without matter fields, here we have to suppress the instantons. We refer to Appendix~\ref{subsec:appendix_instanton} for more discussions on the instanton effects.

\section{Compact Maxwell-Chern-Simons Lattice Hamiltonian}

\label{sec:Hamiltonian}

In the previous work of canonical quantization of lattice Chern-Simons theory~\cite{Sulejmanpasic2024}, there is no proposal of adding the Maxwell term. In this section, we provide a formulation of the Maxwell-Chern-Simons Lattice Hamiltonian, which can be implemented in classical and quantum simulations.
We consider the sum of Eqs.~\eqref{eq:S_CS_full_diagram} and~\eqref{eq:maxwell_villain_S} as the total action,
\begin{align}
    S(A,n,\varphi) = S_{\rm Maxwell}(A,n) + S_{CS}(A,n,\varphi),
\label{eq:S_total}
\end{align}
which will be our starting point to obtain the lattice Hamiltonian for the Maxwell-Chern-Simons theory.

The Chern-Simons term, being topological, does not contribute to the Hamiltonian. 
Therefore, we would expect to see a Maxwell-Chern-Simons lattice Hamiltonian that looks similar to the Hamiltonian of compact QED with instantons suppressed as shown in Eq.~\eqref{qedhamiltonian2}. We would also expect to obtain two additional Hilbert space constraints. However, the expression of the electric field $E$ appearing in the Hamiltonian, the Gauss' law, and the Hilbert space constraints will be modified due to the Chern-Simons term.

We use the transfer matrix method~\cite{Polyakov:1987book}, suppress the instantons by ignoring their configurations in the counting of degrees of freedom, and construct a lattice Hamiltonian based on the action in Eq.~\eqref{eq:S_total} (see Appendix~\ref{subsec:appendix_Maxwell_Chern_Simons} for more details): 
\begin{align}
    \hat{H} = &+ \frac{e^2}{2 a^2} \sum_{\text{plaquettes}} \left[
    \left(\ 
    \begin{tikzpicture}[baseline=-0.5ex]
    \tikzset{mid arrow/.style={
        decoration={markings, mark=at position 0.75 with {\arrow{>}}},
        postaction={decorate},
        thick
    }}
    \draw[dashed] (-0.5,-0.5) -- (-0.5,0.5);
    \draw[mid arrow] (-0.5,0.5) -- (0.5,0.5);
    \node[below] at (0,0.5) {$\hat{p}_1$};
    \end{tikzpicture}
    - \left(\frac{k a^2}{4\pi}\right)\ 
    \begin{tikzpicture}[baseline=-0.5ex]
    \tikzset{mid arrow/.style={
        decoration={markings, mark=at position 0.75 with {\arrow{>}}},
        postaction={decorate},
        thick
    }}
    \draw[mid arrow] (-0.5,-0.5) -- (-0.5,0.5);
    \draw[dashed] (-0.5,0.5) -- (0.5,0.5);
    \node[right] at (-0.5,0) {$\hat{A}_2$};
    \end{tikzpicture}
    \right)^2 +
    \left(\ 
    \begin{tikzpicture}[baseline=-0.5ex]
    \tikzset{mid arrow/.style={
        decoration={markings, mark=at position 0.75 with {\arrow{>}}},
        postaction={decorate},
        thick
    }}
    \draw[dashed] (-0.5,-0.5) -- (0.5,-0.5);
    \draw[mid arrow] (0.5,-0.5) -- (0.5,0.5);
    \node[left] at (0.5,0) {$\hat{p}_2$};
    \end{tikzpicture}\ 
    + \left(\frac{k a^2}{4\pi}\right)\ 
    \begin{tikzpicture}[baseline=-0.5ex]
    \tikzset{mid arrow/.style={
        decoration={markings, mark=at position 0.75 with {\arrow{>}}},
        postaction={decorate},
        thick
    }}
    \draw[mid arrow] (-0.5,-0.5) -- (0.5,-0.5);
    \draw[dashed] (0.5,-0.5) -- (0.5,0.5);
    \node[above] at (0,-0.5) {$\hat{A}_1$};
    \end{tikzpicture}\ 
    \right)^2
    \right] \nonumber \\
    &+ \frac{1}{2 e^2} \sum_{\text{plaquettes}} \left( 
    \begin{tikzpicture}[baseline=-0.5ex]
    \draw[thick,->] (-0.5,-0.5) -- (0.5,-0.5);
    \draw[thick,->] (0.5,-0.5) -- (0.5,0.5);
    \draw[thick,->] (0.5,0.5) -- (-0.5,0.5);
    \draw[thick,->] (-0.5,0.5) -- (-0.5,-0.5);
    \node[left] at (-0.5,0) {$-\hat{A}_2$};
    \node[right] at (0.5,0) {$\hat{A}_2$};
    \node[above] at (0,0.5) {$-\hat{A}_1$};
    \node[below] at (0,-0.5) {$\hat{A}_1$};
    \end{tikzpicture}
    \right)^2
    \nonumber \\
    =& \sum_{x\in\text{sites}} \frac{e^2}{2 a^2}\left[\left(\hat{p}_{x;1}-\frac{ka^2}{4\pi}\hat{A}_{x-\hat{2};2}\right)^2 + \left(\hat{p}_{x;2}+\frac{ka^2}{4\pi}\hat{A}_{x-\hat{1};1}\right)^2\right]  + \frac{1}{2 e^2} \left(\Box \hat{A}_{x;1,2}\right)^2,
\label{eq:Hamiltonian}
\end{align}
where $e^2$ is the coupling constant, $a$ is the lattice spacing, $k$ is the Chern-Simons level, $\hat{A}_i$ is the gauge field operator in the direction $i=1$ or $2$, and $\hat{p}_i$ is the corresponding conjugate momentum operator. They have the standard commutation relation $\left[\hat{A}_{x;i},\hat{p}_{y;j}\right]=i\delta_{x,y}\delta_{i,j}$. We also define $\Box \hat{A}_{x;1,2}\equiv \hat{A}_{x;1} + \hat{A}_{x+\hat{1};2} - \hat{A}_{x+\hat{2};1} - \hat{A}_{x;2}$. The action coefficients and the coupling constant are related by $\beta_0=\beta=\frac{1}{e^2}$. This lattice Hamiltonian lives on a constraint Hilbert space which we will detail below.

In the pure Maxwell lattice theory as we discussed in the previous section, there is an extra integer degree of freedom denoting the total flux through the closed space surface. However, after adding the Chern-Simons term, the total flux is enforced to be zero due to gauge invariant constraints (see Appendix~\ref{subsec:appendix_Maxwell_Chern_Simons} for more details). Therefore, we do not have an integer operator in~\eqref{eq:Hamiltonian}.

\subsection{Hilbert Space and Constraints}

We construct the Hilbert space on a 2D time slice, which is a 2D square lattice. The Hilbert space is constructed with a set of basis $\{|A\rangle: A\in \mathbb{R}^{2 N_1 N_2}\}$, where $A$ denotes a configuration of the gauge fields, $N_1$ and $N_2$ are the lattice sizes in the two spacial directions, respectively, and $2 N_1 N_2$ is the total number of links. i.e., each basis vector is labeled by a configuration in which every link has a variable ranging in $(-\infty, +\infty)$. We note that our formulation provides a natural basis in terms of the gauge field configurations for classical and quantum simulations.

The compactness of the theory is reflected by several constraints on the Hilbert space. 
Any physical state $|\psi\rangle$ in the Hilbert space needs to satisfy the following constraints, where $\hat{A}$ is the gauge field operator, $\hat{p}$ is the conjugate momentum operator, $k$ is the Chern-Simons level, and $a$ is the lattice spacing:
\begin{enumerate}
    \item 
    \begin{align}
        \hat{G} |\psi\rangle = 0,
    \end{align}
    where
    \begin{align}
        \hat{G} &=
        \begin{tikzpicture}[baseline=-0.5ex]
        \draw[thick,->] (0,0) -- (0,-1);
        \draw[thick,->] (0,0) -- (1,0);
        \draw[thick,->] (0,0) -- (0,1);
        \draw[thick,->] (0,0) -- (-1,0);
        \node[above] at (-0.5,0) {$-\hat{p}_1$};
        \node[below] at (0.5,0) {$\hat{p}_1$};
        \node[right] at (0,0.5) {$\hat{p}_2$};
        \node[left] at (0,-0.5) {$-\hat{p}_2$};
        \end{tikzpicture}
        +
        \left(\frac{k a^2}{4\pi}\right)
        \begin{tikzpicture}[baseline=-0.5ex]
        \draw[dashed] (0,0) -- (0,-1);
        \draw[dashed] (0,0) -- (-1,0);
        \draw[thick,->] (0,1) -- (0,0);
        \draw[thick,->] (1,1) -- (0,1);
        \draw[thick,->] (1,0) -- (1,1);
        \draw[thick,->] (0,0) -- (1,0);
        \node[above] at (0.5,1) {$-\hat{A}_1$};
        \node[below] at (0.5,0) {$\hat{A}_1$};
        \node[right] at (1,0.5) {$\hat{A}_2$};
        \node[left] at (0,0.5) {$-\hat{A}_2$};
        \end{tikzpicture}  \nonumber \\
        &= \hat{p}_{x;1} + \hat{p}_{x;2} - \hat{p}_{x-\hat{1};1} - \hat{p}_{x-\hat{2};2} + \frac{k a^2}{4\pi} \left(\Box \hat{A}_{x;1,2}\right)
        ,
    \label{eq:Gauss_law}
    \end{align}
    for any lattice site $x$.
    
    Equivalently, 
    \begin{equation}
        e^{i\lambda\hat{G}} |\psi\rangle = |\psi\rangle, \quad \forall \lambda\in \mathbb{R}, 
    \end{equation}
    which is the Gauss' law or the local gauge transformation.

    \item \begin{equation}
        e^{2\pi i\hat{L}_1} |\psi\rangle = e^{i\theta_1}|\psi\rangle,
        \label{eq:theta_1}
    \end{equation}
    where $\theta_1$ is a constant global phase, and 
    \begin{align}
        \hat{L}_1 &= \begin{tikzpicture}[baseline=-0.5ex]
        \tikzset{mid arrow/.style={
            decoration={markings, mark=at position 0.75 with {\arrow{>}}},
            postaction={decorate},
            thick
        }}
        \foreach \x in {0,1,2} {
            \draw[mid arrow] (\x,-1) -- (\x,0);
            \draw[mid arrow] (\x + 1, 0) -- (\x,0);
        }
        \node[left] at (0,0) {$\cdots$};
        \node[right] at (3,0) {$\cdots$};
        \node[above] at (1.5,0) {$-\frac{k a}{4\pi}\hat{A}_1$};
        \node[right] at (2,-0.5) {$\frac{1}{a}\hat{p}_2$};
        \end{tikzpicture}  \nonumber \\
        &= \sum_{x_1=0}^{N_1-1} \left(\frac{1}{a} \hat{p}_{(x_1,x_2);2} - \frac{k a}{4\pi} \hat{A}_{(x_1,x_2+1);1}\right)
        ,
    \label{eq:constraint_L1}
    \end{align}
    for any $x_2 \in \{0,1,\dots,N_2-1\}$. This loop operator generates one of the large gauge transformations. Note that such a loop can be locally deformed by adding or subtracting the Gauss' law (constraint 1). Therefore, more generally we can write $\hat{L}_1$ for any topologically non-trivial horizontal loop that wraps through the boundary, i.e., any curve with the winding number $(1,0)$. 

    \item \begin{equation}
        e^{2\pi i\hat{L}_2} |\psi\rangle = e^{i\theta_2}|\psi\rangle,
    \end{equation}
    where $\theta_2$ is another constant global phase, and 
    \begin{align}
        \hat{L}_2 &= \begin{tikzpicture}[baseline=-0.5ex]
        \tikzset{mid arrow/.style={
            decoration={markings, mark=at position 0.75 with {\arrow{>}}},
            postaction={decorate},
            thick
        }}
        \foreach \y in {-1.5,-0.5,0.5} {
            \draw[mid arrow] (-1,\y) -- (0,\y);
            \draw[mid arrow] (0, \y) -- (0,\y + 1);
        }
        \node[below] at (0,-1.5) {$\vdots$};
        \node[above] at (0,1.5) {$\vdots$};
        \node[right] at (0,0) {$\frac{k a}{4\pi}\hat{A}_2$};
        \node[above] at (-0.5,0.5) {$\frac{1}{a}\hat{p}_1$};
        \end{tikzpicture}  \nonumber \\
        &= \sum_{x_2=0}^{N_2-1} \left(\frac{1}{a} \hat{p}_{(x_1,x_2);1} + \frac{k a}{4\pi} \hat{A}_{(x_1+1,x_2);2}\right)
        ,
    \label{eq:constraint_L2}
    \end{align}
    for any $x_1 \in \{0,1,\dots,N_1-1\}$. This loop operator generates the other large gauge transformations. Note that such a loop can also be locally deformed by adding or subtracting the Gauss' law (constraint 1). Therefore, more generally we can write $\hat{L}_2$ for any topologically non-trivial vertical loop that wraps through the boundary, i.e., any curve with the winding number $(0,1)$. 
\end{enumerate}

The second and the third constraints compactify the theory. In fact, the two constraints $e^{2\pi i\hat{L}_i} |\psi\rangle = e^{i\theta_i}|\psi\rangle$ allows the spectrum of $\hat{L}_i = m_i + \frac{\theta_i}{2\pi} \in \mathbb{Z} + \frac{\theta_i}{2\pi}$, $i=1,2$.
These two integers $m_1$ and $m_2$ are quantum numbers labeling the eigenfunctions of $\hat{L}_1$ and $\hat{L}_2$, respectively. For example, we can label the eigenfunctions of $\hat{L}_1$ by $|m_1, \theta_1\rangle$, which satisfies $\hat{L}_1|m_1, \theta_1\rangle=\left(\frac{\theta_1}{2\pi}+m_1\right)|m_1, \theta_1\rangle$. Then the function $|\psi\rangle$ that satisfies the constraint $e^{2\pi i\hat{L}_1}|\psi\rangle=e^{i\theta_1}|\psi\rangle$ can be most generally written as $|\psi\rangle=\sum_{m_1\in\mathbb{Z}}c_{m_1}|m_1, \theta_1\rangle$, where $c_{m_1}$ are arbitrary coefficients. However, in order to satisfy the other constraint $e^{2\pi i\hat{L}_2}|\psi\rangle=e^{i\theta_2}|\psi\rangle$, the coefficients are not arbitrary anymore, but need to satisfy $c_{m_1}=c_{m_1+k}$. This fact is due to the commutation relation between $\hat{L}_1$ and $\hat{L}_2$. Therefore, the most general wavefunction satisfying both constraints can be written as $|\psi\rangle=\sum_{m_1\in\{0,\dots,k-1\}}c_{m_1}\sum_{n\in\mathbb{Z}}|m_1+nk, \theta_1\rangle$, i.e. the arbitrary linear combination of $k$ basis wavefunctions $\sum_{n\in\mathbb{Z}}|m_1+nk, \theta_1\rangle$, $m_1=0,\dots,k-1$. Each basis wavefunction $\sum_{n\in\mathbb{Z}}|m_1+nk, \theta_1\rangle$, i.e. eigenfunction of $e^{2\pi i\hat{L}_1}$, is an equal weight summation of eigenfunctions of $\hat{L}_1$.
A large gauge transformation $|\psi\rangle\to e^{2\pi i\hat{L}_j}|\psi\rangle$ changes $m_i\to m_i + k\,\epsilon_{ij}$ and leaves the basis wavefunctions invariant (see the commutator between $\hat{L}_i$ and $\hat{L}_j$ in Eq.~\eqref{eq:commutator_L1_L2}). Therefore, we have a compact theory which is invariant under the large gauge transformation. This compactification is similar in spirit to the Villain approximation, where multiple non-periodic functions are summed with equal weights to become a periodic function. 

After having formulated these three constraints, we need to ensure that the lattice Hamiltonian operator in Eq.~\eqref{eq:Hamiltonian} is well defined on this constrained Hilbert space. This is true because of 
\begin{align}
    [\hat{H}, \hat{G}] &= 0, 
    \label{eq:commutator_H_G}
    \\
    [\hat{H}, \hat{L}_i] &= 0, \quad i=1,2,
    \label{eq:commutator_H_Li}
\end{align}
which can be checked straight-forwardly.
Moreover, we need to check the mutual compatibility between the three constraints, as discussed in the next subsection.

\subsection{Quantization of Chern-Simons Level}
\label{subsec:Hamiltonian_quantization}

To check the compatibility between the three constraints introduced in the previous subsection, we need to ensure that the operators  $e^{i\lambda\hat{G}}$, $e^{2\pi i \hat{L}_1}$, and $e^{2\pi i \hat{L}_2}$ commute with each other. The first operator commutes with the latter two for any $\lambda$ because 
\begin{align}
    [\hat{G}, \hat{L}_i] = 0, \quad i=1,2 \ .
\label{eq:commutator_G_Li}
\end{align}
The latter two, instead, do not commute with each other for arbitrary values of $k$, since
\begin{align}
    [\hat{L}_1, \hat{L}_2] = -\frac{k}{2\pi} i,
\label{eq:commutator_L1_L2}
\end{align}
and therefore
\begin{align}
    e^{2\pi i \hat{L}_1} e^{2\pi i \hat{L}_2} = e^{2\pi i k} e^{2\pi i \hat{L}_2} e^{2\pi i \hat{L}_1}.
    \label{algebra}
\end{align}

The compatibility of the constraints 2 and 3, which is necessary to obtain a non-trivial Hilbert space that contains more than the zero vector,
seems to enforce the quantization of the Chern-Simons coupling constant $k$ in integers:
\begin{align}
    e^{2\pi i k} &= 1  \\
    \implies  k &\in \mathbb{Z}.
\end{align}

The quantization of the Abelian Chern-Simons coupling constant 
$k$ garnered significant interest in the late 1980s and early 1990s. The goal of this section is to clarify that it is incorrect to assert that $k$ must be an integer. In fact, Chern-Simons theory is well-defined for any rational $k =p/q$ where $p$ and $q$ are coprime integers, as demonstrated by Polychronakos~\cite{poly,poly1}. 
This same result was  obtained on the lattice  by Semenoff and Eliezer~\cite{Eliezer:1992sq}, who showed that the theory remains consistent when the Chern-Simons level is taken to be a rational fraction and computed the ground-state degeneracy.
The consistency of the Chern-Simons theory with rational $k$
 was rigorously derived in the context of a Chern-Simons theory coupled to fermions on a torus in the continuum  by Iengo and Lechner~\cite{iengo:1991anyons}. Their analysis provided an exact demonstration of how rational values of 
$k$ naturally emerge in such systems and confirmed that the resulting physical properties align with those of an equivalent theory with an effective coupling $\bar{k} =pq$.
As we will show, also in the Maxwell-Chern-Simons case on the lattice, a theory  with a rational $k =p/q$,  is well defined and equivalent to  a theory with an effective coupling constant $\bar{k} =pq$.

For rational $k$, to obtain gauge-invariant states, we have to enlarge the Hilbert space in order to construct an irreducible representation of the non-commutative algebra in Eq.~\eqref{algebra}.
To this end, let us define the loop operators $h_1$ and $h_2$ as 
\begin{align}
  h_1 &= e^{2\pi i q \hat{L}_1}  , \\
   h_2 &= e^{2\pi i q \hat{L}_2},
    \label{modop}
\end{align}
which corresponds to changing the compactification radius, $2 \pi \rightarrow 2\pi / q$.
Here, $h_1$ and $h_2$ are Casimir operators of the algebra, which satisfy the relation
\begin{equation}
h_1 h_2 = h_2 h_1  e^{2\pi i \bar{k}} , \quad  \bar{k} =pq \in \mathbb{Z} \ ,
\label{rational}
\end{equation}
making constraints 2 and 3 compatible with a rational Chern-Simons coupling constant. 

Note that in our starting lattice action in Eq.~\eqref{eq:S_total}, we require $k\in 2\mathbb{Z}$ in order to make the action periodic in the gauge field variables and thus gauge invariant. Therefore, to be rigorous our Hamiltonian formulation only works for the even-$k$ Maxwell-Chern-Simons theory. Efforts of resolving the pure Chern-Simons lattice theory for any $k$ have been made in Ref.~\cite{Sulejmanpasic2024}. We leave to future work for the extension to odd-$k$ Maxwell-Chern-Simons lattice theory.

Nevertheless, an interesting observation can be made if we regardlessly put an odd $k$ value into our Hamiltonian formulation. The odd-$k$ Maxwell-Chern-Simons theory is a fermionic theory~\cite{Witten:2016spinstructure}. Surprisingly, we can probe the fermionic nature of the odd-$k$ theory in our formulation. Note that $e^{2\pi i \hat{L}_1} e^{2\pi i \hat{L}_2} = e^{\pi i k} e^{2\pi i (\hat{L}_1 + \hat{L}_2)}$. In an odd-$k$ theory, the three operators, $e^{2\pi i \hat{L}_1}$, $e^{2\pi i \hat{L}_2}$, and $e^{2\pi i (\hat{L}_1 + \hat{L}_2)}$, cannot be all imposed to be the identity consistently. More specifically, we have four choices of their values: $(+1, +1, -1)$, $(+1, -1, +1)$, $(-1, +1, +1)$, and $(-1, -1, -1)$. This reflects the fact that there is no standard choice of boundary conditions (spin structures) in a fermionic theory. For a fermionic theory on a torus, there are four spin structures. The result of the theory could depend on the choice of spin structure for an odd $k$ level~\cite{Witten:2016spinstructure}.

\subsection{Degeneracy of States}
\label{subsec:Hamiltonian_degeneracycristina}

One of the most interesting aspects of the Chern-Simons theory is its topological nature: pure Chern-Simons theories are metric independent and therefore independent of the geometry of the manifold on which they are defined, i.e., they depend only on the topology of the manifold. The ground state is degenerate on a manifold with non-trivial topology. On a manifold of genus $g$, the degeneracy is $k^g$ for integer $k$, and $(pq)^g$ for rational $k =p/q$~\cite{Tong:2018lecture}.
As we will show, the ground-state degeneracy is present also in the Maxwell-Chern-Simons theory.
In fact, this  degeneracy of the states can be directly seen from our construction of the constrained Hilbert space.

For integer $k$,  let us consider the following two loop operators:
\begin{align}
\label{eq:op}
   g_1^{1 / k } \quad \text{and} \quad  g_2^{1 / k },
\end{align}
where $g_1=e^{2\pi i \hat{L}_1}$ and $g_2=e^{2\pi i \hat{L}_2}$. These two operators are well defined on the constrained Hilbert space. They commute with the Gauss' law by Eq.~\eqref{eq:commutator_G_Li}. They also commute with  large gauge transformations, for example,
\begin{align}
    g_1^{1 / k } g_2 &= e^{2\pi i k / k} g_2 g_1^{1 / k} \nonumber \\
     &= g_2 g_1^{1 / k } ,
     \label{comrel}
\end{align}
i.e., if $|\psi\rangle$ is a physical state in the constrained Hilbert space, then $g_1^{1 / k } |\psi\rangle$ and $g_2^{1 / k } |\psi\rangle$ are also physical states in the constrained Hilbert space. Furthermore, the new physical states are different from $ |\psi\rangle$ because the two operators do not commute with each other,
\begin{equation}
   g_1^{1 / k }  g_2^{1 / k } = e^{2\pi i /  k} g_2^{1 / k }  g_1^{1 / k } ,
\end{equation}
unless $ k =1$.
Because $e^{i \theta_1}|\psi\rangle = g_1|\psi\rangle$,  where $\theta_1$ is a global phase defined in Eq.~\eqref{eq:theta_1}, 
by applying different powers of the $g_1^{1 / k }$ operator we can obtain $k$ different physical states, 
\begin{align}
    \{|\psi\rangle, g_1^{1 / k }|\psi\rangle, \cdots, g_1^{(k -1)1 / k }|\psi\rangle\}.
\label{eq:k_states}
\end{align}
These states span a $k$-dimensional subspace of the Hilbert space and can be distinguished by the $g_2^{1 / k }$ operator resulting in a $k$-fold degeneracy of the ground state for each state in the spectrum.

In a modern language, $g_1^{1 / k }$ is the generator of a discrete 1-form $\mathbb{Z}_k$ symmetry, and $g_2^{1 / k }$ is the generator of another discrete 1-form $\mathbb{Z}_k$ symmetry. The non-commuting nature of these two symmetries results in the $k$-fold degeneracy.

For rational $k= p/q$, let us define the operators:
\begin{align}
h_1^{1 / \bar{k} } \quad \text{and} \quad  h_2^{1 / \bar{k} },
\label{eq:oph}
\end{align}
with $h_1$ and $h_2$ defined in Eq.~\eqref{modop} and $\bar{k} =pq$. These two operators obey the same algebraic relations as  $g_1^{1 /k} \quad \text{and} \quad  g_2^{1 / k }$  with $\bar{k}$ replacing $k$. 
As for integer $k$, by applying different powers of the $h_1^{1 / \bar{k} }$ operator we can obtain $\bar{k} = pq$ different physical states, 
\begin{align}
    \{|\psi\rangle, h_1^{1 / \bar{k} }|\psi\rangle, \cdots, h_1^{(\bar{k} -1)1 / \bar{k} }|\psi\rangle \}\ ,
\end{align}
which span a $\bar{k}$-dimensional subspace of the Hilbert space and  which can be distinguished by the $h_2^{1 / \bar{k}} $ operator.  We have, thus, a $\bar{k} = pq$-fold degeneracy of the ground state for each state in the spectrum. This  confirms the degeneracy of states  obtained for the Maxwell-Chern-Simons theory in the continuum in Refs.~\cite{poly,poly1}, and, on the lattice, for the ground-state of pure Chern-Simons theories in~\cite{Eliezer:1992sq}. 

Note that the two operators defined in Eq.~\eqref{eq:op} commute with the Hamiltonian by Eq.~\eqref{eq:commutator_H_Li}. 
Therefore, if $|\psi\rangle$ is an energy eigenstate of the Hamiltonian, all $k$ distinguishable physical states in Eq.~\eqref{eq:k_states} have the same energy. In conclusion, there is a $k$-fold degeneracy for each state in the spectrum. The same holds for the two operators defined in Eq.~\eqref{eq:oph}, leading  to a $\bar{k} = pq$-fold degeneracy for each state in the spectrum.

In what follows, for simplicity, we will concentrate on the case of integer $k$. The generalization  to rational $k$ is straightforward.

\subsection{Wilson Loop Operator}
\label{sec:wilson}

In the earlier discussion of the degeneracy of states, we saw that the two loop operators in Eq.~\eqref{eq:op}
span the finite-dimensional Hilbert space. As usual, we can construct a Wilson loop operator 
\begin{equation}
    \hat{W} = \exp\left(i Q a \sum_{\text{loop}} \hat{A}\right),
\label{eq:W_loop}
\end{equation}
where $Q$ is the charge of the Wilson loop. The Wilson loop commutes with the gauge transformation operators $\hat{G}$, $e^{2\pi i \hat{L}_1}$, and $e^{2\pi i \hat{L}_2}$ if the charge $Q$ is an integer. In particular, the loop can be chosen to be non-contractible. We define two non-contractible Wilson loops
\begin{align}
    \hat{W}_1 = \exp\left(i Q a \sum_{x_1=0}^{N_1-1} \hat{A}_{(x_1,0);1}\right), \quad \hat{W}_2 = \exp\left(i Q a \sum_{x_2=0}^{N_2-1} \hat{A}_{(0,x_2);2}\right).
\end{align}

However, the Wilson loop operator $\hat{W}$ does not commute with the Hamiltonian. When inserting $\hat{W}$, we shift the ground state to another topological sector. Let us evaluate the commutator between a non-contractible Wilson loop operator 
and the operators $e^{2\pi i \hat{L}_1 / k }$, $e^{2\pi i \hat{L}_2 / k}$ defined in Eq.~\eqref{eq:op},
\begin{align}
\label{eq:W_1_commutation}
    \hat{W}_1 e^{2\pi i \hat{L}_1 / k } = e^{2\pi i \hat{L}_1 / k } \hat{W}_1, \quad 
    \hat{W}_1 e^{2\pi i \hat{L}_2 / k } = e^{-2\pi i Q / k} e^{2\pi i \hat{L}_2 / k } \hat{W}_1, \\
\label{eq:W_2_commutation}
    \hat{W}_2 e^{2\pi i \hat{L}_1 / k } = e^{-2\pi i Q / k} e^{2\pi i \hat{L}_1 / k } \hat{W}_2, \quad 
    \hat{W}_2 e^{2\pi i \hat{L}_2 / k } = e^{2\pi i \hat{L}_2 / k } \hat{W}_2.
\end{align}
We can see from these commutation relations that on the degenerate ground state subspace, a horizontal non-contractible Wilson loop $\hat{W}_1$ with charge $Q$ is equivalent to $e^{-2\pi i Q \hat{L}_1 / k }$, which has exactly same commutation relations as shown in Eq.~\eqref{eq:W_1_commutation}. A vertical non-contractible Wilson loop $\hat{W}_2$ is equivalent to $e^{2\pi i Q \hat{L}_2 / k }$, which has exactly same commutation relations as shown in Eq.~\eqref{eq:W_2_commutation}. In other words, the projections of $\hat{W}_1$, $\hat{W}_2$ on the degenerate ground state subspace is proportional to $\hat{W}_1^{\text{TOP}}=e^{-2\pi i Q \hat{L}_1 / k }$, $\hat{W}_2^{\text{TOP}}=e^{2\pi i Q \hat{L}_2 / k }$, respectively, up to a multiplicative renormalization factor.\footnote{The multiplicative renormalization factor is a term $e^{-mL}$ in the lowest order, where $m$ is the effective mass and $L$ is the length of the loop. In the following analyses, this factor stays as a constant in each scenario, so it has no effect in the analyses.} Here we use ``TOP'' to highlight the topological nature: The ground state sub-Hilbert space is also the subspace of the topological order, and the projected Wilson loops are also 1-form symmetry charge operators. We note that the Wilson loop here is different from the framed Wilson loop introduced in Ref.~\cite{Sulejmanpasic2023}, but we still capture the framing topological feature of the Wilson loop. This is because its projection on the degenerate ground state subspace, $\hat{W}_i^{\text{TOP}}=e^{\pm 2\pi i Q \hat{L}_i / k}$, has dangling links indicating the framing as shown in Eqs.~\eqref{eq:constraint_L1} and~\eqref{eq:constraint_L2}.
Furthermore, this ground state projection can be used to show the perimeter scaling law of the Wilson loop operator $\hat{W}$, as explained in the following.

\begin{figure}[t!]
    \centering
        \begin{tikzpicture}[baseline=-0.5ex]
        \tikzset{
        mid arrow/.style={
            decoration={markings, mark=at position 0.5 with {\arrow{>}}},
            postaction={decorate},
            thick
        },
        grid line/.style={thin, gray},
        }
        \foreach \x in {0,4} {
            \foreach \z in {0,4} {
                \draw[grid line] (\x,0,\z) -- (\x,5,\z);
            }
        }
        \foreach \y in {0,5} {
            \foreach \z in {0,4} {
                \draw[grid line] (0,\y,\z) -- (4,\y,\z);
            }
        }
        \foreach \y in {0,5} {
            \foreach \x in {0,4} {
                \draw[grid line] (\x,\y,0) -- (\x,\y,4);
            }
        }
        \draw[mid arrow] (0,1,2) -- (4,1,2);
        \draw[mid arrow] (4,3,2) -- (0,3,2);
        \node[above] at (2.5,1,2) {$\hat{W}_1$};
        \node[right] at (4,1,2) {$\tau_0$};
        \node[above] at (2.5,3,2) {$\hat{W}_1^{-1}$};
        \node[right] at (4,3,2) {$\tau_1$};
    \end{tikzpicture}
    \begin{tikzpicture}[baseline=-1.5cm]
        \draw[white] (-0.5,0) -- (0,0);
        \draw[thick,dashed,->] (0,0) -- (1,0);
        \draw[white] (1,0) -- (1.5,0);
    \end{tikzpicture}
    \begin{tikzpicture}[baseline=-0.5ex]
        \tikzset{
        mid arrow/.style={
            decoration={markings, mark=at position 0.5 with {\arrow{>}}},
            postaction={decorate},
            thick
        },
        grid line/.style={thin, gray},
        }
        \foreach \x in {0,4} {
            \foreach \z in {0,4} {
                \draw[grid line] (\x,0,\z) -- (\x,5,\z);
            }
        }
        \foreach \y in {0,5} {
            \foreach \z in {0,4} {
                \draw[grid line] (0,\y,\z) -- (4,\y,\z);
            }
        }
        \foreach \y in {0,5} {
            \foreach \x in {0,4} {
                \draw[grid line] (\x,\y,0) -- (\x,\y,4);
            }
        }
        \draw[mid arrow] (4,2,2) arc[start angle=0, end angle=180, x radius=2cm, y radius=1cm];
        \draw[mid arrow] (0,2,2) arc[start angle=180, end angle=360, x radius=2cm, y radius=1cm];
    \end{tikzpicture}
    \caption{An example of a contractible Wilson loop displaying the perimeter law. Note that with periodic boundary conditions, the two straight lines in the left panel represent two non-contractible Wilson loops $\hat{W}_1$ and $\hat{W}_1^{-1}$. Thus, they form one contractible loop. This implies that the left panel can be smoothly deformed into the right panel, which helps to visualize the topology.
    The upward direction is the positive time direction. Evaluating $\langle\hat{W}_1^{-1}(\tau_1)\hat{W}_1(\tau_0)\rangle$ gives a constant value independent of $|\tau_1-\tau_0|$ when $\tau_1 \gg \tau_0$, which implies the perimeter law. 
    } 
    \label{fig:perimeter_law}
\end{figure}
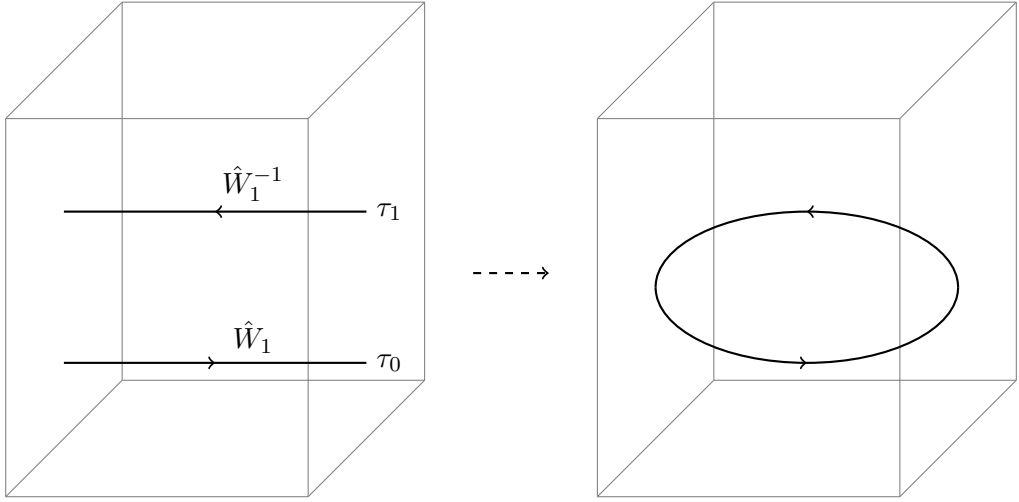

As in the continuum Maxwell-Chern-Simons theory, the Wilson loop operator always exhibits a perimeter law. For example, in Fig.~\ref{fig:perimeter_law}, we consider a contractible Wilson loop consisting of two non-contractible Wilson loops across the spatial boundary, $\hat{W}_1$ and $\hat{W}_1^{-1}$, with a temporal distance of $|\tau_1 - \tau_0|$ between each other. The area of the Wilson loop is $|\tau_1 - \tau_0|\cdot L$, which is linearly dependent on $|\tau_1 - \tau_0|$. Here, $L$ is the system size in the spatial direction. The perimeter of the Wilson loop is $2L$, which is independent of $|\tau_1 - \tau_0|$. We can distinguish the scaling law of the Wilson loop by looking at how its expectation value $\langle\hat{W}_1^{-1}(\tau_1)\hat{W}_1(\tau_0)\rangle$ scales with $|\tau_1 - \tau_0|$. If $\hat{W}_1(\tau_0)$ is applied to one of the ground states which is an eigenstate of $e^{2\pi i \hat{L}_2 / k}$, it shifts the ground state to another topological sector (i.e. another eigenstate of $e^{2\pi i \hat{L}_2 / k}$ but with a different eigenvalue) by Eq.~\eqref{eq:W_1_commutation} , and also excites the system to some higher energy state. For example, we start with one ground state and eigenstate of $e^{2\pi i \hat{L}_2 / k}$, $|\text{GS}_{\alpha}\rangle$. Suppose we get $\hat{W}_1(\tau_0)|\text{GS}_{\alpha}\rangle = |\text{GS}_{\beta}\rangle + \sum_i|E_i\rangle$, where $|\text{GS}_{\beta}\rangle$ is another ground state and eigenstate of $e^{2\pi i \hat{L}_2 / k}$, and $|E_i\rangle$ is the excited state with energy $E_i$ (all states unnormalized). At time $\tau_1$, the resulted state evolves into $|\text{GS}_{\beta}\rangle + \sum_i e^{-E_i(\tau_1-\tau_0)}|E_i\rangle$. Contract the time-evolved state with $\langle \text{GS}_{\alpha}|\hat{W}_1^{-1}(\tau_1)$ we get the expectation value $\langle\hat{W}_1^{-1}(\tau_1)\hat{W}_1(\tau_0)\rangle = \langle\text{GS}_{\beta}|\text{GS}_{\beta}\rangle + \sum_i e^{-E_i(\tau_1-\tau_0)} \langle E_i|E_i\rangle$.
Because the system is gapped, the excited state components decay upon time evolution. When $|\tau_1 - \tau_0|\to\infty$, the Wilson loop expectation value $\langle\hat{W}_1^{-1}(\tau_1)\hat{W}_1(\tau_0)\rangle$ converges to a constant value $\langle\text{GS}_{\beta}|\text{GS}_{\beta}\rangle$ which is independent of $|\tau_1 - \tau_0|$. Therefore, the Wilson loop expectation value agrees with the perimeter law.

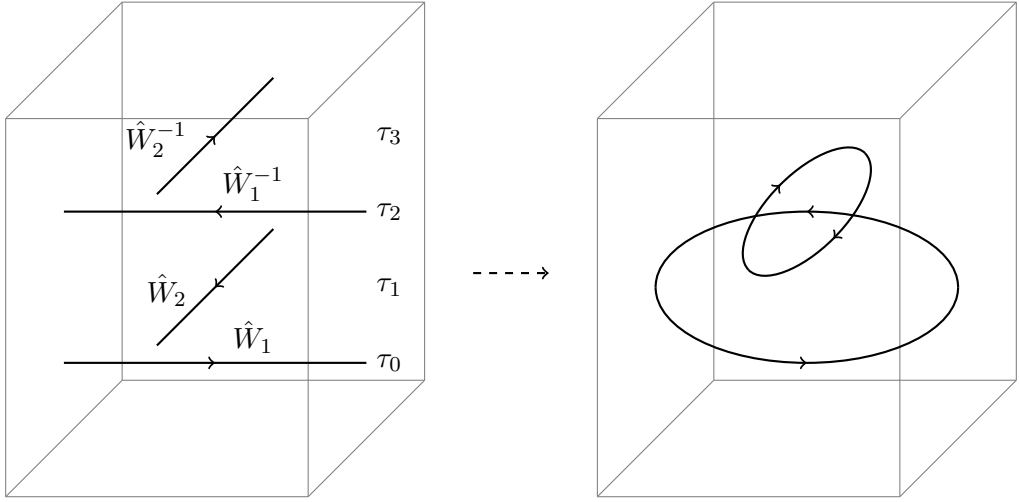
\begin{figure}[t!]
    \centering
        \begin{tikzpicture}[baseline=-0.5ex]
        \tikzset{
        mid arrow/.style={
            decoration={markings, mark=at position 0.5 with {\arrow{>}}},
            postaction={decorate},
            thick
        },
        grid line/.style={thin, gray},
        }
        \foreach \x in {0,4} {
            \foreach \z in {0,4} {
                \draw[grid line] (\x,0,\z) -- (\x,5,\z);
            }
        }
        \foreach \y in {0,5} {
            \foreach \z in {0,4} {
                \draw[grid line] (0,\y,\z) -- (4,\y,\z);
            }
        }
        \foreach \y in {0,5} {
            \foreach \x in {0,4} {
                \draw[grid line] (\x,\y,0) -- (\x,\y,4);
            }
        }
        \draw[mid arrow] (0,1,2) -- (4,1,2);
        \draw[mid arrow] (2,2,0) -- (2,2,4);
        \draw[mid arrow] (4,3,2) -- (0,3,2);
        \draw[mid arrow] (2,4,4) -- (2,4,0);
        \node[above] at (2.5,1,2) {$\hat{W}_1$};
        \node[left] at (1.8,2,2.1) {$\hat{W}_2$};
        \node[above] at (2.5,3,2) {$\hat{W}_1^{-1}$};
        \node[left] at (1.8,4,2.1) {$\hat{W}_2^{-1}$};
        \node[right] at (4,1,2) {$\tau_0$};
        \node[right] at (4,2,2) {$\tau_1$};
        \node[right] at (4,3,2) {$\tau_2$};
        \node[right] at (4,4,2) {$\tau_3$};
    \end{tikzpicture}
    \begin{tikzpicture}[baseline=-1.5cm]
        \draw[white] (-0.5,0) -- (0,0);
        \draw[thick,dashed,->] (0,0) -- (1,0);
        \draw[white] (1,0) -- (1.5,0);
    \end{tikzpicture}
    \begin{tikzpicture}[baseline=-0.5ex]
        \tikzset{
        mid arrow/.style={
            decoration={markings, mark=at position 0.5 with {\arrow{>}}},
            postaction={decorate},
            thick
        },
        grid line/.style={thin, gray},
        }
        \foreach \x in {0,4} {
            \foreach \z in {0,4} {
                \draw[grid line] (\x,0,\z) -- (\x,5,\z);
            }
        }
        \foreach \y in {0,5} {
            \foreach \z in {0,4} {
                \draw[grid line] (0,\y,\z) -- (4,\y,\z);
            }
        }
        \foreach \y in {0,5} {
            \foreach \x in {0,4} {
                \draw[grid line] (\x,\y,0) -- (\x,\y,4);
            }
        }
        \draw[mid arrow] (4,2,2) arc[start angle=0, end angle=180, x radius=2cm, y radius=1cm];
        \draw[mid arrow] (0,2,2) arc[start angle=180, end angle=360, x radius=2cm, y radius=1cm];
        \draw[mid arrow, rotate around={45:(2,3,0)}] (2,3,0) arc[start angle=360, end angle=180, x radius=1.09cm, y radius=0.5cm];
        \draw[mid arrow, rotate around={45:(2,3,4)}] (2,3,4) arc[start angle=180, end angle=0, x radius=1.09cm, y radius=0.5cm];
    \end{tikzpicture}
    \caption{Illustration of linked Wilson loops. The upward direction is the positive time direction. The pair of non-contractible Wilson loops $\hat{W}_1$ and $\hat{W}_1^{-1}$ together form a contractible loop. The pair of non-contractible Wilson loops $\hat{W}_2$ and $\hat{W}_2^{-1}$ together form another contractible loop. These two contractible loops are linked with each other and have a non-zero linking number between them. The left panel can be smoothly deformed into the right panel, which helps to visualize the topology. Evaluating $\hat{W}_2^{-1}(\tau_3)\hat{W}_1^{-1}(\tau_2)\hat{W}_2(\tau_1)\hat{W}_1(\tau_0)$ on the ground state Hilbert space will give a non-trivial topological phase factor of $e^{2\pi i Q_1 Q_2/k}$, where $k$ is the Chern-Simons level, $Q_1$ is the charge of $\hat{W}_1$, and $Q_2$ is the charge of $\hat{W}_2$. This non-trivial topological phase is due to the non-zero linking number between the two loops.} 
    \label{fig:chained}
\end{figure}

In addition to the perimeter scaling law, the Wilson loop operators also display non-trivial topological features.
For example, we can consider two linked Wilson loops, as shown in Fig.~\ref{fig:chained}. The pair of non-contractible loops $\hat{W}_1$ and $\hat{W}_1^{-1}$ form one contractible Wilson loop, and the other pair of non-contractible loops $\hat{W}_2$ and $\hat{W}_2^{-1}$ form a second contractible Wilson loop. The two contractible Wilson loops are linked with each other. To unlink the two loops, we can exchange $\hat{W}_2$ and $\hat{W}_1^{-1}$, which are perpendicular to each other. If $\hat{W}_1$ has charge $Q_1$ and $\hat{W}_2$ has charge $Q_2$, this exchange generates a commutator with a topological phase factor $e^{2\pi i Q_1 Q_2/k}$, as shown in the following:
\begin{align}
    \hat{W}_2^{-1}(\tau_3) \hat{W}_1^{-1}(\tau_2) \hat{W}_2(\tau_1) \hat{W}_1(\tau_0) 
    &\sim e^{2\pi i Q_1 Q_2/k}\, \hat{W}_2^{-1}(\tau_3) \hat{W}_2(\tau_2) \hat{W}_1^{-1}(\tau_1) \hat{W}_1(\tau_0) \nonumber \\
    &\sim e^{2\pi i Q_1 Q_2/k},
\end{align}
where the times $\tau_0<\tau_1<\tau_2<\tau_3$ are well separated, $k$ is the Chern-Simons level, and in the last step $\hat{W}_2^{-1}(\tau_3)$ almost cancels with $\hat{W}_2(\tau_2)$, $\hat{W}_1^{-1}(\tau_1)$ almost cancels with $\hat{W}_1(\tau_0)$, by a similar argument as above when we discussed the perimeter law. 
Therefore, this example demonstrates that two linked Wilson loops with linking number 1 have a non-trivial ground state expectation value with a topological phase factor of $e^{2\pi i Q_1 Q_2/k}$. 

\subsection{Anyon Statistics}
\label{sec:anyon}

In Sec.~\ref{sec:wilson}, we have seen that our formulation supports Wilson loops, which display topological features. Here, we extend the discussion to open Wilson lines, which means that in Eq.~\eqref{eq:W_loop} we integrate the gauge field on an open line instead of a loop, i.e., 
\begin{equation}
\hat{W}_\text{open}=\exp\left(iQa\sum_\text{line}\hat{A}\right).
\end{equation}
In order to have $\hat{W}_\text{open}$ commute with the large gauge transformations, $e^{2\pi i \hat{L}_1}$ and $e^{2\pi i \hat{L}_2}$, we need $Q$ to be an integer, which is called the charge of the open Wilson line. Although a single open Wilson line is not locally gauge invariant (i.e., the Gauss' law in Eq.~\eqref{eq:Gauss_law} does not commute with the open Wilson line at its two ends), the combination of an open Wilson line $\hat{W}_\text{open}$ with its inverse $\hat{W}_\text{open}^{-1}$ is gauge invariant, because the gauge non-invariant parts at their ends cancel with each other. For example, the combined operator $\hat{W}_{\text{open},1}\hat{W}_{\text{open},2}\hat{W}_{\text{open},1}^{-1}\hat{W}_{\text{open},2}^{-1}$ is gauge invariant: After a gauge transformation, the four open Wilson line operators generate $c$-number phases, and those phases generated by the operators and by their inverse operators cancel, respectively.

We are interested in open Wilson lines because they can be viewed as the world lines of anyons, which are excitations with fractional statistics~\cite{Leinaas:1977anyon,wilczek:1982anyon}. Although our formulation does not involve any matter fields, we can still probe the mutual statistics and self statistics of the anyons through the open Wilson lines. We briefly demonstrate the ideas below. The operators $\hat{W}$ in this subsection all refer to open Wilson lines, unless otherwise stated.

\subsubsection{Mutual Statistics}

\begin{figure} [t]
    \centering
    \begin{tikzpicture}
        \tikzset{
        mid arrow 1/.style={
            decoration={markings, mark=at position 0.5 with {\arrow{>}}},
            postaction={decorate},
            thick, red
        },
        line 1/.style={thick, red},
        mid arrow 2/.style={
            decoration={markings, mark=at position 0.5 with {\arrow{>}}},
            postaction={decorate},
            thick, blue
        },
        line 2/.style={thick, blue},
        }
        \draw[mid arrow 1] (1.5,1.5) -- (5,1.5);
        \node[above] at (3.5,1.5) {$\hat{W}_1$};
    
        \draw[mid arrow 2] (0,1.5) -- (0,0);
        \node[right] at (0,1) {$\hat{W}_2$};
        
        \draw[line 2] (0,0) -- (0,3) -- (3,3) -- (3,0) -- cycle;
    
        \fill (0,1.5) circle (2pt);
        \draw (1.5,1.5) node {x};
        \draw (5,1.5) node {x};

        \foreach \xx in {9} {
            \foreach \yy in {1.5} {
                \foreach \s in {0.05} {
                    \foreach \x in {0,0.5,1,1.5,2,2.5} {
                        \draw[line 1, dashed] (\x-1.5+\xx,-0.5+\yy) -- (\x-1.5+\xx,0+\yy);
                        \draw[line 1] (\x-1.5+\xx,0+\yy) -- (\x-1+\xx,0+\yy);
                        \draw[line 2, dashed] (-0.5+\xx+\s,\x-1.5+\yy+\s) -- (0+\xx+\s,\x-1.5+\yy+\s);
                        \draw[line 2] (0+\xx+\s, \x-1.5+\yy+\s) -- (0+\xx+\s,\x -1+\yy+\s);
                    }
                }
                \node[above] at (1+\xx,0+\yy) {$\hat{W}_1^{\text{TOP}}$};
                \node[left] at (-0.5+\xx,1+\yy) {$\hat{W}_2^{\text{TOP}}$};
            }
        }
    
    \end{tikzpicture}
    \caption[Illustration of mutual statistics between anyons]{ Illustration of mutual statistics between anyons. In the left panel, $\hat{W}_1$ is an open Wilson line with charge $Q_1$, and $\hat{W}_2$ is a Wilson loop with charge $Q_2$. We illustrate the process of a charge-$Q_2$ anyon circling around a charge-$Q_1$ anyon as follows~\cite{Levin:2003anyon}. Suppose that initially there is a charge-$Q_2$ anyon (illustrated by the black dot). Then, we first create a charge-$Q_1$ anyon-anti-anyon pair located at two ends (``x'') with the open Wilson line $\hat{W}_1$ (red line).  Second, we move the charge-$Q_2$ anyon around the charge-$Q_1$ anyon with $\hat{W}_2$ (blue line). Third, we annihilate the charge-$Q_1$ anyon-anti-anyon pair with $\hat{W}_1^{-1}$. Finally, we move back the charge-$Q_2$ anyon with $\hat{W}_2^{-1}$ to cancel the dynamical phase created by its movement.
    In the right panel, we zoom into the left panel to look at the intersection of the red and blue lines. $\hat{W}_1^{\text{TOP}}$ and $\hat{W}_2^{\text{TOP}}$ are the projections of the operators $\hat{W}_1$ and $\hat{W}_2$ on the degenerate ground-state Hilbert space, respectively. From the microscopic details of $\hat{W}_1^{\text{TOP}}$ and $\hat{W}_2^{\text{TOP}}$ on the lattice (i.e., the solid links are proportional to the gauge field $\hat{A}$, the dashed links are proportional to the canonical conjugate momentum $\hat{p}$, similar to $\hat{L}_i$ in Eqs.~\eqref{eq:constraint_L1} and~\eqref{eq:constraint_L2}), we can see that $\hat{W}_1^{\text{TOP}}$ and $\hat{W}_2^{\text{TOP}}$ do not commute with each other due to the fact that they share common links.
    }
    \label{fig:anyon_mutual}
\end{figure}
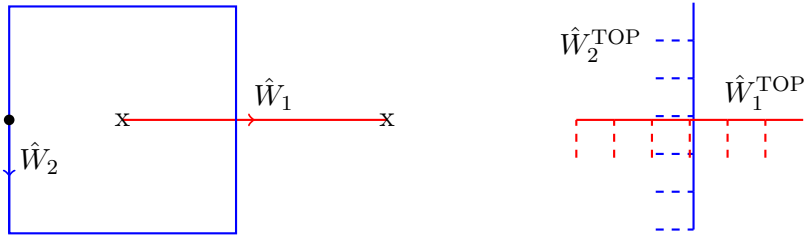

First, we look at the mutual statistics between two types of anyons, one with charge $Q_1$ and the other with charge $Q_2$. To probe their mutual statistics, we want to move one anyon around the other and compare the phases between the initial and final states. We can move the charge-$Q_1$ anyon by applying a Wilson line $\hat{W}_1$ with charge $Q_1$, and the charge-$Q_2$ anyon by applying a Wilson line $\hat{W}_2$ with charge $Q_2$. To make the movements, we apply the operator
\begin{equation}
    \hat{W}_2^{-1}(\tau_4) \hat{W}_1^{-1}(\tau_3) \hat{W}_2(\tau_2) \hat{W}_1(\tau_1) ,
    \label{eq:opanyon}
\end{equation}
to a state with a charge-$Q_2$ anyon present, see Fig.~\ref{fig:anyon_mutual} and Ref.~\cite{Levin:2003anyon}. Here, we consider the times  $\tau_1<\tau_2<\tau_3<\tau_4$, and the spacings between these times are large enough so that we only need to consider the ground-state properties of the system (i.e., the system is gapped and correlations from the the excited states are exponentially suppressed for long times). The presence of the initial charge-$Q_2$ anyon can be created by some Wilson line on the ground state at $\tau\to-\infty$, and we can annihilate it by the inverse Wilson line at $\tau\to\infty$ to go back to the degenerate ground-state Hilbert space. In the following discussion, we will neglect the creation and annihilation of this initial charge-$Q_2$ anyon, and only briefly discuss it in footnote 1.

As shown in the left panel of Fig.~\ref{fig:anyon_mutual}, we probe the mutual statistics of the anyons by applying the operator $\hat{W}_2^{-1}(\tau_4) \hat{W}_1^{-1}(\tau_3) \hat{W}_2(\tau_2) \hat{W}_1(\tau_1)$ from Eq.~\eqref{eq:opanyon}. First, we create a pair of anyon and anti-anyon (illustrated by the black x's) of charge $Q_1$ with the open Wilson line $\hat{W}_1(\tau_1)$ (illustrated by the red line). Second, we circle the existing charge-$Q_2$ anyon (illustrated by the black dot) around the charge-$Q_1$ anyon with $\hat{W}_2(\tau_2)$ (illustrated by the blue line). Third, we annihilate the charge-$Q_1$ anyon-anti-anyon pair with $\hat{W}_1^{-1}(\tau_3)$. Finally, we cancel the dynamical phase of the charge-$Q_2$ anyon with $\hat{W}_2^{-1}(\tau_4)$. The net effect of these operations results in a charge-$Q_2$ anyon circling around a charge-$Q_1$ anyon. Therefore, the ground state expectation value $\langle\hat{W}_2^{-1}(\tau_4) \hat{W}_1^{-1}(\tau_3) \hat{W}_2(\tau_2) \hat{W}_1(\tau_1)\rangle$ has a topological phase, showing the mutual statistics between charge-$Q_1$ and charge-$Q_2$ anyons.\footnote{More precisely, we compare the phase between $\langle\hat{W}_0^{-1}(+\infty) \hat{W}_2^{-1}(\tau_4) \hat{W}_1^{-1}(\tau_3) \hat{W}_2(\tau_2) \hat{W}_1(\tau_1) \hat{W}_0(-\infty) \rangle$ and $\langle\hat{W}_0^{-1}(+\infty) \hat{W}_2^{-1}(\tau_4) \hat{W}_2(\tau_3) \hat{W}_1^{-1}(\tau_2) \hat{W}_1(\tau_1) \hat{W}_0(-\infty) \rangle$, where $\hat{W}_0(-\infty)$ prepares the initial charge-$Q_2$ anyon and $\hat{W}_0^{-1}(+\infty)$ annihilate it back to the ground state. Note that in the latter process each operator cancels with its inverse, and thus the phase is zero. Therefore, this relative phase that shows the mutual statistics of the anyons is equivalent to the phase of $\langle\hat{W}_2^{-1}(\tau_4) \hat{W}_1^{-1}(\tau_3) \hat{W}_2(\tau_2) \hat{W}_1(\tau_1)\rangle$.}

Next, we compute the ground state expectation value $\langle\hat{W}_2^{-1}(\tau_4) \hat{W}_1^{-1}(\tau_3) \hat{W}_2(\tau_2) \hat{W}_1(\tau_1)\rangle$. For this, we consider a charge-$Q$ open Wilson line
\begin{align}
    \hat{W} = \exp\left(i Q a \sum_{(x,i)} \hat{A}_{x;i}\right),
\label{eq:anyon_W}
\end{align}
and a ``dangling Wilson line''
\begin{align}
    \quad \hat{W}^{\text{TOP}} 
    &= \exp\left[\frac{2\pi i Q}{k} \sum_{(x,i)} \left(\frac{ka}{4\pi}\hat{A}_{x;i}-\epsilon_{ij}\frac{1}{a}\hat{p}_{x-\hat{j};j}\right)\right]  \nonumber \\
    &= \exp\left[i Q \sum_{(x,i)} \left(\frac{a}{2}\hat{A}_{x;i}-\epsilon_{ij}\frac{2\pi}{k a}\hat{p}_{x-\hat{j};j}\right)\right],
\label{eq:anyon_W_GS}
\end{align}
which is a truncated line segment of the loop operator $e^{\pm 2\pi i Q \hat{L}_i/ k}$, where $\hat{L}_i$ are the large gauge transformation operators in Eqs.~\eqref{eq:constraint_L1} and~\eqref{eq:constraint_L2}. Here,  $(x,i)\in\{\text{(site,direction)}\}$ represent the sites in a line and the corresponding directions along the line. For visualization, see the correspondence between the left and right panels in Figs.~\ref{fig:anyon_mutual} and~\ref{fig:anyon_self}, where the solid links contain the gauge field $\hat{A}$ and the dashed links contain the canonical conjugate momentum $\hat{p}$. The relative signs between $\hat{A}$ and $\hat{p}$ in $\hat{W}^{\text{TOP}}$ follow the ones in $\hat{L}_i$ in Eqs.~\eqref{eq:constraint_L1} and~\eqref{eq:constraint_L2}.

When the temporal spacings are large, we can compute $\langle\hat{W}_2^{-1}(\tau_4) \hat{W}_1^{-1}(\tau_3) \hat{W}_2(\tau_2) \hat{W}_1(\tau_1)\rangle$ by projecting everything onto the degenerate ground-state subspace. By an argument similar to the one we used in the previous discussion on the perimeter law in Sec.~\ref{sec:wilson}, $\hat{W}$ is equivalent to its projection $\hat{W}^{\text{TOP}}$ on the degenerate ground-state subspace when its temporal spacings between other operators are large enough, which ensures that the excited-state components exponentially decay out. Although we have only showed this with Wilson loops, but not with open Wilson lines, we assume the proportional relation also holds for open Wilson lines, i.e.,
\begin{align}
    \langle\hat{W}_2^{-1}(\tau_4) \hat{W}_1^{-1}(\tau_3) \hat{W}_2(\tau_2) \hat{W}_1(\tau_1)\rangle 
    \sim \langle\hat{W}_2^{\text{TOP}^{-1}}\hat{W}_1^{\text{TOP}^{-1}} \hat{W}_2^{\text{TOP}} \hat{W}_1^{\text{TOP}}\rangle .
\label{eq:W_W_GS}
\end{align}
We note that the operators in both sides of Eq.~\eqref{eq:W_W_GS} are gauge invariant due to the gauge phase cancellation between the open Wilson lines and their corresponding inverse operators.

The right hand side of Eq.~\eqref{eq:W_W_GS} can be directly computed:
\begin{align}
    \langle\hat{W}_2^{\text{TOP}^{-1}}\hat{W}_1^{\text{TOP}^{-1}} \hat{W}_2^{\text{TOP}} \hat{W}_1^{\text{TOP}}\rangle 
    &= e^{\frac{2\pi i}{k}Q_1 Q_2} \langle\hat{W}_2^{\text{TOP}^{-1}} \hat{W}_2^{\text{TOP}} \hat{W}_1^{\text{TOP}^{-1}} \hat{W}_1^{\text{TOP}}\rangle \nonumber \\
    &= e^{\frac{2\pi i}{k}Q_1 Q_2},
\label{eq:W_W_GS_2}
\end{align}
where we used the commutation relation 
\begin{align}
    \hat{W}_1^{\text{TOP}^{-1}} \hat{W}_2^{\text{TOP}} = e^{\frac{2\pi i}{k}Q_1 Q_2} \hat{W}_2^{\text{TOP}} \hat{W}_1^{\text{TOP}^{-1}},
\end{align}
which can be seen by the overlapping red solid link ($\textcolor{red}{\hat{A}_{x-\hat{1};1}}$) and blue dashed link ($\textcolor{blue}{\hat{p}_{x-\hat{1};1}}$) with the commutator $[\textcolor{red}{\hat{A}_{x-\hat{1};1}},\textcolor{blue}{\hat{p}_{x-\hat{1};1}}]=i$, and the blue solid link ($\textcolor{blue}{\hat{A}_{x-\hat{2};2}}$) and the red dashed link ($\textcolor{red}{\hat{p}_{x-\hat{2};2}}$) with the commutator $[\textcolor{blue}{\hat{A}_{x-\hat{2};2}},\textcolor{red}{\hat{p}_{x-\hat{2};2}}]=i$, where the red variables are in $\hat{W}_1^{\text{TOP}^{-1}}$ and the blue variables are in $\hat{W}_2^{\text{TOP}}$, and $x$ is the intersection site of the red and blue solid lines, shown in the right panel in Fig.~\ref{fig:anyon_mutual}. In particular, the phase comes from
\begin{align}
    & \exp\left[- i Q_1 \left(\frac{a}{2}\textcolor{red}{\hat{A}_{x-\hat{1};1}}-\frac{2\pi}{k a}\textcolor{red}{\hat{p}_{x-\hat{2};2}}\right)\right] \exp\left[i Q_2 \left(\frac{a}{2}\textcolor{blue}{\hat{A}_{x-\hat{2};2}}+\frac{2\pi}{k a}\textcolor{blue}{\hat{p}_{x-\hat{1};1}}\right)\right] \nonumber \\
    =&\ e^{\frac{2\pi i}{k}Q_1 Q_2} \exp\left[i Q_2 \left(\frac{a}{2}\textcolor{blue}{\hat{A}_{x-\hat{2};2}}+\frac{2\pi}{k a}\textcolor{blue}{\hat{p}_{x-\hat{1};1}}\right)\right] \exp\left[- i Q_1 \left(\frac{a}{2}\textcolor{red}{\hat{A}_{x-\hat{1};1}}-\frac{2\pi}{k a}\textcolor{red}{\hat{p}_{x-\hat{2};2}}\right)\right],
\end{align}
due to the commutator
\begin{align}
    \left[- i Q_1 \left(\frac{a}{2}\textcolor{red}{\hat{A}_{x-\hat{1};1}}-\frac{2\pi}{k a}\textcolor{red}{\hat{p}_{x-\hat{2};2}}\right),\ i Q_2 \left(\frac{a}{2}\textcolor{blue}{\hat{A}_{x-\hat{2};2}}+\frac{2\pi}{k a}\textcolor{blue}{\hat{p}_{x-\hat{1};1}}\right)\right] = \frac{2\pi i}{k}Q_1 Q_2,
\end{align}
where the minus sign in the first term comes from the inverse ($-1$) in $\hat{W}_1^{\text{TOP}^{-1}}$.

In summary, we have computed that $\langle\hat{W}_2^{-1}(\tau_4) \hat{W}_1^{-1}(\tau_3) \hat{W}_2(\tau_2) \hat{W}_1(\tau_1)\rangle$ has a non-trivial topological phase $e^{\frac{2\pi i}{k}Q_1 Q_2}$, which demonstrates the mutual statistics between charge-$Q_1$ and charge-$Q_2$ anyons.

\subsubsection{Self Statistics}

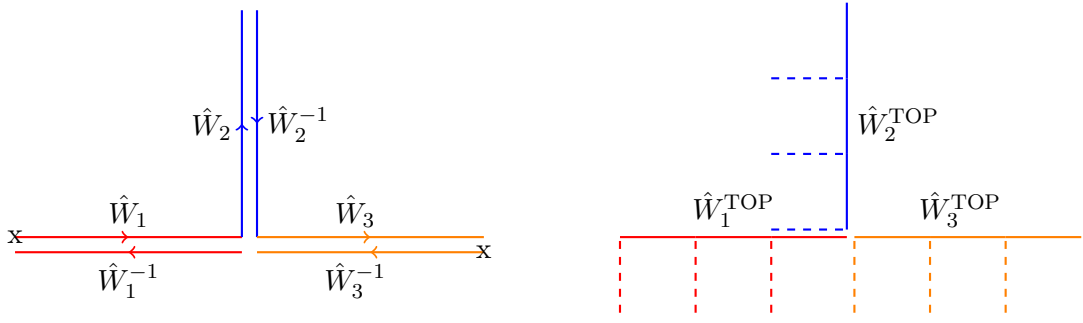
\begin{figure} [t]
    \centering
    \begin{tikzpicture}[baseline=-0.5ex]
        \tikzset{
        mid arrow 1/.style={
            decoration={markings, mark=at position 0.5 with {\arrow{>}}},
            postaction={decorate},
            thick, red
        },
        line 1/.style={thick, red},
        mid arrow 2/.style={
            decoration={markings, mark=at position 0.5 with {\arrow{>}}},
            postaction={decorate},
            thick, blue
        },
        line 2/.style={thick, blue},
        mid arrow 3/.style={
            decoration={markings, mark=at position 0.5 with {\arrow{>}}},
            postaction={decorate},
            thick, orange
        },
        line 3/.style={thick, orange},
        }


        \foreach \x in {-8} {
        \foreach \yy in {0.2} {
            \draw[mid arrow 1] (\x-3,0) -- (\x,0);
            \draw[mid arrow 1] (\x,-\yy) -- (\x-3,-\yy);
            \draw[mid arrow 2] (\x,0) -- (\x,3);
            \draw[mid arrow 2] (\x+\yy,3) -- (\x+\yy,0);
            \draw[mid arrow 3] (\x+\yy,0) -- (\x+\yy+3,0);
            \draw[mid arrow 3] (\x+\yy+3,-\yy) -- (\x+\yy,-\yy);
            \node[above] at (\x-1.5,0) {$\hat{W}_1$};
            \node[below] at (\x-1.5,-\yy) {$\hat{W}_1^{-1}$};
            \node[left] at (\x,1.5) {$\hat{W}_2$};
            \node[right] at (\x+\yy,1.5) {$\hat{W}_2^{-1}$};
            \node[above] at (\x+1.5,0) {$\hat{W}_3$};
            \node[below] at (\x+1.5,-\yy) {$\hat{W}_3^{-1}$};
            \draw (\x-3,0) node {x};
            \draw (\x+\yy+3,-\yy) node {x};
        }}
        
        \foreach \x in {0,1,2} {
            \draw[line 1, dashed] (\x-3,-1) -- (\x-3,0);
            \draw[line 1] (\x-3,0) -- (\x-2,0);
            \draw[line 2, dashed] (-1,\x+0.1) -- (0,\x+0.1);
            \draw[line 2] (0, \x+0.1) -- (0,\x + 1+0.1);
            \draw[line 3, dashed] (\x+0.1,-1) -- (\x+0.1,0);
            \draw[line 3] (\x+0.1,0) -- (\x+1+0.1,0);
        }

        \foreach \x in {0} {
            \node[above] at (\x-1.5,0) {$\hat{W}_1^{\text{TOP}}$};
            \node[right] at (\x,1.5) {$\hat{W}_2^{\text{TOP}}$};
            \node[above] at (\x+1.5,0) {$\hat{W}_3^{\text{TOP}}$};
        }
    \end{tikzpicture}
    \caption[Illustration of self statistics of one type of anyon]{Illustration of self statistics of one type of anyon. In the left panel, $\hat{W}_1$, $\hat{W}_2$, and $\hat{W}_3$ are three open Wilson lines with charge $Q$. We illustrate the process of two charge-$Q$ anyons exchanging their positions as follows~\cite{Kawagoe:2020anyon}. Suppose initially there are two identical anyons located at the ``x''s. We first move the left anyon to the middle by $\hat{W}_1$ (red line), second we move it upward by $\hat{W}_2$ (blue line), third we move the right anyon to the middle by $\hat{W}_3^{-1}$ (orange line), fourth we move it leftward by $\hat{W}_1^{-1}$ (orange line), fifth we move the original left anyon back to middle by $\hat{W}_2^{-1}$ (blue line), and finally we move it rightward by $\hat{W}_3$ (red line). Note that the displacements between parallel lines in the figure are for visual clarity. We have exchanged the position of the two anyons without them getting close to each other, and all dynamical phases created by anyon movements are cancelled.
    In the right panel, we zoom in to look at the intersection of the red, blue, and orange lines. $\hat{W}_1^{\text{TOP}}$, $\hat{W}_2^{\text{TOP}}$, and $\hat{W}_3^{\text{TOP}}$ are the projections of the operators $\hat{W}_1$, $\hat{W}_2$, and $\hat{W}_3$ on the degenerate ground-state Hilbert space, respectively. From the microscopic details of $\hat{W}_1^{\text{TOP}}$, $\hat{W}_2^{\text{TOP}}$, and $\hat{W}_3^{\text{TOP}}$ on the lattice (solid links are proportional to the gauge field $\hat{A}$, dashed links are proportional to the canonical conjugate momentum $\hat{p}$, similar to $\hat{L}_i$ in Eqs.~\eqref{eq:constraint_L1} and~\eqref{eq:constraint_L2}), we can see that $\hat{W}_1^{\text{TOP}}$ and $\hat{W}_2^{\text{TOP}}$ do not commute with each other due to the fact that they share a common link.}
    \label{fig:anyon_self}
\end{figure}

Next, we study the anyon self statistics using open Wilson lines. To probe the self statistics, we want to exchange the position of two anyons with the same charge $Q$ and compare the phase between the initial and final states. To make the movements, we apply the composite operator
\begin{equation}
\hat{W}_3(\tau_6)\hat{W}_2^{-1}(\tau_5)\hat{W}_1^{-1}(\tau_4)\hat{W}_3^{-1}(\tau_3)\hat{W}_2(\tau_2)\hat{W}_1(\tau_1),
\end{equation}
to a state with two charge-$Q$ anyons present at the end of the open Wilson lines $\hat{W}_1$ and $\hat{W}_3$, see Fig.~\ref{fig:anyon_self} and Ref.~\cite{Kawagoe:2020anyon}. As before, we consider times $\tau_1<\tau_2<\tau_3<\tau_4<\tau_5<\tau_6$, and the spacing between these times are large enough so that we only need to consider the ground-state properties. Similar to the earlier discussion on the mutual statistics, the creation and annihilation of the anyons are performed at $\tau\to-\infty$ and $\tau\to\infty$, respectively, and they are not important to our discussion.

As shown in Fig.~\ref{fig:anyon_self}, we apply $\hat{W}_3(\tau_6)\hat{W}_2^{-1}(\tau_5)\hat{W}_1^{-1}(\tau_4)\hat{W}_3^{-1}(\tau_3)\hat{W}_2(\tau_2)\hat{W}_1(\tau_1)$ to operate the anyons as follows: Initially, we have two anyons locating at the two crosses respectively, and we want to exchange their location without having them close to each other. First, we move the anyon on the left to the middle with the open Wilson line $\hat{W}_1(\tau_1)$. Second, we move it upward with $\hat{W}_2(\tau_2)$, in order to make space for the other anyon. Third, we move the other anyon on the right to the middle with $\hat{W}_3^{-1}(\tau_3)$. Forth, we move it to the left cross with $\hat{W}_1^{-1}(\tau_4)$. Fifth, we move the original left anyon downward back to the middle with $\hat{W}_2^{-1}(\tau_5)$. Finally, we move it to the right cross with $\hat{W}_3(\tau_6)$. With these six steps, we exchange the locations of the two anyons.

Therefore, the expectation value $\langle\hat{W}_3(\tau_6)\hat{W}_2^{-1}(\tau_5)\hat{W}_1^{-1}(\tau_4)\hat{W}_3^{-1}(\tau_3)\hat{W}_2(\tau_2)\hat{W}_1(\tau_1)\rangle$ comparing to the trivial process $\langle\hat{W}_1^{-1}(\tau_6)\hat{W}_2^{-1}(\tau_5)\hat{W}_3(\tau_4)\hat{W}_3^{-1}(\tau_3)\hat{W}_2(\tau_2)\hat{W}_1(\tau_1)\rangle$ has a topological phase showing the self statistics between two charge-$Q$ anyons. The former anyon-exchanged ground state expectation value can be computed similarly with the projections of the open Wilson lines on the degenerate ground state subspace. On the right pannel in Fig.~\ref{fig:anyon_self}, we show the corresponding projections $\hat{W}_i^{\text{TOP}}$ at the junction point of the three open Wilson lines. The formula of the open Wilson lines and their projections are written in Eqs.~\eqref{eq:anyon_W} and~\eqref{eq:anyon_W_GS}, respectively.

When the temporal spacings are large, we can compute the ground state expectation value $\langle\hat{W}_3(\tau_6)\hat{W}_2^{-1}(\tau_5)\hat{W}_1^{-1}(\tau_4)\hat{W}_3^{-1}(\tau_3)\hat{W}_2(\tau_2)\hat{W}_1(\tau_1)\rangle$ by projecting everything onto the degenerate ground state subspace, similar to what we did above in the discussion on the mutual statistics in Eqs.~\eqref{eq:W_W_GS} and~\eqref{eq:W_W_GS_2}. We have
\begin{align}
    &\langle\hat{W}_3(\tau_6)\hat{W}_2^{-1}(\tau_5)\hat{W}_1^{-1}(\tau_4)\hat{W}_3^{-1}(\tau_3)\hat{W}_2(\tau_2)\hat{W}_1(\tau_1)\rangle \nonumber \\
    &\sim \langle\hat{W}_3^{\text{TOP}}\hat{W}_2^{\text{TOP}^{-1}}\hat{W}_1^{\text{TOP}^{-1}}\hat{W}_3^{\text{TOP}^{-1}}\hat{W}_2^{\text{TOP}}\hat{W}_1^{\text{TOP}}\rangle  \nonumber \\
    &\sim e^{\frac{\pi i}{k}Q^2} \langle\hat{W}_1^{\text{TOP}^{-1}}\hat{W}_2^{\text{TOP}^{-1}}\hat{W}_3^{\text{TOP}}\hat{W}_3^{\text{TOP}^{-1}}\hat{W}_2^{\text{TOP}}\hat{W}_1^{\text{TOP}}\rangle  \nonumber \\
    &\sim e^{\frac{\pi i}{k}Q^2}, 
\end{align}
where we use the fact that $\hat{W}_3^{\text{TOP}}$ commutes with the other open Wilson lines' projections because of no overlapping links (see the right panel in Fig.~\ref{fig:anyon_self}), and we use the commutation relation
\begin{align}
    \hat{W}_2^{\text{TOP}^{-1}}\hat{W}_1^{\text{TOP}^{-1}} = e^{\frac{\pi i}{k}Q^2} \hat{W}_1^{\text{TOP}^{-1}}\hat{W}_2^{\text{TOP}^{-1}},
\end{align}
which can be seen by the overlapping red solid link ($\textcolor{red}{\hat{A}_{x-\hat{1};1}}$) and blue dashed link ($\textcolor{blue}{\hat{p}_{x-\hat{1};1}}$) with commutator $[\textcolor{red}{\hat{A}_{x-\hat{1};1}},\textcolor{blue}{\hat{p}_{x-\hat{1};1}}]=i$, where red variables are in $\hat{W}_1^{\text{TOP}^{-1}}$ and blue variables are in $\hat{W}_2^{\text{TOP}^{-1}}$, $x$ is the junction site of the three solid lines, shown in the right panel in Fig.~\ref{fig:anyon_self}. In particular, the phase comes from
\begin{align}
    \exp\left[-i Q \frac{2\pi}{k a}\textcolor{blue}{\hat{p}_{x-\hat{1};1}}\right] \exp\left[- i Q \frac{a}{2}\textcolor{red}{\hat{A}_{x-\hat{1};1}}\right] = e^{\frac{\pi i}{k}Q^2} \exp\left[- i Q \frac{a}{2}\textcolor{red}{\hat{A}_{x-\hat{1};1}}\right] \exp\left[- i Q \frac{2\pi}{k a}\textcolor{blue}{\hat{p}_{x-\hat{1};1}}\right],
\end{align}
due to the commutator
\begin{align}
    \left[- i Q \frac{2\pi}{k a}\textcolor{blue}{\hat{p}_{x-\hat{1};1}},\ - i Q \frac{a}{2}\textcolor{red}{\hat{A}_{x-\hat{1};1}}\right] = \frac{\pi i}{k}Q^2,
\end{align}
where the minus sign in both terms come from the inverse ($-1$) in $\hat{W}_i^{\text{TOP}^{-1}}$.

In summary, we have computed that $\langle\hat{W}_3(\tau_6)\hat{W}_2^{-1}(\tau_5)\hat{W}_1^{-1}(\tau_4)\hat{W}_3^{-1}(\tau_3)\hat{W}_2(\tau_2)\hat{W}_1(\tau_1)\rangle$ has a non-trivial topological phase $e^{\frac{\pi i}{k}Q^2}$, which demonstrates the self statistics between two charge-$Q$ anyons.

Note that for the minimally charged anyons with $Q=1$, the self statistics has the exchange phase $e^{\frac{\pi i}{k}}$. We know the bosonic statistics has the phase ($+1$) and the fermionic statistics has the phase ($-1$). Therefore, for $k\ge 2$, this phase $e^{\frac{\pi i}{k}}$ demonstrates a fractional statistics (i.e. neither bosonic nor fermionic statistics) of anyons.

We discussed earlier in Sec.~\ref{subsec:Hamiltonian_quantization} that an odd-$k$ Chern-Simons level implies the fermionic nature of the theory. Now from the self statstics, we can directly see that the theory contains fermions when $k$ is odd. Consider the anyons with charge $Q=k$. Their self statistics has the exchange phase $e^{\frac{\pi i}{k}Q^2}=e^{i \pi k}$. When $k$ is even, the charge-$k$ anyons are bosons. When $k$ is odd, the charge-$k$ anyons are fermions.

From the calculation of the anyon statistics above, we demonstrate that our lattice Hamiltonian formulation captures the non-trivial braiding of Wilson lines, and therefore realizes the framing anomaly~\cite{Polyakov:1988framing,Witten:1989framing,Kitaev:2006framing,Sulejmanpasic2023,Sulejmanpasic2024}. In our formulation, the projections of the Wilson lines on the degenerate ground state subspace, as shown in Eq.~\eqref{eq:anyon_W_GS}, explicitly include a well-defined point-splitting regularization. The dangling dashed links shown in Figs.~\ref{fig:anyon_mutual} and~\ref{fig:anyon_self}, on which the canonical conjugate momentum operators $\hat{p}$ are sitting, define the framing of the Wilson loops. This agrees with the observation in Ref.~\cite{Sulejmanpasic2023,Sulejmanpasic2024} that the physical operators are ribbons, or framed Wilson loops. There, the authors construct the framed Wilson loops by two displaced Wilson loops~\cite{Sulejmanpasic2023,Sulejmanpasic2024}. Here, with the Maxwell term introduced, we can define independent canonical conjugate momentum operators $\hat{p}$, and we construct the framed Wilson loops by the ``dangling Wilson loops'' as shown in Figs.~\ref{fig:anyon_mutual} and~\ref{fig:anyon_self}.

\section{Analytical Solution}
\subsection{Constant Mode}
\label{sec:mode}

We know from the continuum model that, when the Maxwell-Chern-Simons theory is defined on a torus, in the Hodge decomposition of the gauge field $A$, there are harmonic parts which corresponds to global constant modes~\cite{moore,poly}. These modes  are topological modes with  flat connections corresponding to the non-trivial homology cycles of the torus. They define the topological sector of the theory and they are, thus, the relevant objects to study its topological properties.
Quantizing the theory in the  $A_0 = 0$ gauge, we can decompose $A$ in the sum of non-flat local modes and global flat modes:
\begin{equation}
\bar{A}(\bar{x},t) = \tilde{A}(\bar{x},t) + \bar{a}(t) = A_{\rm nf}
+ A_{\rm f} \ .
\end{equation}
The flat modes satisfy trivially the Gauss' law constraint and  have to be quantize independently.
Since in the Hodge decomposition $A_{\rm nf}$ and $A_{\rm f}$ are orthogonal, the Hamiltonian can be written as
\begin{equation}
    H = H_{\rm nf} +  H_{\rm f} \ , 
\end{equation}
and the Hilbert space is the direct product of two Hilbert spaces:
\begin{equation}
    \mathcal{H} = \mathcal{H}_{\rm nf} \otimes  \mathcal{H}_{\rm f} \ . 
\end{equation}
The wave function is, thus, given by:
\begin{equation}
    \Psi = \Psi_{\rm nf} \otimes  \Psi_{\rm f} \ . 
\end{equation}

To study the constant modes solution on the lattice, we start from  Hamiltonian Eq.~\eqref{eq:Hamiltonian}.
For constant modes, only the first term survives, since the magnetic field $B$ acting on constant modes is zero.  So no Villain approximation is required to solve the theory in the topological sector.

The Hamiltonian Eq.~\eqref{eq:Hamiltonian} for constants modes is:
\begin{equation}
    H ={e^2 \over 2 a^2} S \left[ \left(p_1 - {k a^2 \over 4 \pi} a_2\right)^2 + \left(p_2 + {k a^2 \over 4 \pi} a_1\right)^2 \right] \ ,
    \label{conmod}
\end{equation}
with $S = N_1 N_2$ the (dimensionless) area of the flat lattice torus, $k=p/q$ and $p_i$ are the constant modes' canonical momenta in the zero-lattice-momentum sector. Note that, since $k$ is rational we have $(a_1,a_2) \simeq (a_1 + 2\pi q/a,a_2 + 2\pi q /a)$.
Equation~\eqref{conmod} is equivalent to the Hamiltonian of a charged particle of mass $m = (a^2 / e^2  S)$ moving in a fictitious magnetic field: $B = k a^2/2\pi$ on a torus of area $A= (2\pi)^2 q^2/ a^2$, where $a_1$ and $a_2$ are the coordinates and $p_i = -i {\partial \over \partial a_i}$ (note that the  area of the torus with coordinates $a_1$ and $a_2$ has dimensions of a $(lenght)^{-2}$, so all other dimensions are scaled accordingly):
\begin{equation}
    H ={1 \over 2 m} S \left[ \left(p_1  - {\cal A}_1\right)^2 + \left(p_2 + {\cal A}_2\right)^2 \right] \ ,
    \label{magnetic}
\end{equation}
with, in the symmetric gauge, ${\cal A}_1 =-B a_2/2$, ${\cal A}_2 = B a_1/2$. In the Landau gauge, ${\cal A}_1 =0$, ${\cal A}_2 = B a_1$, which is used in our following solution.
 
This is a well known problem whose solution was found in Ref.~\cite{haldane}.
We will, thus, present  only the results. A detailed derivation can be found in Ref.~\cite{wiese}.
On the infinite plane, the spectrum consists of  Landau levels determined by the eigenvalue of a shifted harmonic oscillator, with frequencies $\omega = k e^2 / 2 \pi$, given by the Chern-Simons mass in our case~\cite{Tong:2018lecture}. The states  are infinitely degenerate since the energy does not depend on one of the momenta (in the present case $p_2$ in the Landau gauge).
On the torus, although the Hamiltonian which describes the system is the same the energy spectrum has a finite degeneracy given by the number of fluxes $N_\phi$ piercing the surface
\begin{equation}
   \Phi = N_\phi \phi_0 \ ,
\end{equation}
where $\Phi$ is the total flux and $\phi_0 = 2 \pi$ is the flux quantum. For consistency of the theory on the torus $N_\phi$ must be an integer.
In our case we have  $N_\phi = pq \in Z$, giving exactly the same degeneracy $\bar{k}$ we found in Sec.~\ref{subsec:Hamiltonian_degeneracycristina} for a genus 1 surface. This degeneracy can be computed looking at the algebra of magnetic translation operators (see Ref.~\cite{wiese}).

In what follows we will consider the case of integer $k$ with $(a_1,a_2) \simeq (a_1 + 2\pi /a,a_2 + 2\pi /a)$.
The vector potential is periodic up to a gauge transformation:
\begin{align}
    {\cal A}_i(a_1+2\pi /a, a_2) &= {\cal A}_i(a_1,a_2) -\partial_i \phi_1 (a_2) \ , \\
    {\cal A}_i(a_1, a_2 +2\pi /a) &= {\cal A}_i(a_1,a_2) -\partial_i \phi_2 (a_1) \ ,
\label{periodicity}
\end{align}
with:
\begin{align}
    \phi_1( a_2) &= \theta_1 - {2\pi \over a} B a_2 \ ,\\
    \phi_2(a_1) &= \theta_2  \ ,
\label{gauge}
\end{align}
where $\theta_1$ and $\theta_2$ are fixed phases defined in Sec.~\ref{subsec:Hamiltonian_degeneracycristina}.
The phases $\phi_i$ satisfy the cocycle consistency condition:
\begin{equation}
   \phi_2 \left(a_1 + 2 {\pi \over a}\right) + \phi_1 (a_2) - \phi_1 \left(a_2 + 2 {\pi \over a}\right) - \phi_2 (a_1) = 2 \pi k \ .
\end{equation}
The periodicity in the torus with coordinates $a_i$, is equivalent to a large gauge transformation.
The wave function transforms, thus, as:
\begin{align}
\label{wavefun}
   \Psi(a_1+2\pi /a, a_2) &=  e^{\phi_1 (a_2)}\Psi(a_1,a_2) = e^{\theta_1 - {2\pi \over a} Ba_2}\Psi(a_1,a_2) \ , \\
    \Psi(a_1, a_2+2\pi /a) &=  e^{\phi_2 (a_1)}\Psi(a_1,a_2) = e^{\theta_2 }\Psi(a_1,a_2) \ .
\label{wavefun2}
\end{align}
These phases $\theta_i$ can be reabsorbed in a redefinition of the magnetic translation generators and distinguish different superselection sector of the theory~\cite{wiese}. In our discussion we  will set $\theta_i = 0$. 

Reintroducing the original gauge and positing the complex topological components~\cite{moore,iengo}:
\begin{equation}
    b =  {1\over 2 \pi} (ia_1 + a_2) \ ; \ \ \ \bar{b} = {1\over 2 \pi} (-ia_1 + a_2) \ ,
\end{equation}
we can rewrite Eq.~\eqref{conmod} as:
\begin{align}
    H &= {e^2 S \over 2 a^2}   \left( -{1 \over  \pi} { \partial \over \partial b } + {k a^2 \over 2} \bar{b}\right) \left({1 \over  \pi} { \partial \over \partial \bar{b} } + {k a^2 \over 2}  b\right) + {\rm const} \\ 
    &= {e^2 S \over 2 a^2}  B^+ B + {\rm const} \ ,
    \label{conmod1}
\end{align}
with
\begin{equation}
    B \equiv {1 \over  \pi} { \partial \over \partial \bar{b} } + {k a^2 \over 2}  b , \quad [B, B^+] = {k a^2 \over \pi } \ .
\end{equation}
Equation~\eqref{conmod1} is the Hamiltonian of an harmonic oscillators and the ground state $\Psi^0$ is given by $B \Psi^0 = 0$.
Imposing the periodicity conditions Eqs.~\eqref{wavefun} and~\eqref{wavefun2} we see that there are $k$ independent ground states  $\Psi^0_s$
with $s = 0,...,k - 1$. 
The ground-state wave function is:
\begin{equation}
    \Psi^0_s = e^{{ \pi k a^2 \over 2}b^2} e^{{- \pi k a^2 \over 2 } b \bar{b}}  \theta \left[\begin{matrix} s/k & \\ 0 & \end{matrix} \right](k b | i k ) \ ,
    \label{gswf}
 \end{equation}
 expressed in terms of the Jacobi $\theta$-function.
 In terms of the  $a_i$ fields the ground-state wave function is:
 \begin{equation}
    \Psi^0_s = e^{{ k a^2 \over 8 \pi } (ia_1 + a_2)^2} e^{{- k a^2 \over 8 \pi } (a^2_1 + a^2_2)} \sum_{m \in Z} e^{{-\pi \over k }(m k + s)^2}e^{a(m k + s) (-a_1 + i a_2)}
    \ .
     \label{gswf1}
 \end{equation}
These $k$ independent ground states are consistent with our earlier discussion on the $k$-fold degeneracy of states in Sec.~\ref{subsec:Hamiltonian_degeneracycristina}. 
 
The excited states can be obtained applying $B^+$ to the ground state wave function Eq.~\eqref{gswf1}, and have energies of the order $\omega$, the Chern-Simons mass.  In the limit in which the Chern-Simons mass goes to infinity (strong coupling limit), we recover the topological theory.

\subsection{Energy Spectrum}
\label{sec:solution}

In this section, we present the solution of all lattice momentum modes to the Hamiltonian in Eq.~\eqref{eq:Hamiltonian}. Since the Hamiltonian is quadratic, we can use the Fourier transform to decompose it into independent momentum sectors. We rewrite the Hamiltonian as 
\begin{equation}
    \hat{H} = \sum_{\substack{x=(x_1,x_2) \\ x_1 \in \{0,1,\dots,N_1-1\} \\ x_2 \in \{0,1,\dots,N_2-1\} }} \frac{e^2}{2 a^2} \sum_{\mu,\nu\in\{1,2\}}\left(\hat{p}_{x;\mu}-\epsilon_{\mu\nu}\frac{k a^2}{4\pi}\hat{A}_{x-\hat{\nu};\nu}\right)^2 + \frac{1}{2 e^2} \left(\Box\hat{A}_{x;1,2}\right)^2.
\end{equation}

We apply the following Fourier transform
\begin{align}
    \hat{A}_{q;\mu} &= \frac{1}{\sqrt{N_1 N_2}} \sum_{\substack{x=(x_1,x_2) \\ x_1 \in \{0,1,\dots,N_1-1\} \\ x_2 \in \{0,1,\dots,N_2-1\} }} e^{i\Vec{q}\cdot\Vec{x}} \hat{A}_{x;\mu}, \\
    \hat{p}_{q;\mu} &= \frac{1}{\sqrt{N_1 N_2}} \sum_{\substack{x=(x_1,x_2) \\ x_1 \in \{0,1,\dots,N_1-1\} \\ x_2 \in \{0,1,\dots,N_2-1\} }} e^{i\Vec{q}\cdot\Vec{x}} \hat{p}_{x;\mu},
\end{align}
with inverse Fourier transform
\begin{align}
    \hat{A}_{x;\mu} &= \frac{1}{\sqrt{N_1 N_2}} \sum_{\substack{q=(q_1,q_2) \\ q_1 \in \{0,\frac{2\pi}{N_1},\dots,\frac{2\pi}{N_1}(N_1-1)\} \\ q_2 \in \{0,\frac{2\pi}{N_2},\dots,\frac{2\pi}{N_2}(N_2-1)\} }} e^{- i\Vec{q}\cdot\Vec{x}} \hat{A}_{q;\mu}, \\
    \hat{p}_{x;\mu} &= \frac{1}{\sqrt{N_1 N_2}} \sum_{\substack{q=(q_1,q_2) \\ q_1 \in \{0,\frac{2\pi}{N_1},\dots,\frac{2\pi}{N_1}(N_1-1)\} \\ q_2 \in \{0,\frac{2\pi}{N_2},\dots,\frac{2\pi}{N_2}(N_2-1)\} }} e^{- i\Vec{q}\cdot\Vec{x}} \hat{p}_{q;\mu}.
\end{align}
The commutation relation between the operators is 
\begin{align}
    \label{eq:A_p_commutator}
    [\hat{A}_{x;\mu}, \hat{p}_{y;\nu}] &= i \delta_{x,y} \delta_{\mu,\nu} \\
    \label{eq:A_p_commutator_fourier_transformed}
    [\hat{A}_{q;\mu}, \hat{p}_{s;\nu}] &= i \delta_{q+s, 2\pi\mathbb{Z}} \delta_{\mu,\nu}
\end{align}

The Fourier transformed Hamiltonian looks like
\begin{align}
    \hat{H} = \sum_{\substack{q=(q_1,q_2) \\ q_1 \in \{0,\frac{2\pi}{N_1},\dots,\frac{2\pi}{N_1}(N_1-1)\} \\ q_2 \in \{0,\frac{2\pi}{N_2},\dots,\frac{2\pi}{N_2}(N_2-1)\} }}  \frac{e^2}{2 a^2} \left[
    \left(
    \hat{p}_{q;1} - \frac{k a^2}{4\pi} e^{i q_2} \hat{A}_{q;2}
    \right)\left(
    \hat{p}_{-q;1} - \frac{k a^2}{4\pi} e^{-i q_2} \hat{A}_{-q;2}
    \right)\right.& \nonumber \\
    +\left.\left(
    \hat{p}_{q;2} + \frac{k a^2}{4\pi} e^{i q_1} \hat{A}_{q;1}
    \right)\left(
    \hat{p}_{-q;2} + \frac{k a^2}{4\pi} e^{-i q_1} \hat{A}_{-q;1}
    \right)
    \right]& \nonumber \\
    + \frac{1}{2 e^2} \left[
    \left(1-e^{-i q_2}\right) \hat{A}_{q;1}
    - \left(1-e^{-i q_1}\right) \hat{A}_{q;2}
    \right]\left[
    \left(1-e^{i q_2}\right) \hat{A}_{-q;1}
    - \left(1-e^{i q_1}\right) \hat{A}_{-q;2}
    \right].&
    \label{eq:Hamiltonian_fourier_transformed}
\end{align}

\begin{figure}[htp!]
\centering

\begin{subfigure}{0.45\textwidth}
  \centering
  \begin{tikzpicture}
    \draw (0, 0) node[inner sep=0] {\includegraphics[width=\linewidth]{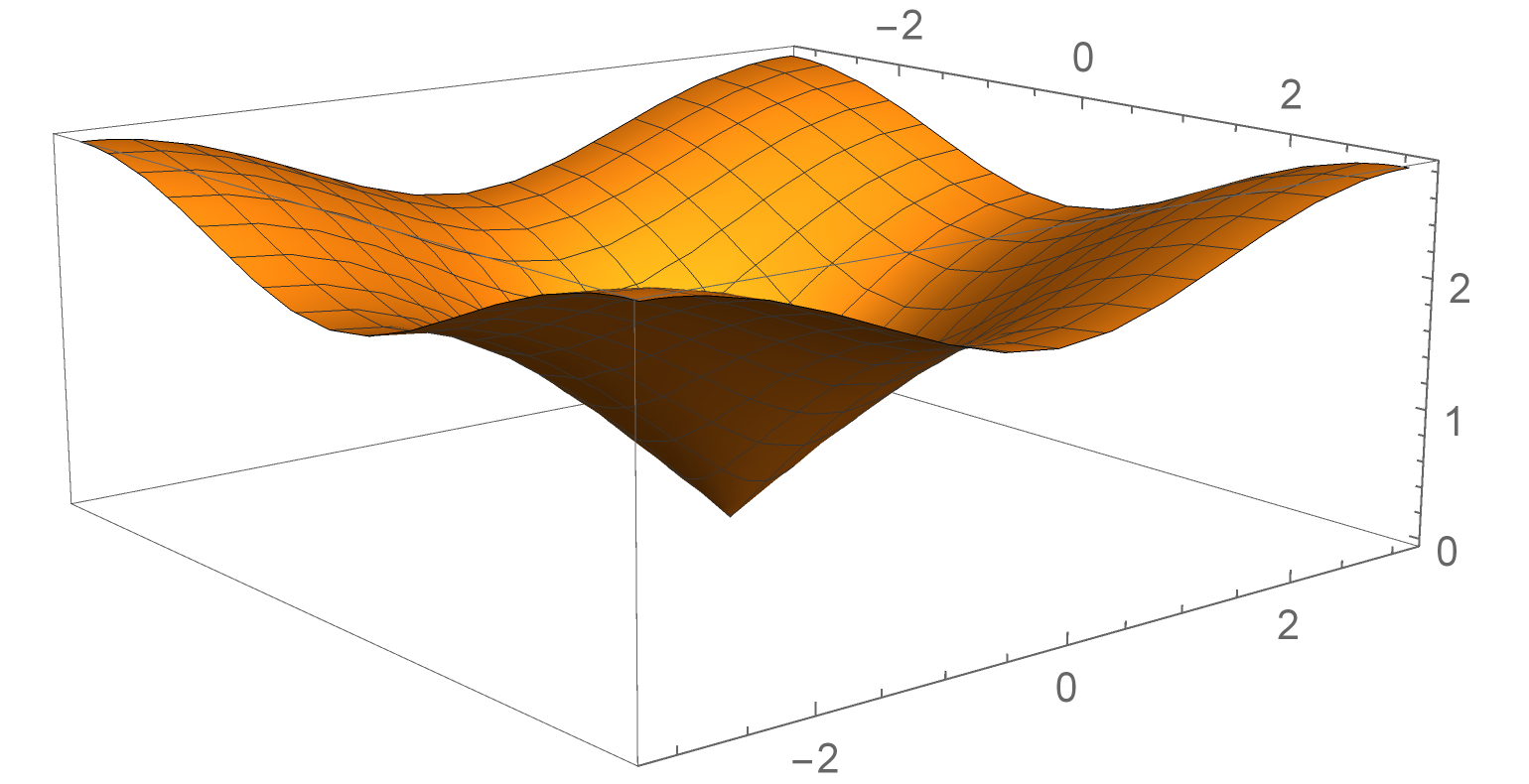}};
    \draw (2.5, 0.2) node {$\Delta E$};
    \draw (-2.1, -1.4) node {$q_1$};
    \draw (2, -1.5) node {$q_2$};
  \end{tikzpicture}
  \caption{}
  \label{fig:sub1}
\end{subfigure}
\hfill
\begin{subfigure}{0.45\textwidth}
  \centering
  \begin{tikzpicture}
    \draw (0, 0) node[inner sep=0] {\includegraphics[width=\linewidth]{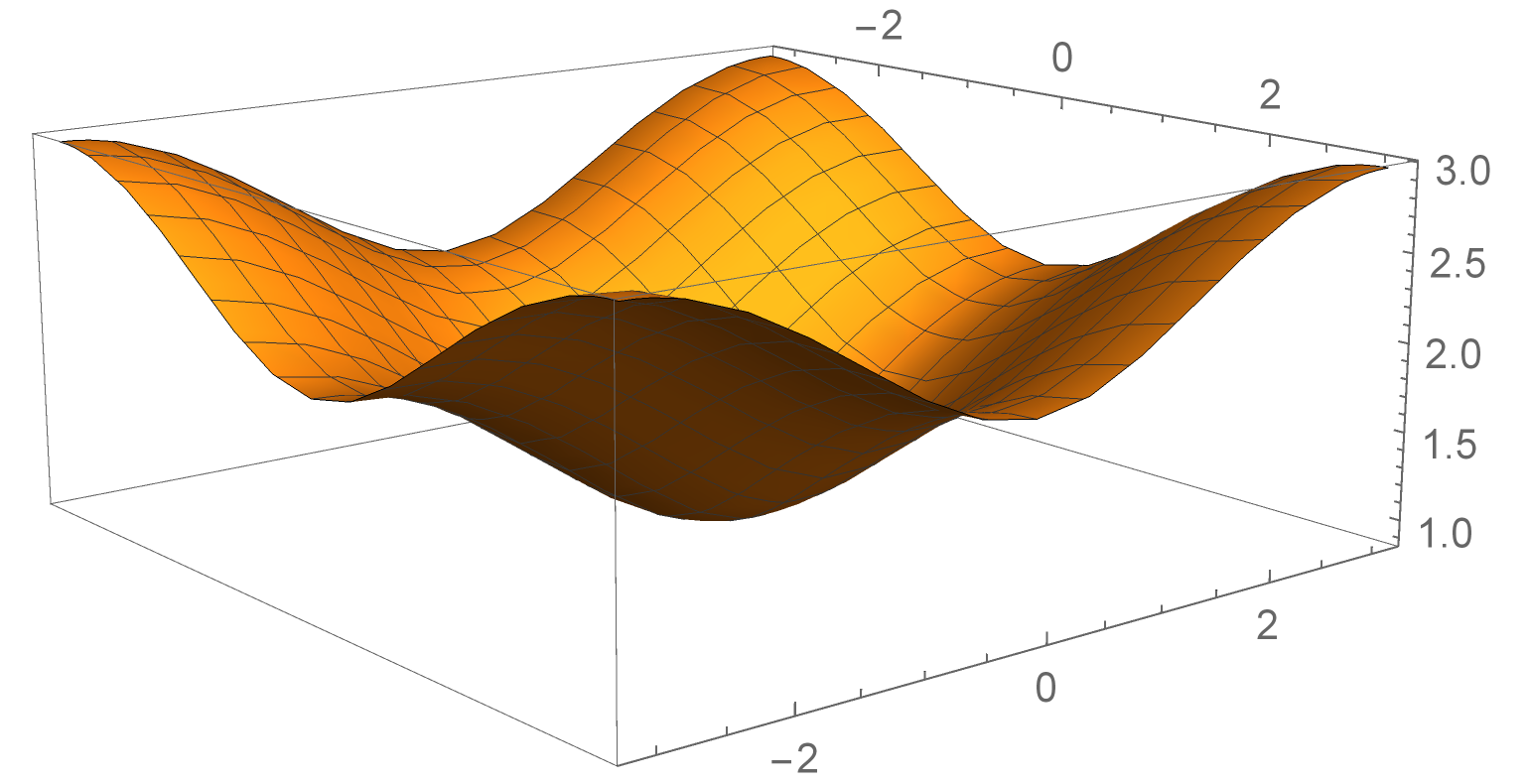}};
    \draw (2.4, 0.2) node {$\Delta E$};
    \draw (-2.1, -1.4) node {$q_1$};
    \draw (2, -1.5) node {$q_2$};
  \end{tikzpicture}
  \caption{}
  \label{fig:sub2}
\end{subfigure}

\begin{subfigure}{0.45\textwidth}
  \centering
  \begin{tikzpicture}
    \draw (0, 0) node[inner sep=0] {\includegraphics[width=\linewidth]{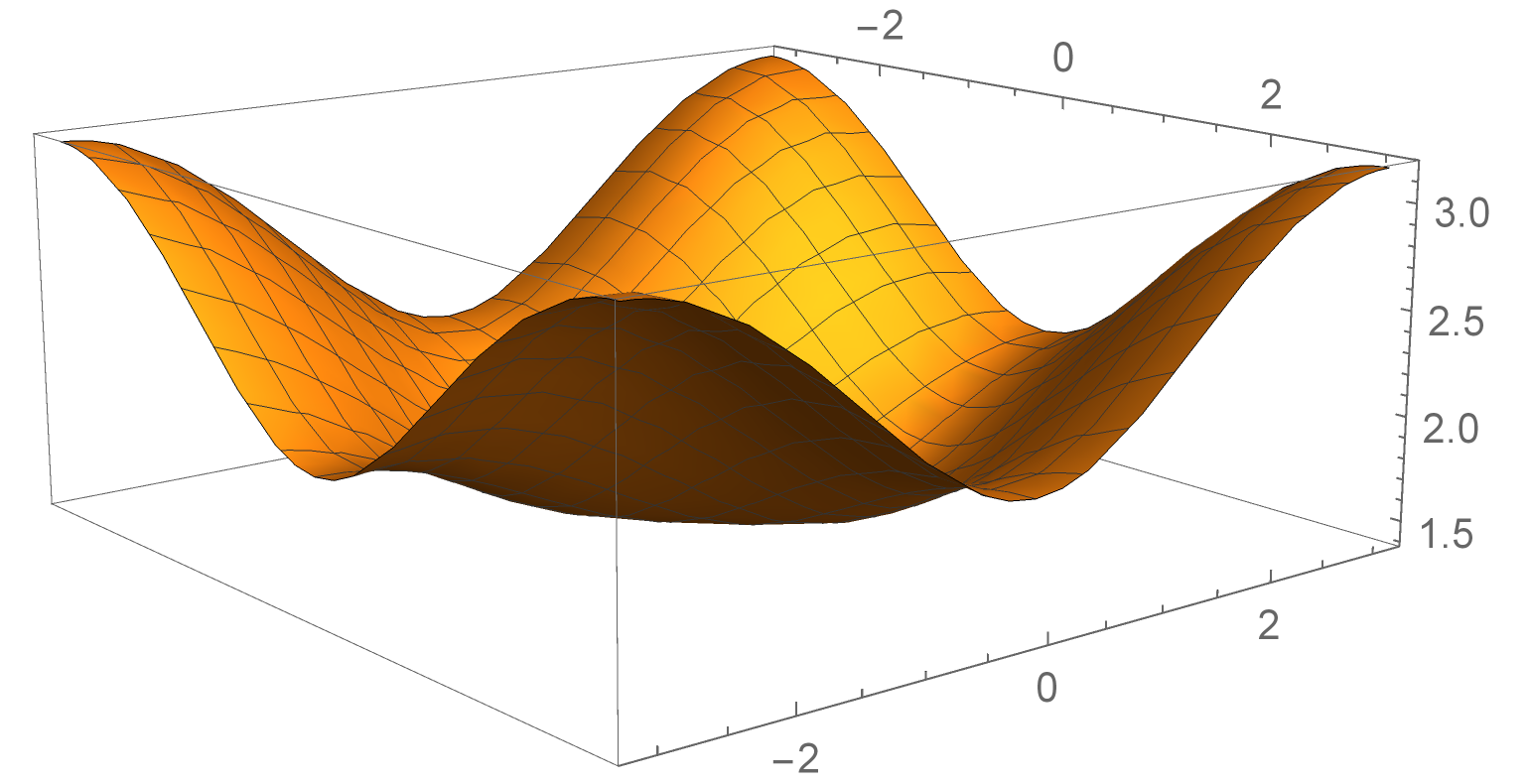}};
    \draw (2.4, 0.2) node {$\Delta E$};
    \draw (-2.1, -1.4) node {$q_1$};
    \draw (2, -1.5) node {$q_2$};
  \end{tikzpicture}
  \caption{}
  \label{fig:sub3}
\end{subfigure}
\hfill
\begin{subfigure}{0.45\textwidth}
  \centering
  \begin{tikzpicture}
    \draw (0, 0) node[inner sep=0] {\includegraphics[width=\linewidth]{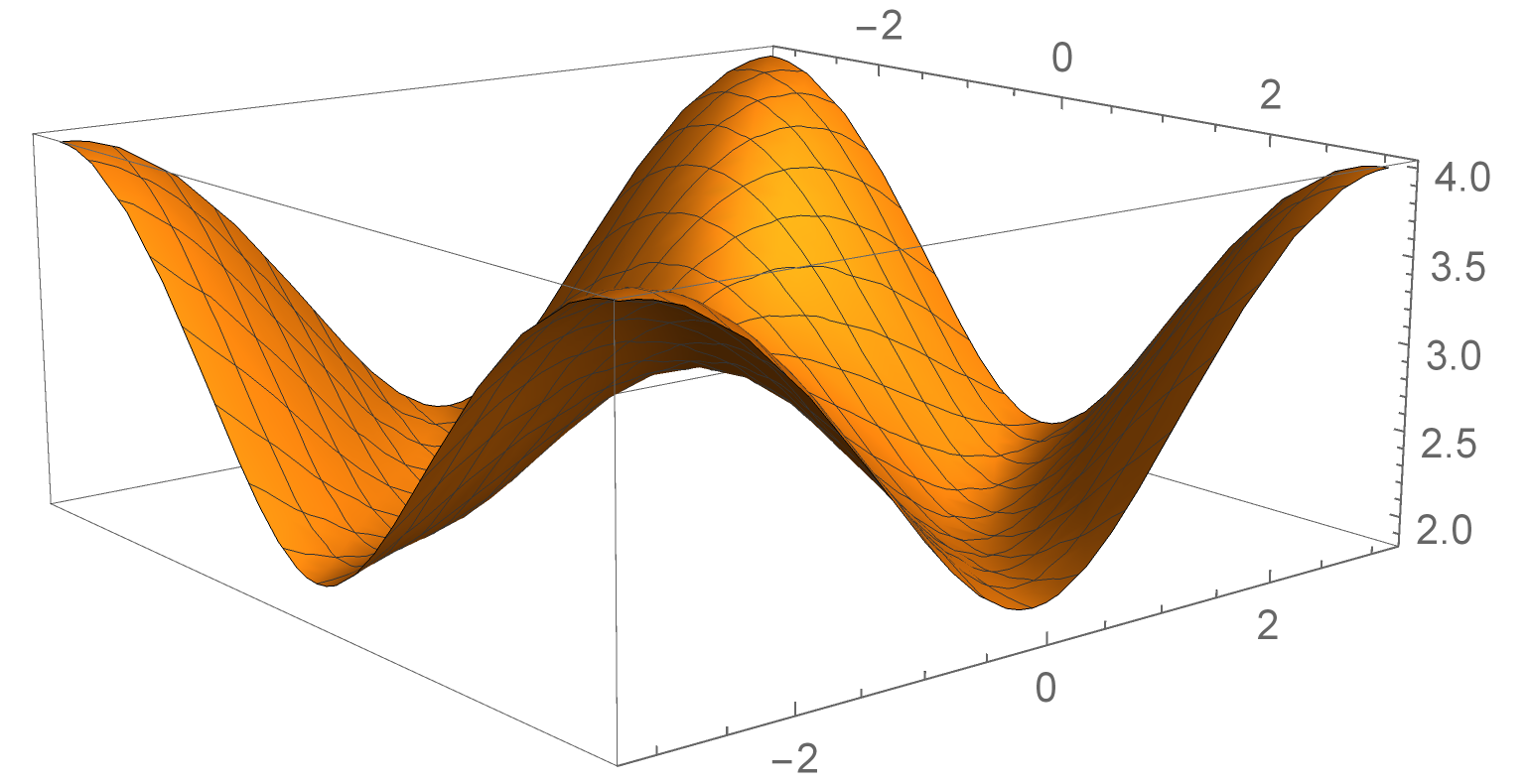}};
    \draw (2.4, 0.2) node {$\Delta E$};
    \draw (-2.1, -1.4) node {$q_1$};
    \draw (2, -1.5) node {$q_2$};
  \end{tikzpicture}
  \caption{}
  \label{fig:sub4}
\end{subfigure}

\caption{3D plot of the band structure of the analytical solution to the Maxwell-Chern-Simons Hamiltonian. The vertical axis is the energy of the first excited state $\Delta E$ above the ground state. Two horizontal axes are the two components of the lattice momentum $q=(q_1, q_2)$. Each subfigure corresponds to a different coupling strength $\lambda = \left(k e^2 a/4\pi\right)^2$. (a) $\lambda = 0$, this corresponds to the pure Maxwell theory, in which we can see a linear-dispersion gapless photon. (b) $\lambda = 0.2$, the Chern-Simons term gaps out the theory and gives the photon a mass. (c) $\lambda = 0.5$, this is the special $\lambda$ value beyond which the band bottom is no longer at $q=(0,0)$. (d) $\lambda = 2.0$, for large $\lambda$ value the band looks like the one for a pure Chern-Simons theory, in which the band bottom are two lines specified by $q_1 + q_2 = \pi + 2\pi\mathbb{Z}$. These two lines correspond to the ``staggered momentum modes'' mentioned in Ref.~\cite{Sulejmanpasic2024}.}
\label{fig:band}
\end{figure}
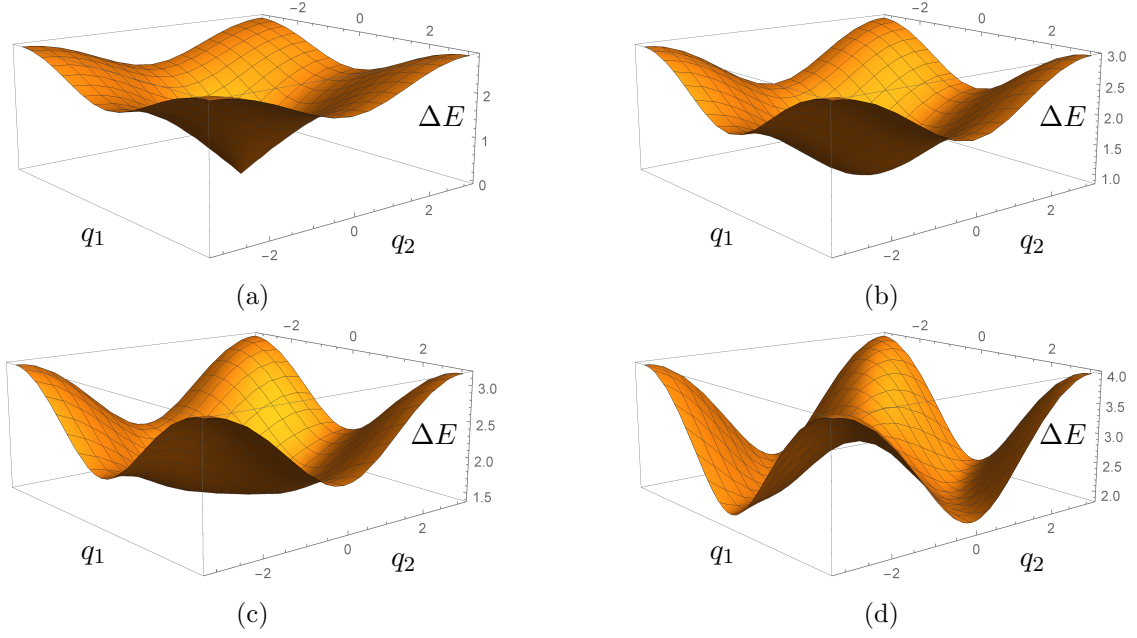

After some change of variables and applying the Gauss' law Eq.~\eqref{eq:Gauss_law} (see more details in the Appendix~\ref{subsec:appendix_analytical_solution}), the Hamiltonian in the lattice momentum $q=(q_1,q_2)$ sector corresponds to a simple harmonic oscillator with angular frequency
\begin{align}
    \label{eq:solution_omega}
    \omega^2 = \frac{1}{a^2} \left[2(1-\cos q_1) + 2(1-\cos q_2)\right] + \left(\frac{k e^2}{4\pi}\right)^2\left[2+2\cos(q_1 + q_2)\right].
\end{align}
The energy of the corresponding first excited state above the ground state is 
\begin{align}
    \Delta E = \omega = \sqrt{\frac{1}{a^2} \left[2(1-\cos q_1) + 2(1-\cos q_2)\right] + \left(\frac{k e^2}{4\pi}\right)^2\left[2+2\cos(q_1 + q_2)\right]},
\end{align}
which is plotted in Fig.~\ref{fig:band}. The different subplots in Fig.~\ref{fig:band} correspond to different values of a combined parameter $\left(ke^2a/4\pi\right)^2$. Each state in the band is $k$-fold degenerate. 

There are two interesting limits of this energy spectrum. First we note that the dimensionless lattice momentum $q = a\Tilde{q}$, where $\Tilde{q}$ is the dimensionful momentum. Then by taking the continuum limit $a\to 0$, we get 
\begin{equation}
    \omega^2 \to |\Tilde{q}|^2 + \left(\frac{k e^2}{2\pi}\right)^2,
\end{equation}
which describes a massive particle with mass equals to $k e^2/2\pi$. i.e., the photon becomes massive, and the theory opens a gap. The $|\Tilde{q}|^2$ part comes from the first term $\frac{2}{a^2}(1-\cos\, a\Tilde{q}_i)$ in Eq.~\eqref{eq:solution_omega}. This energy dispersion agrees with the one in the continuous Maxwell-Chern-Simons theory~\cite{Deser:1981wh}.

Another limit is when $k \to \infty$. The Chern-Simons term dominates. The band structure looks similar to the one of the pure continuous Chern-Simons theory, with the band bottom located at two lines specified by $q_1 + q_2 = \pi + 2\pi\mathbb{Z}$. These two lines correspond to the ``staggered momentum modes'' mentioned in Ref.~\cite{Sulejmanpasic2024}.

One interesting observation is that the band bottom changes location, although this does not mean a phase transition because the band is showing the first excited state but not the ground state.

There is also a distinction between the $q=(0,0)$ zero mode and the other modes. The Gauss' law (constraint 1) acts trivially on the zero mode, and therefore leaves a free degree of freedom. The Constraint 2 and 3 requires both the momentum and the coordinate of this free degree of freedom to be quantized, which is equivalent to a single particle on a circle with finite sites. As shown in the Appendix~\ref{subsec:appendix_analytical_solution}, there are exactly $k$ sites on the circle. Therefore, the Hilbert space for the zero mode is $k$ dimensional, and results in the $k$-fold degeneracy of every states. This is consistent with our earlier discussion on constant modes in Sec.~\ref{sec:mode}.

\section{Conclusion}
\label{sec:conclusion}

In this paper, we derive the Hamiltonian formulation of Maxwell-Chern-Simons theory on a lattice in 2+1D. Comparing to the well-known 2+1D compact lattice QED Hamiltonian, the modification in our Hamiltonian formulation is as follows. First, the instantons in the theory are suppressed as a hard constraint. Similar to the Witten effect in 3+1D~\cite{Witten:1979monopole}, an instanton (i.e., a magnetic monopole in 2+1D) must have an electric charge under the Chern-Simons action; thus, our pure gauge theory cannot allow instantons. We further show that the instanton suppression can turn the 2+1D compact lattice QED Hamiltonian into an unusual form (see Appendix~\ref{subsec:appendix_Maxwell_Instanton_Suppressed}). This instanton-suppressed compact lattice QED Hamiltonian has quadratic terms in the gauge field $\hat{A}$, and it is indeed a compact theory by the Villain approximation with two 1-form constraints. The two 1-form constraints ensure that the theory is invariant under large gauge transformations in the two spacial directions. Furthermore, this instanton-suppressed quadratic Hamiltonian is analytically solvable, showing a gapless theory with massless photons. Unlike the instanton-allowed compact lattice QED, which is gapped in 2+1D, the instanton-suppressed version corresponds the Maxwell theory in the continuum, which is gapless.

The second modification is to introduce the Chern-Simons term into this instanton-suppressed Hamiltonian. The effect of this step is that a gauge field operator $\hat{A}$ on a shifted site and in the perpendicular direction is attached to the conjugate momentum operator $\hat{p}$, wherever $\hat{p}$ appears. Therefore, we see the modification in the quadratic $\hat{p}$ terms in the Hamiltonian in Eq.~\eqref{eq:Hamiltonian}, the Gauss' law constraint in Eq.~\eqref{eq:constraint_L1}, and the two 1-form constraints in Eqs.~\eqref{eq:constraint_L1} and~\eqref{eq:constraint_L2}.

We demonstrate the topological features of our Hamiltonian formulation. The compatibility between the constraints requires the quantization of the Chern-Simons level $k$. The existence of two 1-form operators generates the $k$-fold degeneracy of the spectrum. The Wilson loop expectation values in the theory always show a perimeter law. Linked Wilson loops have non-trivial topological phases evaluated on the degenerate Hilbert subspace. Anyons as the excitations at the ends of open Wilson lines display fractional mutual statistics and self statistics. These topological features agree with the ones in a continuum Maxwell-Chern-Simons theory, showing our Hamiltonian formulation is the correct lattice theory for the continuum theory.

The continuum limit of our lattice Hamiltonian formulation can also be directly seen from the analytical solution of the theory. Since the Hamiltonian in Eq.~\eqref{eq:Hamiltonian} is quadratic, we perform the Fourier transform and solve its band structure in the Brillouin zone. We can see the correct band gap $k e^2/2\pi$ and the massive photon dispersion that agree with the continuum Maxwell-Chern-Simons theory.

In our work, we add the Maxwell term to the Chern-Simon theory, which allows us to construct the independent canonical conjugate momentum operator $\hat{p}$. Our formulation accepts the gauge field configurations as the natural basis for the Hilbert space. This is in contract to the recent work that presents a canonical quantization of lattice Chern-Simons theory~\cite{Sulejmanpasic2024}, which does not have Maxwell action and the canonical quantization unavoidably has non-trivial commutation between operators on different spacial locations. This ``fuzzy'' space makes it difficult to construct a good orthonormal basis for the Hilbert space and apply numerical methods. Therefore, our lattice Hamiltonian formulation opens the door to many possible numerical methods to study this interesting topological theory. For future study, it is possible to design quantum algorithms for simulating with our lattice Hamiltonian formulation. Numerical simulations with tensor network ansatz or machine-learning approach are also applicable with our formulation. It will also be an important direction to further extend the simulations for our model coupled to matter fields.

\vspace{10pt}

\textbf{Note} \quad We noticed another preprint~\cite{JingYuanChen2024} on the same topic, which appeared a few months after ours. In Ref.~\cite{JingYuanChen2024}, the authors study the same Lagrangian as ours and derived a Hamiltonian formulation without the final step of gauge fixing in Appendix~\ref{subsec:appendix_Maxwell_Chern_Simons}. By doing so, they extend the discussions of the Maxwell-Chern-Simons theory to sectors beyond the zero total magnetic flux sector. This zero total magnetic flux sector is what we focus on in this work, because we study the pure Maxwell-Chern-Simons theory without matter. We construct a local and uniform Hamiltonian formulation in this sector in perspective of Hamiltonian simulation. In Ref.~\cite{JingYuanChen2024}, efforts have been made to construct a local and uniform Hamiltonian formulation for space-time manifolds with arbitrary topology and arbitrary flux sectors. The focus of Ref.~\cite{JingYuanChen2024} is to derive the chiral properties of the theory, in contrast to our emphasis on future quantum computing implementations.

\vspace{10pt}

\textbf{Acknowledgement} \quad The authors acknowledge insightful discussion with Max Metlitski, Arianna Crippa, Emil Rosanowski, Aleksey Cherman, Michael DeMarco, Ho Tat Lam, Haoyu Guo, Robert Jones, and Jing-Yuan Chen. This project is supported by the Deutsche Forschungsgemeinschaft (DFG, German Research Foundation) as part of the CRC 1639 NuMeriQS -- project no.\ 511713970. DL acknowledges support from the National Science Foundation under Cooperative Agreement PHY-2019786 (The NSF AI Institute for Artificial Intelligence and Fundamental Interactions,
http://iaifi.org/). This work is funded by the European Union’s Horizon Europe Framework Programme (HORIZON) under the ERA Chair scheme with grant agreement no.\ 101087126. 
This project has received funding from the European Union’s Horizon 2020 Research and Innovation Programme under the Marie Skłodowska-Curie Grant Agreement No.\ 101034267.
This work is supported with funds from the Ministry of Science, Research and Culture of the State of Brandenburg within the Centre for Quantum Technology and Applications (CQTA). 
\begin{center}
    \includegraphics[width = 0.12\textwidth]{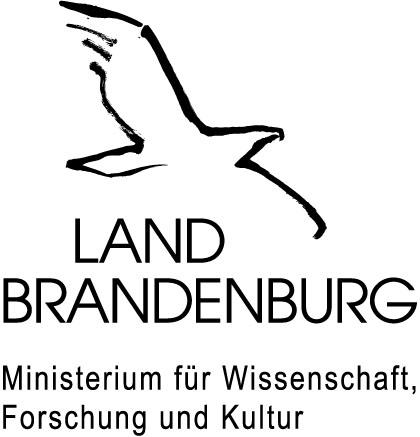}
\end{center}

\clearpage

\section{Appendices}

\subsection{Review of Instantons in Compact Maxwell Theory}
\label{subsec:appendix_instanton}

In this section we briefly review the effect of instantons on a 2+1D compact U(1) gauge theory (i.e. Maxwell theory)~\cite{Polyakov:1975monopole,Polyakov:1977monopole}. The theory contains instantons if it has a finite cut-off scale, which is our lattice spacing here. Instantons are discontinuities in the gauge field configuration where the total flux through a plane suddenly changes by $2\pi$. In 2+1D, an instanton configuration has a finite action. Therefore, when computing the contribution to the partition function $e^{-S(A)}$, instanton configurations have non-zero weight. Because we need to take into account the instanton configurations, the theory is no longer Gaussian quadratic around the $A=0$ minimum, i.e. no longer a free theory. Moreover, because an instanton changes the flux permanently after its appearance, its long-range feature has significant influence on the correlation functions. In particular, it makes the Wilson loops scale as the Area law, and confines charged particles~\cite{Polyakov:1975monopole}.

The effect of instantons can be seen clearly using the duality between the U(1) gauge theory and the XY model~\cite{Wen:2004book}. In the duality, the flux in the U(1) gauge theory maps to the charge in the XY model. Therefore, the instanton operator, as creation/annihilation operator of the flux, maps to $e^{\pm i \theta}$ in the XY model. Summing over instanton configurations in the partition function maps to the following Lagrangian in the XY model:
\begin{align}
\label{eq:XY_model}
    \mathcal{L} = \frac{\chi}{2} (\partial_\mu \theta)^2 - K \cos \theta,
\end{align}
where $\chi$ and $K$ are coefficients related to the coupling strength and instanton fugacity, respectively.

Because the $\cos\theta$ term is always relevant in 2+1D~\cite{Busiello:1985sineGordon}, the instanton operator gaps out the theory. The $\cos\theta$ term also breaks the U(1)$_{XY}$ symmetry, leaving the theory with a unique ground state and massive excitations. The linear potential between charges in the U(1) gauge theory maps to the linear potential between vortices in the XY model, which is due to the $\cos\theta$ energy cost at the $\theta=\pi$ or $-\pi$ branch-cut line between the vortices~\cite{Wen:2004book}. 

When the Chern-Simons action is present, an instanton will carry charge, which is similar to the Witten effect in 3+1D~\cite{Witten:1979monopole}. Moreover, when the Chern-Simons level $k$ is an odd integer, the instanton operator will be fermionic. We do not consider the matter field in this work, and therefore we would like to suppress the instantons.

When instantons are suppressed, it is equivalent to set $K=0$ in Eq.~\eqref{eq:XY_model}. The U(1)$_{XY}$ symmetry is spontaneously broken and we have massless Goldstone modes, which are the massless photons in the U(1) gauge theory. However, we can no longer view the gauge field $A\in[0,2\pi/a)$ on each link for the instanton-suppressed compact U(1) gauge theory. The equivalence between $A=0$ and $A=2\pi/a$ makes the flux only well defined by mod $2\pi$, which naturally allows the existence of instantons. We have to use the Villain approximation to formulate an instanton-suppressed theory.

\subsection{Villain Approximation}

A theory with compact variables means that the partition function $Z$ is periodic in these compact variables $\theta_i$. For such a theory, it is natural to include in the action terms like $\cos(\sum_i c_i \theta_i)$, where $c_i$ are the coefficients that match the periodicity of the compact variable $\theta_i$. However, these cosine terms are nonlinear and are difficult to handle analytically. The Villain approximation simplifies the analysis by replacing the cosine terms with a periodic Gaussian potential~\cite{Villain1975},

\begin{equation}
    e^{\beta \cos(\theta)} \approx \sum_{n=-\infty}^{\infty} e^{- \frac{\Tilde{\beta}}{2} (\theta + 2\pi n)^2},
\end{equation}
where $\Tilde{\beta}$ is a function of $\beta$, and---without loss of generality---we write $\theta$ as $2\pi$ periodic here.

\subsection{Single Quantum Rotor}

We will go step by step to see how to construct a Hamiltonian formulation for a compact lattice field theory. We start with the action of a single quantum rotor, i.e., a $0+1$-D compact quantum field. A careful construction of its Hamiltonian formulation will show the spirit of the later steps. 

We use Euclidean time formulation, discretize the time, and set up periodic boundary condition. The action of the rotor is
\begin{equation}
    S(\theta) = -\frac{\beta}{d\tau} \sum_{i=0}^{T-1} \cos(\theta_{i+1} - \theta_i),
\end{equation}
where $\beta$ is a coefficient, $d\tau$ is the time spacing, $\theta_i \in [0, 2\pi)$ is the rotor angle at the $i$-th time slice, $\theta_T = \theta_0$, $T$ is the total number of time slices.

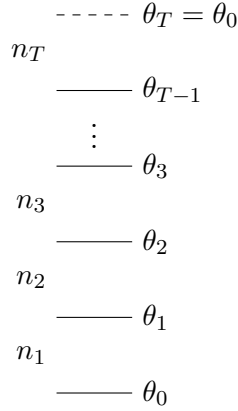
\begin{figure}[htp!]
    \centering
        \begin{tikzpicture}
            \foreach \y/\i in {5/{{T-1}}, 4/3, 3/2, 2/1, 1/0} {
                \draw (0,\y) -- (1,\y) node[right] {$\theta_\i$};
            }
            \draw[dashed] (0,6) -- (1,6) node[right] {$\theta_T = \theta_0$};
            \node at (0.5,4.5) {$\vdots$};
            \foreach \y/\i in {5.5/{T}, 3.5/3, 2.5/2, 1.5/1} {
                \node[left] at (0,\y) {$n_\i$};
            }
        \end{tikzpicture}
    \caption{Illustration of the variables in a single quantum rotor model, i.e. a $0+1$-D compact quantum field. The vertical direction is time. There are $T$ number of discrete time slices, and have periodic boundary condition.}
    \label{fig:rotor}
\end{figure}

The partition function of the rotor is
\begin{equation}
    Z = \sum_{\theta_0, \theta_1, \dots, \theta_{T-1}\in [0,2\pi)} e^{-S(\theta)},
\end{equation}
where we have use the summation symbol to represent the integral over continuous variables $\theta$'s.

We can apply the Villain approximation to the cosine terms in the action. By doing this, we introduce $T$ number of integer degrees of freedom, $n_1, n_2, \dots, n_T$, each between a pair of $\theta$'s, as illustrated in Fig.~\ref{fig:rotor}. The approximated action with these extra degrees of freedom looks like
\begin{equation}
    S(n, \theta) = \frac{\Tilde{\beta}}{2 d\tau} \sum_{i=0}^{T-1} (\theta_{i+1} - \theta_i + 2\pi n_{i+1})^2,
\end{equation}
where $\Tilde{\beta}$ is a function of $\beta$. When $d\tau \to 0$, $\beta/d\tau \to \infty$, the asymptotic behavior of the Villain approximation says $\Tilde{\beta} \to \beta$~\cite{Kleinert1989book}.

The approximated partition function with the extra degrees of freedom looks like
\begin{equation}
    Z = \sum_{\substack{\theta_0, \theta_1, \dots, \theta_{T-1} \in [0, 2\pi) \\ n_1, n_2, \dots, n_T \in \mathbb{Z}}} e^{-S(n, \theta)}.
\end{equation}

The approximated action $S(n,\theta)$ holds discrete transformations that keeps it invariant. For every $i\in{0, 1, \dots, T-1}$, the following transformation
\begin{align*}
    \theta_i & \to \theta_i + 2\pi \\
    n_i & \to n_i - 1 \\
    n_{i+1} & \to n_{i+1} + 1
\end{align*}
keeps the approximated action, and thus the approximated partition function, invariant.

Writing down the transformation invariance of the approximated action explicitly, and taking $i=1$, we have
\begin{equation}
    S(n_1 - 1, n_2 + 1, \theta_1 + 2\pi, *) = S(n_1, n_2, \theta_1, *),
\end{equation}
where $*$ represents all the other variables in $S(n, \theta)$, which are fixed.

We can apply this transformation multiple times, until we make $n_1=0$. That is to say,
\begin{equation}
    S(0, n_2 + n_1, \theta_1 + 2\pi n_1, *) = S(n_1, n_2, \theta_1, *).
\end{equation}

Note that $\theta_1 + 2\pi n_1$ takes value in $[2\pi n_1, 2\pi (n_1+1))$. When we sum over $n_1\in\mathbb{Z}$, it is equivalent to sum over $\theta_1\in(-\infty,+\infty)$. In other words, by absorbing $n_1$, $\theta_1$ is lifted from $[0,2\pi)$ to $(-\infty,+\infty)$. More rigorously, we define new variables
\begin{align}
    \Tilde{\theta}_1 &= \theta_1 + 2\pi n_1, \quad \Tilde{\theta}_1\in(-\infty,+\infty),\\
    \Tilde{n}_2 &= n_2 + n_1, \quad \Tilde{n}_2\in\mathbb{Z},
\end{align}
with the inverse
\begin{align}
    n_1 &= \left\lfloor \frac{\Tilde{\theta}_1}{2\pi} \right\rfloor, \\
    \theta_1 &= \Tilde{\theta}_1 - 2\pi \left\lfloor \frac{\Tilde{\theta}_1}{2\pi} \right\rfloor, \quad \theta_1\in[0, 2\pi), \\
    n_2 &=\Tilde{n}_2 - \left\lfloor \frac{\Tilde{\theta}_1}{2\pi} \right\rfloor,
\end{align}
where $\left\lfloor \cdot \right\rfloor$ is the largest integer less than or equal to a given number.

The approximate partition function now looks like
\begin{align}
    Z &= \sum_{\substack{\theta_0, \theta_1, \dots, \theta_{T-1} \in [0, 2\pi) \\ n_1, n_2, \dots, n_T \in \mathbb{Z}}} e^{-S(n_1, n_2, \theta_1, *)}  \nonumber \\
    &= \sum_{\substack{\theta_0, \theta_1, \dots, \theta_{T-1} \in [0, 2\pi) \\ n_1, n_2, \dots, n_T \in \mathbb{Z}}} e^{-S(0, n_2 + n_1, \theta_1 + 2\pi n_1, *)}  \nonumber \\
    &= \sum_{\substack{\theta_0, \theta_2, \dots, \theta_{T-1} \in [0, 2\pi) \\ \Tilde{\theta}_1\in(-\infty,+\infty), \Tilde{n}_2\in\mathbb{Z} \\ n_3, n_4, \dots, n_T \in \mathbb{Z}}} e^{-S(0, \Tilde{n}_2, \Tilde{\theta}_1, *)}  \nonumber \\
    &= \sum_{\substack{\theta_0, \theta_2, \dots, \theta_{T-1} \in [0, 2\pi) \\ \Tilde{\theta}_1\in(-\infty,+\infty), \Tilde{n}_2\in\mathbb{Z} \\ n_3, n_4, \dots, n_T \in \mathbb{Z}}} e^{-\Tilde{S}(\Tilde{n}_2, \Tilde{\theta}_1, *)},
\end{align}
where $\Tilde{S}(\Tilde{n}_2, \Tilde{\theta}_1, *) = S(0, \Tilde{n}_2, \Tilde{\theta}_1, *)$, i.e. $n_1$ is dropped from the dependencies.

We can repeat this process to absorb $\Tilde{n}_2$ by
\begin{align}
    \Tilde{\theta}_2 &= \theta_2 + 2\pi \Tilde{n}_2, \quad \Tilde{\theta}_2\in(-\infty,+\infty), \\
    \Tilde{n}_3 &= n_3 + \Tilde{n}_2, \quad \Tilde{n}_3\in\mathbb{Z}, \\
    &\dots \nonumber
\end{align}
Until
\begin{align}
    \Tilde{\theta}_{T-1} &= \theta_{T-1} + 2\pi \Tilde{n}_{T-1}, \quad \Tilde{\theta}_{T-1}\in(-\infty,+\infty), \\
    \Tilde{n}_T &= n_T + \Tilde{n}_{T-1} = \sum_{i=1}^T n_i, \quad \Tilde{n}_T\in\mathbb{Z}, 
\end{align}
which makes the approximate partition function into 
\begin{equation}
    Z = \sum_{\substack{\theta_0 \in [0, 2\pi) \\ \Tilde{\theta}_1, \Tilde{\theta}_2, \dots, \Tilde{\theta}_{T-1}\in(-\infty,+\infty), \\ \Tilde{n}_T \in \mathbb{Z}}} e^{-\frac{\Tilde{\beta}}{2 d\tau} \left[ (\Tilde{\theta}_1 - \theta_0)^2 + (\Tilde{\theta}_2 - \Tilde{\theta}_1)^2 + \cdots + (\Tilde{\theta}_{T-1} - \Tilde{\theta}_{T-2})^2 + (\theta_0 - \Tilde{\theta}_{T-1} + 2\pi \Tilde{n}_T)^2\right]}.
\end{equation}
Note that the $\theta_0$ is not lifted, and the extra integer degree of freedom on the last layer, $\Tilde{n}_T$, cannot be absorbed. It cannot be absorbed because $\Tilde{n}_T = \sum_{i=1}^T n_i$ has a physical meaning of the total flux in the Polyakov loop, and thus is gauge invariant.

\begin{figure}[htp!]
    \centering
        \begin{tikzpicture}
            \draw (0,1) -- (1,1) node[right] {$\theta_0 \in [0,2\pi)$};
            \foreach \y/\i in {5/{{T-1}}, 4/3, 3/2, 2/1} {
                \draw (0,\y) -- (1,\y) node[right] {$\theta_\i \in (-\infty,+\infty)$};
            }
            \draw[dashed] (0,6) -- (1,6) node[right] {$\theta_T = \theta_0$};
            \node at (0.5,4.5) {$\vdots$};
            \node[left] at (0,5.5) {$\Tilde{n}_T$};
        \end{tikzpicture}
    \caption{Illustration of the variables in a single quantum rotor model, i.e. a $0+1$-D compact quantum field. The vertical direction is time. There are $T$ number of discrete time slices, and have periodic boundary condition. After absorbing most of the extra integer degrees of freedom introduced by the Villain approximation, most of the compact variables are lifted to $(-\infty,+\infty)$. However, the variable in the bottom layer, $\theta_0$, is not lifted. Also, the integer in the last layer, $\Tilde{n}_T$, cannot be absorbed.}
    \label{fig:rotor2}
\end{figure}

Now with this quadratic formulation of the partition function, we can apply the standard method to get the Hamiltonian formulation. More explicitly, we set up the Hilbert space, write down the transfer matrix, construct the Hamiltonian operator, and recognize the non-absorb-able degree of freedom as constraint on the Hilbert space.

When setting up the Hilbert space, we choose the set of basis to be the configurations of a middle time layer. In our example here, the set of basis is $\{|\Tilde{\theta}\rangle: \Tilde{\theta} \in (-\infty, +\infty)\}$. Note that even if we start from a compact model, after the Villain approximation and absorption of degrees of freedom, we end up with a seemingly non-compact set of basis. We do not need to worry at this moment, as later the constraint on the Hilbert space will bring us back to a compact model.

We also note that the first time layer (i.e. the bottom layer in Fig.~\ref{fig:rotor2}) has fewer configurations. Because of the periodicity of the variable $\theta_0$, we can duplicate the partition function to extend its configuration space to match the one of the middle layers. In a non-rigorous way, we can do
\begin{equation}
    \sum_{\theta_0\in[0,2\pi)} = \frac{1}{|\mathbb{Z}|} \sum_{\Tilde{\theta}_0\in(-\infty,+\infty)}.
\end{equation}
This $|\mathbb{Z}|$ factor will later be cancelled out when we set the constraint on the Hilbert space. We will return the degree of freedom we borrowed here.

Given the Hilbert space, we want to write down the transfer matrix $\hat{T}$. For two neighboring middle time layers, we have the matrix elements of the transfer matrix as
\begin{equation}
    \langle \Tilde{\theta}_{i+1} | \hat{T} | \Tilde{\theta}_i \rangle = e^{- \frac{\Tilde{\beta}}{2 d\tau} (\Tilde{\theta}_{i+1} - \Tilde{\theta}_i)^2},
\end{equation}
where $\Tilde{\theta}_{i+1}, \Tilde{\theta}_i \in (-\infty, +\infty)$ are configurations on the two neighboring middle time layers.

For the last time layer (i.e. the top layer in Fig.~\ref{fig:rotor2}), we have a different transfer matrix $\hat{T}_{top}$ with matrix elements as
\begin{equation}
    \langle \Tilde{\theta}_{0} | \hat{T}_{top} | \Tilde{\theta}_{T-1} \rangle = e^{- \frac{\Tilde{\beta}}{2 d\tau} (\Tilde{\theta}_{0} - \Tilde{\theta}_{T-1} + 2\pi \Tilde{n}_T)^2}.
\end{equation}

In terms of the transfer matrices, the partition function looks like
\begin{equation}
    Z = \frac{1}{|\mathbb{Z}|} \sum_{\Tilde{n}_T \in \mathbb{Z}} Tr(\hat{T}_{top} \hat{T}^{T-1}),
\end{equation}
where the $|\mathbb{Z}|$ factor was borrowed earlier to match the configuration spaces.

Now we construct the Hamiltonian operator. For a general function $f(\hat{p})$ of the momentum operator $\hat{p}$, its matrix elements can be calculated by inserting the complete momentum basis $\mathbb{I}=\int_{-\infty}^{\infty}\frac{dp}{2\pi}|p\rangle\langle p|$. Therefore, for any $\Tilde{\theta}', \Tilde{\theta}\in(-\infty,+\infty)$,
\begin{equation}
    \langle \Tilde{\theta}' | f(\hat{p}) | \Tilde{\theta} \rangle = \int_{-\infty}^{\infty}\frac{dp}{2\pi} e^{i p (\theta ' - \theta)} f(p),
\end{equation}
from which we can do the Fourier transform to get back the function $f(\cdot)$,
\begin{equation}
    f(p) = \int_{-\infty}^{\infty} d\Tilde{\theta} e^{-i p \Tilde{\theta}} \langle \Tilde{\theta} | f(\hat{p}) | 0 \rangle.
\end{equation}

Do Fourier transform on the transfer matrix elements:
\begin{align}
    \int_{-\infty}^{\infty} d\Tilde{\theta} e^{-i p \Tilde{\theta}} \langle \Tilde{\theta} | \hat{T} | 0 \rangle &= \int_{-\infty}^{\infty} d\Tilde{\theta} e^{-i p \Tilde{\theta}} e^{- \frac{\Tilde{\beta}}{2 d\tau} \Tilde{\theta}^2}  \nonumber \\
    &\propto e^{- \frac{d\tau}{2\Tilde{\beta}} p^2},
\end{align}
and
\begin{align}
    \int_{-\infty}^{\infty} d\Tilde{\theta} e^{-i p \Tilde{\theta}} \langle \Tilde{\theta} | \hat{T}_{top} | 0 \rangle &= \int_{-\infty}^{\infty} d\Tilde{\theta} e^{-i p \Tilde{\theta}} e^{- \frac{\Tilde{\beta}}{2 d\tau} (\Tilde{\theta} + 2\pi \Tilde{n}_T)^2}  \nonumber \\
    &\propto e^{i p 2\pi \Tilde{n}_T} e^{- \frac{d\tau}{2\Tilde{\beta}} p^2} .
\end{align}

The partition function now becomes
\begin{align}
    Z &= \frac{1}{|\mathbb{Z}|} \sum_{\Tilde{n}_T \in \mathbb{Z}} Tr(e^{i \hat{p} 2\pi \Tilde{n}_T} e^{- \frac{d\tau}{2\Tilde{\beta}} \hat{p}^2 T} ) \nonumber \\
    &= Tr(\hat{C} e^{-\hat{H}\tau}),
\end{align}
where $\hat{H}=\frac{1}{2\Tilde{\beta}}\hat{p}^2$ is the Hamiltonian operator, $\tau = d\tau T$ is the total Euclidean time, and 
\begin{align}
    \hat{C} &= \frac{1}{|\mathbb{Z}|} \sum_{\Tilde{n}_T \in \mathbb{Z}} e^{i \hat{p} 2\pi \Tilde{n}_T} \nonumber \\
    &= \frac{1}{|\mathbb{Z}|} \sum_{\Tilde{n}_T \in \mathbb{Z}} (e^{2\pi i \hat{p}}) ^{\Tilde{n}_T} \nonumber \\
    &= \begin{cases} 
        \mathbb{I} & \text{if } e^{2\pi i \hat{p}} = \mathbb{I}, \\
        0 & \text{if } e^{2\pi i \hat{p}} \neq \mathbb{I}.
        \end{cases}
\end{align}
Note that the $|\mathbb{Z}|$ factor is cancelled. This sum over non-absorb-able degree of freedom $\Tilde{n}_T$ sets up a constraint on the physical Hilbert space. In order to have a non-zero partition function, the allowed physical states $|\psi\rangle$ in the Hilbert space need to satisfy
\begin{equation}
    e^{2\pi i \hat{p}} |\psi\rangle = |\psi\rangle,
\end{equation}
which tells us that the momentum $\hat{p}$ is quantized, which is expected for a single quantum rotor. This constraint also tells us that the physical wave functions are periodic $\psi(\theta + 2\pi) = \psi(\theta)$. We are back to a compact variable $\theta$.

In conclusion, for a single quantum rotor, this road map of constructing lattice Hamiltonian finally leads to the following result:
\begin{itemize}
    \item Hilbert space basis: $\{|\Tilde{\theta}\rangle: \Tilde{\theta} \in (-\infty, +\infty)\}$,
    \item Hamiltonian operator: $\hat{H}=\frac{1}{2\Tilde{\beta}}\hat{p}^2$,
    \item Constraint on physical states: $e^{2\pi i \hat{p}} |\psi\rangle = |\psi\rangle$.
\end{itemize}

\subsection{General Road Map from Lattice Action to Lattice Hamiltonian}

From the example of deriving the Hamiltonian for a single quantum rotor, we can summarize the procedure to go from a time-discrete action to the Hamiltonian, involving the Villain approximation. In the following sections, we will apply similar procedures to the pure compact lattice Maxwell theory with instantons allowed, to the one with instantons suppressed, and finally to the compact lattice Maxwell-Chern-Simons theory we are interested in. The following steps are conceptual summary of the procedures in the following sections. 

\begin{itemize}
    \item Start from a compact lattice action.
    \item Apply Villain approximation: add an integer degree of freedom for every cosine term in the action. Turn the cosine terms into quadratic.
    \item Gauge fixing the variables to get rid of all gauge degrees of freedom. Absorb some integer degrees of freedom introduced earlier, to lift some compact variables into $(-\infty, +\infty)$.
    \item Set up the Hilbert space with the set of basis to be the configurations in a middle time layer.
    \item Usually the first time layer (bottom layer) has fewer configurations. Duplicate the partition function to match the configuration space with the one in a middle layer. The degrees of freedoms borrowed here will be returned later.
    \item Usually the last time layer (top layer) has extra degrees of freedom. Summing over these degrees of freedom becomes constraints on the Hilbert space. Check that the number of extra degrees of freedom in the top layer equals to the borrowed degrees of freedom by the bottom layer.
    \item Write down the transfer matrix elements for a pair of neighboring middle layers, and for the top layer.
    \item Fourier transform the transfer matrix elements for middle layers to see the Hamiltonian operator.
    \item Fourier transform the transfer matrix elements for the top layer to see the constraints on the physical Hilbert space.
\end{itemize}

\subsection{Pure Maxwell Lattice Theory With Instantons}
\label{subsec:appendix_Maxwell_Instanton_Allowed}

The second example we will show here is the 2+1D U(1) lattice gauge theory. One could allow the existence of instantons in the theory. One could also completely suppress the instantons. In this subsection, we show how to derive the lattice Hamiltonian for the former theory. In next subsection, we will show the instanton-suppressed version.

We have a cubic lattice with periodic boundary condition on all three directions. We set the up direction to be the positive time direction. The U(1) gauge field $A$ lives on the links of the cubic lattice. The lattice action is
\begin{align}
S(A) = &-\frac{\beta_0}{d\tau} \sum_{\substack{\text{time-like} \\ \text{plaquettes}}} 
\left[ \cos \left(a \cdot \begin{tikzpicture}[baseline=-0.5ex]
\draw[thick,->] (0.5,-0.5) -- (-0.5,-0.5);
\draw[thick,->] (0.5,0.5) -- (0.5,-0.5);
\draw[thick,->] (-0.5,0.5) -- (0.5,0.5);
\draw[thick,->] (-0.5,-0.5) -- (-0.5,0.5);
\node[left] at (-0.5,0) {$\frac{d\tau}{a}A_0$};
\node[right] at (0.5,0) {$-\frac{d\tau}{a}A_0$};
\node[above] at (0,0.5) {$A_1$};
\node[below] at (0,-0.5) {$-A_1$};
\end{tikzpicture} \right) + \cos \left(a \cdot \begin{tikzpicture}[baseline=-0.5ex]
\draw[thick,->] (0,0.5,-0.5) -- (0,-0.5,-0.5);
\draw[thick,->] (0,0.5,0.5) -- (0,0.5,-0.5);
\draw[thick,->] (0,-0.5,0.5) -- (0,0.5,0.5);
\draw[thick,->] (0,-0.5,-0.5) -- (0,-0.5,0.5);
\node[left] at (0,0,0.5) {$\frac{d\tau}{a}A_0$};
\node[right] at (0,0,-0.5) {$-\frac{d\tau}{a}A_0$};
\node[above] at (-0.2,0.5,0) {$A_2$};
\node[below] at (0.2,-0.5,0) {$-A_2$};
\end{tikzpicture} \right) \right] \nonumber \\
&- \frac{\beta d\tau}{a^2} \sum_{\substack{\text{space-like} \\ \text{plaquettes}}} \left[ \cos \left(a \cdot \begin{tikzpicture}[baseline=-0.5ex]
\draw[thick,->] (0.5,0,-0.5) -- (-0.5,0,-0.5);
\draw[thick,->] (0.5,0,0.5) -- (0.5,0,-0.5);
\draw[thick,->] (-0.5,0,0.5) -- (0.5,0,0.5);
\draw[thick,->] (-0.5,0,-0.5) -- (-0.5,0,0.5);
\node[below] at (0,0,0.5) {$A_1$};
\node[above] at (0,0,-0.5) {$-A_1$};
\node[left] at (-0.5,0.1) {$-A_2$};
\node[right] at (0.5,-0.1) {$A_2$};
\end{tikzpicture} \right) \right],
\end{align}
where $\beta_0$ and $\beta$ are the coefficients in the time and space direction respectively, $d\tau$ and $a$ are the lattice spacing in the time and space direction respectively, $A_0$ is the gauge field in the time direction, and $A_1$, $A_2$ are the gauge fields in the two spacial direction.

We can also define the lattice difference of a field
\begin{align}
    \Delta_0 A_{x;\mu} &= \frac{A_{x+\hat{0};\mu} - A_{x;\mu}}{d\tau}, \\
    \Delta_i A_{x;\mu} &= \frac{A_{x+\hat{i};\mu} - A_{x;\mu}}{a},
\end{align}
where $\Delta_0$ represents the time difference, $\Delta_{i\in\{1,2\}}$ represents the spacial difference, $d\tau$ and $a$ are the lattice spacing in the time and space direction, respectively.

Using the lattice difference notation, the action is 
\begin{align}
    S(A) = &-\frac{\beta_0}{d\tau} \sum_{\substack{x\in\text{sites} \\ i\in\{1,2\}}} \cos(a d\tau (\Delta_0 A_{x;i} - \Delta_i A_{x;0}))
    - \frac{\beta d\tau}{a^2} \sum_{x\in\text{sites}} \cos(a^2(\Delta_1 A_{x;2} - \Delta_2 A_{x;1})).
\end{align}

The partition function is 
\begin{align}
    Z = \sum_{A} e^{-S(A)},
\end{align}
where the summation is over the gauge equivalent classes, i.e. all possible configurations of $A$ quotient by the gauge transformations. In particular, $A_0\in[0,2\pi/d\tau)$, $A_{1,2}\in[0,2\pi/a)$, and the configurations related by local gauge transformations are viewed as equivalent and are not duplicated in the sum.

We follow our road map and apply the Villain approximation, replace the cosine terms by quadratic terms, and introduce integer degrees of freedom on every plaquettes. The approximated partition function is
\begin{align}
    Z = \sum_{A, n} e^{-S(A, n)},
\end{align}
where 
\begin{align}
    S(A, n) &= \frac{\beta_0}{2 d\tau} \sum_{\substack{x\in\text{sites} \\ i\in\{1,2\}}} (a d\tau (\Delta_0 A_{x;i} - \Delta_i A_{x;0}) + 2\pi n_{x;0,i})^2 \nonumber \\
    &\quad + \frac{\beta d\tau}{2 a^2} \sum_{x\in\text{sites}} (a^2(\Delta_1 A_{x;2} - \Delta_2 A_{x;1}) + 2\pi n_{x;1,2})^2,
\end{align}
where $n_{x;\mu,\nu}\in\mathbb{Z}$ is the integer degree of freedom on the plaquette at location $x$ extended in $\mu,\nu$ directions. The sum over $A$ has the same meaning as before.

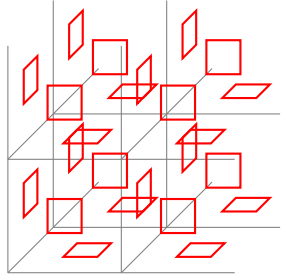
\begin{figure}[htp!]
    \centering
        \begin{tikzpicture}[scale=1.5, x={(1cm,0cm)}, y={(0cm,1cm)}, z={(0.4cm,0.4cm)}]
        \tikzset{
            grid line/.style={thin, gray},
            small square/.style={red, thick}
        }
    
        \foreach \x in {0,1} {
            \foreach \y in {0,1} {
                \draw[grid line] (\x,0, \y) -- (\x,2, \y);
                \draw[grid line] (0,\x, \y) -- (2,\x, \y);
                \draw[grid line] (\y,\x, 0) -- (\y,\x, 2);
            }
        }
    
        \foreach \a in {0.35} {
            \foreach \b in {0.65} {
                \foreach \x in {0,1} {
                    \foreach \y in {0,1} {
                        \foreach \z in {0,1} {
                            \draw[small square] (\x+\a,\y+\a, \z) -- (\x+\b,\y+\a, \z) -- (\x+\b,\y+\b, \z) -- (\x+\a,\y+\b, \z) -- cycle;
            
                            \draw[small square] (\x,\y+\a, \z+\a) -- (\x,\y+\b, \z+\a) -- (\x,\y+\b, \z+\b) -- (\x,\y+\a, \z+\b) -- cycle;
            
                            \draw[small square] (\x+\a,\y, \z+\a) -- (\x+\b,\y, \z+\a) -- (\x+\b,\y, \z+\b) -- (\x+\a,\y, \z+\b) -- cycle;
                        }
                    }
                }
            }
        }
    \end{tikzpicture}
    \caption{Illustration of the integer degrees of freedom in the cubic lattice. On every plaquette, the Villain approximation introduces an integer variable, shown as the red squares in the figure.}
    \label{fig:ndof}
\end{figure}

Written in the graphic representation, it is 
\begin{align}
S(A,n) &= \frac{\beta_0}{2 d\tau} \sum_{\substack{\text{time-like} \\ \text{plaquettes}}} 
\left[  \left(a \cdot \begin{tikzpicture}[baseline=-0.5ex]
\draw[thick,->] (0.5,-0.5) -- (-0.5,-0.5);
\draw[thick,->] (0.5,0.5) -- (0.5,-0.5);
\draw[thick,->] (-0.5,0.5) -- (0.5,0.5);
\draw[thick,->] (-0.5,-0.5) -- (-0.5,0.5);
\node[left] at (-0.5,0) {$\frac{d\tau}{a}A_0$};
\node[right] at (0.5,0) {$-\frac{d\tau}{a}A_0$};
\node[above] at (0,0.5) {$A_1$};
\node[below] at (0,-0.5) {$-A_1$};
\end{tikzpicture} 
+ 2\pi\ 
\begin{tikzpicture}[baseline=-0.5ex]
\draw[thick,dashed] (0.5,-0.5) -- (-0.5,-0.5);
\draw[thick,dashed] (0.5,0.5) -- (0.5,-0.5);
\draw[thick,dashed] (-0.5,0.5) -- (0.5,0.5);
\draw[thick,dashed] (-0.5,-0.5) -- (-0.5,0.5);
\draw[thick] (-0.35,-0.35) rectangle (0.35,0.35);
\node at (0,0) {$n_{0,1}$};
\end{tikzpicture} 
\right)^2 \right. \nonumber \\
&\left. \quad\qquad\qquad\qquad +  \left(a \cdot \begin{tikzpicture}[baseline=-0.5ex]
\draw[thick,->] (0,0.5,-0.5) -- (0,-0.5,-0.5);
\draw[thick,->] (0,0.5,0.5) -- (0,0.5,-0.5);
\draw[thick,->] (0,-0.5,0.5) -- (0,0.5,0.5);
\draw[thick,->] (0,-0.5,-0.5) -- (0,-0.5,0.5);
\node[left] at (0,0,0.5) {$\frac{d\tau}{a}A_0$};
\node[right] at (0,0,-0.5) {$-\frac{d\tau}{a}A_0$};
\node[above] at (-0.2,0.5,0) {$A_2$};
\node[below] at (0.2,-0.5,0) {$-A_2$};
\end{tikzpicture} 
+ 2\pi\ 
\begin{tikzpicture}[baseline=-0.5ex]
\draw[thick,dashed] (0,0.5,-0.5) -- (0,-0.5,-0.5);
\draw[thick,dashed] (0,0.5,0.5) -- (0,0.5,-0.5);
\draw[thick,dashed] (0,-0.5,0.5) -- (0,0.5,0.5);
\draw[thick,dashed] (0,-0.5,-0.5) -- (0,-0.5,0.5);
\draw[thick] (0,-0.25,-0.25) -- (0,-0.25,0.25);
\draw[thick] (0,-0.25,0.25) -- (0,0.25,0.25);
\draw[thick] (0,0.25,0.25) -- (0,0.25,-0.25);
\draw[thick] (0,0.25,-0.25) -- (0,-0.25,-0.25);
\node[right] at (0.15,0,0) {$n_{0,2}$};
\end{tikzpicture} 
\right)^2 \  \right] \nonumber \\
&\quad + \frac{\beta d\tau}{2 a^2} \sum_{\substack{\text{space-like} \\ \text{plaquettes}}} \left[  \left(a \cdot \begin{tikzpicture}[baseline=-0.5ex]
\draw[thick,->] (0.5,0,-0.5) -- (-0.5,0,-0.5);
\draw[thick,->] (0.5,0,0.5) -- (0.5,0,-0.5);
\draw[thick,->] (-0.5,0,0.5) -- (0.5,0,0.5);
\draw[thick,->] (-0.5,0,-0.5) -- (-0.5,0,0.5);
\node[below] at (0,0,0.5) {$A_1$};
\node[above] at (0,0,-0.5) {$-A_1$};
\node[left] at (-0.5,0.1) {$-A_2$};
\node[right] at (0.5,-0.1) {$A_2$};
\end{tikzpicture} 
+ 2\pi\ 
\begin{tikzpicture}[baseline=-0.5ex]
\draw[thick,dashed] (0.5,0,-0.5) -- (-0.5,0,-0.5);
\draw[thick,dashed] (0.5,0,0.5) -- (0.5,0,-0.5);
\draw[thick,dashed] (-0.5,0,0.5) -- (0.5,0,0.5);
\draw[thick,dashed] (-0.5,0,-0.5) -- (-0.5,0,0.5);
\draw[thick] (-0.25,0,-0.25) -- (-0.25,0,0.25);
\draw[thick] (-0.25,0,0.25) -- (0.25,0,0.25);
\draw[thick] (0.25,0,0.25) -- (0.25,0,-0.25);
\draw[thick] (0.25,0,-0.25) -- (-0.25,0,-0.25);
\node[above] at (0,0.15,0) {$n_{1,2}$};
\end{tikzpicture} 
\right)^2 \  \right],
\label{eq:appendix_maxwell_allow_S}
\end{align}
which has the following gauge redundancies:
\begin{align}
    S\left(
    \begin{tikzpicture}[baseline=-0.5ex]
        \draw[thick] (-1,0) -- (1,0);
        \draw[thick] (0,-1) -- (0,1);
        \draw[thick] (0,0,-1) -- (0,0,1);
        \fill (0,0) circle (3pt);
        \node at (-1.1,-0.3) {$A_1 - \frac{\lambda}{a}$};
        \node at (-0.8,-0.8) {$A_2 - \frac{\lambda}{a}$};
        \node at (0.8,-0.8) {$A_0 - \frac{\lambda}{d\tau}$};
        \node at (1,-0.3) {$A_1 + \frac{\lambda}{a}$};
        \node at (1.1,0.3) {$A_2 + \frac{\lambda}{a}$};
        \node at (0.8,0.8) {$A_0 + \frac{\lambda}{d\tau}$};
    \end{tikzpicture}
    , *
    \right)
    =
    S\left(
    \begin{tikzpicture}[baseline=-0.5ex]
        \draw[thick] (-1,0) -- (1,0);
        \draw[thick] (0,-1) -- (0,1);
        \draw[thick] (0,0,-1) -- (0,0,1);
        \fill (0,0) circle (3pt);
        \node at (-1.1,-0.3) {$A_1$};
        \node at (-0.6,-0.6) {$A_2$};
        \node at (0.4,-0.8) {$A_0$};
        \node at (1,-0.3) {$A_1$};
        \node at (0.7,0.3) {$A_2$};
        \node at (0.4,0.8) {$A_0$};
    \end{tikzpicture}
    , *
    \right) 
    \label{eq:appendix_maxwell_allow_local}
\end{align}
\begin{align}
    S\left(
    \begin{tikzpicture}[baseline=-0.5ex]
        \draw[thick,->] (0,-0.5) -- (0,0.5);
        \draw[thick] (0.15,-0.35,0) rectangle (0.85,0.35,0);
        \draw[thick] (-0.15,-0.35,0) rectangle (-0.85,0.35,0);
        \draw[thick] (0,0.25,0.25) -- (0,0.25,0.75);
        \draw[thick] (0,0.25,0.75) -- (0,-0.25,0.75);
        \draw[thick] (0,-0.25,0.75) -- (0,-0.25,0.25);
        \draw[thick] (0,-0.25,0.25) -- (0,0.25,0.25);
        \draw[thick] (0,0.25,-0.25) -- (0,0.25,-0.75);
        \draw[thick] (0,0.25,-0.75) -- (0,-0.25,-0.75);
        \draw[thick] (0,-0.25,-0.75) -- (0,-0.25,-0.25);
        \draw[thick] (0,-0.25,-0.25) -- (0,0.25,-0.25);
        \node at (-0.5,0.8) {$A_0+\frac{2\pi}{d\tau}$};
        \node at (0.9,0.8) {$n_{0,2}-1$};
        \node at (1.5,0) {$n_{0,1}-1$};
        \node at (-0.9,-0.8) {$n_{0,2}+1$};
        \node at (-1.5,0) {$n_{0,1}+1$};
    \end{tikzpicture}
    , *
    \right) 
    =
    S\left(
    \begin{tikzpicture}[baseline=-0.5ex]
        \draw[thick,->] (0,-0.5) -- (0,0.5);
        \draw[thick] (0.15,-0.35,0) rectangle (0.85,0.35,0);
        \draw[thick] (-0.15,-0.35,0) rectangle (-0.85,0.35,0);
        \draw[thick] (0,0.25,0.25) -- (0,0.25,0.75);
        \draw[thick] (0,0.25,0.75) -- (0,-0.25,0.75);
        \draw[thick] (0,-0.25,0.75) -- (0,-0.25,0.25);
        \draw[thick] (0,-0.25,0.25) -- (0,0.25,0.25);
        \draw[thick] (0,0.25,-0.25) -- (0,0.25,-0.75);
        \draw[thick] (0,0.25,-0.75) -- (0,-0.25,-0.75);
        \draw[thick] (0,-0.25,-0.75) -- (0,-0.25,-0.25);
        \draw[thick] (0,-0.25,-0.25) -- (0,0.25,-0.25);
        \node at (-0.2,0.8) {$A_0$};
        \node at (0.5,0.8) {$n_{0,2}$};
        \node at (1.3,0) {$n_{0,1}$};
        \node at (-0.5,-0.8) {$n_{0,2}$};
        \node at (-1.3,0) {$n_{0,1}$};
    \end{tikzpicture}
    , *
    \right) 
    \label{eq:appendix_maxwell_allow_A_0}
\end{align}
\begin{align}
    S\left(
    \begin{tikzpicture}[baseline=-0.5ex]
        \draw[thick,->] (0,0) -- (1,0);
        \draw[thick] (0.15,0.15,0) rectangle (0.85,0.85,0);
        \draw[thick] (0.15,-0.15,0) rectangle (0.85,-0.85,0);
        \draw[thick] (0.25,0,0.25) -- (0.25,0,0.75);
        \draw[thick] (0.25,0,0.75) -- (0.75,0,0.75);
        \draw[thick] (0.75,0,0.75) -- (0.75,0,0.25);
        \draw[thick] (0.75,0,0.25) -- (0.25,0,0.25);
        \draw[thick] (0.25,0,-0.25) -- (0.25,0,-0.75);
        \draw[thick] (0.25,0,-0.75) -- (0.75,0,-0.75);
        \draw[thick] (0.75,0,-0.75) -- (0.75,0,-0.25);
        \draw[thick] (0.75,0,-0.25) -- (0.25,0,-0.25);
        \node at (1.7,0) {$A_1+\frac{2\pi}{a}$};
        \node at (1.7,0.4) {$n_{1,2}-1$};
        \node at (1.5,0.8) {$n_{0,1}+1$};
        \node at (-0.7,-0.4) {$n_{1,2}+1$};
        \node at (-0.5,-0.8) {$n_{0,1}-1$};
    \end{tikzpicture}
    , *
    \right) 
    =
    S\left(
    \begin{tikzpicture}[baseline=-0.5ex]
        \draw[thick,->] (0,0) -- (1,0);
        \draw[thick] (0.15,0.15,0) rectangle (0.85,0.85,0);
        \draw[thick] (0.15,-0.15,0) rectangle (0.85,-0.85,0);
        \draw[thick] (0.25,0,0.25) -- (0.25,0,0.75);
        \draw[thick] (0.25,0,0.75) -- (0.75,0,0.75);
        \draw[thick] (0.75,0,0.75) -- (0.75,0,0.25);
        \draw[thick] (0.75,0,0.25) -- (0.25,0,0.25);
        \draw[thick] (0.25,0,-0.25) -- (0.25,0,-0.75);
        \draw[thick] (0.25,0,-0.75) -- (0.75,0,-0.75);
        \draw[thick] (0.75,0,-0.75) -- (0.75,0,-0.25);
        \draw[thick] (0.75,0,-0.25) -- (0.25,0,-0.25);
        \node at (1.4,0) {$A_1$};
        \node at (1.5,0.4) {$n_{1,2}$};
        \node at (1.3,0.8) {$n_{0,1}$};
        \node at (-0.5,-0.4) {$n_{1,2}$};
        \node at (-0.3,-0.8) {$n_{0,1}$};
    \end{tikzpicture}
    , *
    \right) 
    \label{eq:appendix_maxwell_allow_A_1}
\end{align}
\begin{align}
    S\left(
    \begin{tikzpicture}[baseline=-0.5ex]
        \draw[thick,->] (0,0,0.5) -- (0,0,-0.5);
        \draw[thick] (0,0.25,0.25) -- (0,0.25,-0.25);
        \draw[thick] (0,0.25,-0.25) -- (0,0.75,-0.25);
        \draw[thick] (0,0.75,-0.25) -- (0,0.75,0.25);
        \draw[thick] (0,0.75,0.25) -- (0,0.25,0.25);
        \draw[thick] (0,-0.25,0.25) -- (0,-0.25,-0.25);
        \draw[thick] (0,-0.25,-0.25) -- (0,-0.75,-0.25);
        \draw[thick] (0,-0.75,-0.25) -- (0,-0.75,0.25);
        \draw[thick] (0,-0.75,0.25) -- (0,-0.25,0.25);
        \draw[thick] (0.25,0,0.25) -- (0.25,0,-0.25);
        \draw[thick] (0.25,0,-0.25) -- (0.75,0,-0.25);
        \draw[thick] (0.75,0,-0.25) -- (0.75,0,0.25);
        \draw[thick] (0.75,0,0.25) -- (0.25,0,0.25);
        \draw[thick] (-0.25,0,0.25) -- (-0.25,0,-0.25);
        \draw[thick] (-0.25,0,-0.25) -- (-0.75,0,-0.25);
        \draw[thick] (-0.75,0,-0.25) -- (-0.75,0,0.25);
        \draw[thick] (-0.75,0,0.25) -- (-0.25,0,0.25);
        \node at (0.8,0.5) {$A_2+\frac{2\pi}{a}$};
        \node at (1.5,0) {$n_{1,2}+1$};
        \node at (0.8,-0.5) {$n_{0,2}-1$};
        \node at (-1.5,0) {$n_{1,2}-1$};
        \node at (-0.8,0.5) {$n_{0,2}+1$};
    \end{tikzpicture}
    , *
    \right) 
    =
    S\left(
    \begin{tikzpicture}[baseline=-0.5ex]
        \draw[thick,->] (0,0,0.5) -- (0,0,-0.5);
        \draw[thick] (0,0.25,0.25) -- (0,0.25,-0.25);
        \draw[thick] (0,0.25,-0.25) -- (0,0.75,-0.25);
        \draw[thick] (0,0.75,-0.25) -- (0,0.75,0.25);
        \draw[thick] (0,0.75,0.25) -- (0,0.25,0.25);
        \draw[thick] (0,-0.25,0.25) -- (0,-0.25,-0.25);
        \draw[thick] (0,-0.25,-0.25) -- (0,-0.75,-0.25);
        \draw[thick] (0,-0.75,-0.25) -- (0,-0.75,0.25);
        \draw[thick] (0,-0.75,0.25) -- (0,-0.25,0.25);
        \draw[thick] (0.25,0,0.25) -- (0.25,0,-0.25);
        \draw[thick] (0.25,0,-0.25) -- (0.75,0,-0.25);
        \draw[thick] (0.75,0,-0.25) -- (0.75,0,0.25);
        \draw[thick] (0.75,0,0.25) -- (0.25,0,0.25);
        \draw[thick] (-0.25,0,0.25) -- (-0.25,0,-0.25);
        \draw[thick] (-0.25,0,-0.25) -- (-0.75,0,-0.25);
        \draw[thick] (-0.75,0,-0.25) -- (-0.75,0,0.25);
        \draw[thick] (-0.75,0,0.25) -- (-0.25,0,0.25);
        \node at (0.5,0.5) {$A_2$};
        \node at (1.2,0) {$n_{1,2}$};
        \node at (0.5,-0.5) {$n_{0,2}$};
        \node at (-1.2,0) {$n_{1,2}$};
        \node at (-0.5,0.5) {$n_{0,2}$};
    \end{tikzpicture}
    , *
    \right) 
    \label{eq:appendix_maxwell_allow_A_2}
\end{align}
for $\forall \lambda\in\mathbb{R}$. The $*$ in the parentheses represents all other variables, which are the same on the left and right hand sides in the equations. Note that comparing to the Eqs.~\eqref{eq:A_0_transform},~\eqref{eq:A_1_transform}, and~\eqref{eq:A_2_transform} in the main text, here the $S$ itself is invariant.

Next, we do gauge fixing by applying the invariant transformations above. We want to get rid of all gauge degrees of freedom, and to fix the gauge field and integer field configurations into a canonical form, where we can explicitly read out the true physical degrees of freedom. 

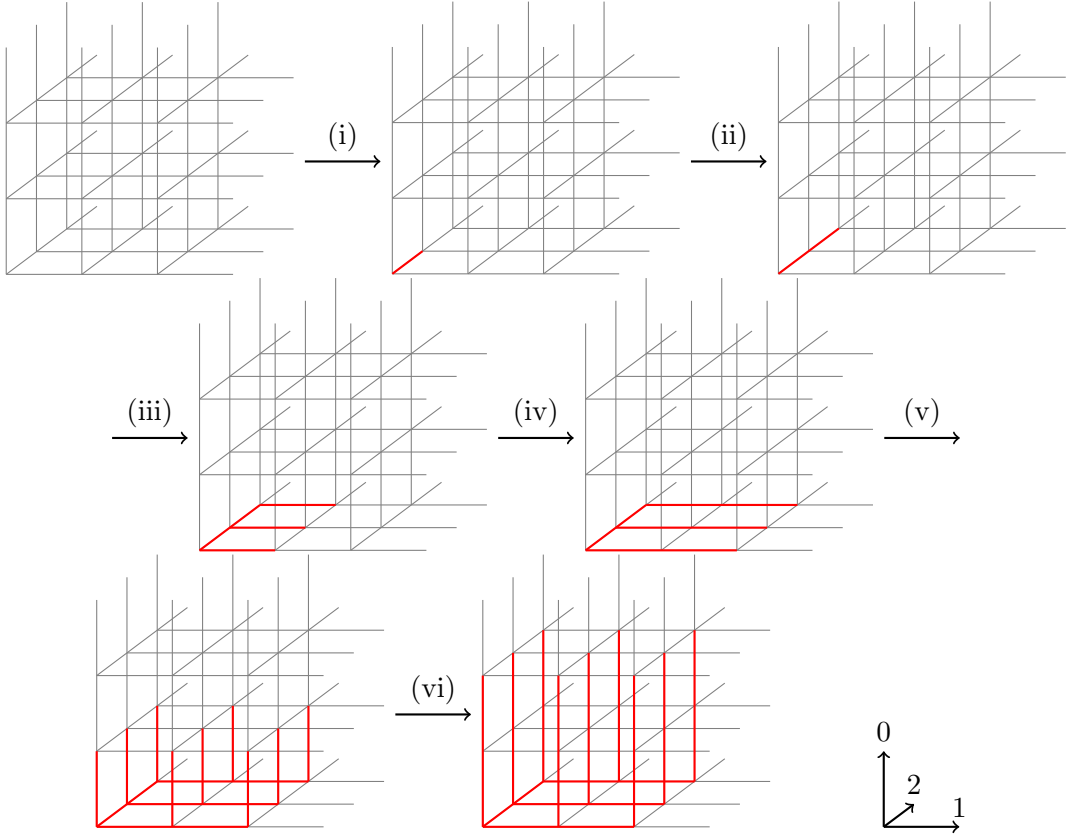
\begin{figure}[htp!]
    \centering
    \begin{tikzpicture}[scale=1.0, x={(1cm,0cm)}, y={(0cm,1cm)}, z={(0.4cm,0.3cm)}]
        \tikzset{
            grid line/.style={thin, gray},
            zero line/.style={red, thick}
        }
    
        \foreach \x in {0,1,2} {
            \foreach \y in {0,1,2} {
                \draw[grid line] (\x,0, \y) -- (\x,3, \y);
                \draw[grid line] (0,\x, \y) -- (3,\x, \y);
                \draw[grid line] (\y,\x, 0) -- (\y,\x, 3);
            }
        }
    
    \end{tikzpicture}
    \begin{tikzpicture}[baseline=-1.5cm]
        \draw[thick,->] (0,0) -- (1,0);
        \node[above] at (0.5,0) {(i)};
    \end{tikzpicture}
    \begin{tikzpicture}[scale=1.0, x={(1cm,0cm)}, y={(0cm,1cm)}, z={(0.4cm,0.3cm)}]
        \tikzset{
            grid line/.style={thin, gray},
            zero line/.style={red, thick}
        }
    
        \foreach \x in {0,1,2} {
            \foreach \y in {0,1,2} {
                \draw[grid line] (\x,0, \y) -- (\x,3, \y);
                \draw[grid line] (0,\x, \y) -- (3,\x, \y);
                \draw[grid line] (\y,\x, 0) -- (\y,\x, 3);
            }
        }

        \draw[zero line] (0,0,0) -- (0,0,1);
    
    \end{tikzpicture}
    \begin{tikzpicture}[baseline=-1.5cm]
        \draw[thick,->] (0,0) -- (1,0);
        \node[above] at (0.5,0) {(ii)};
    \end{tikzpicture}
    \begin{tikzpicture}[scale=1.0, x={(1cm,0cm)}, y={(0cm,1cm)}, z={(0.4cm,0.3cm)}]
        \tikzset{
            grid line/.style={thin, gray},
            zero line/.style={red, thick}
        }
    
        \foreach \x in {0,1,2} {
            \foreach \y in {0,1,2} {
                \draw[grid line] (\x,0, \y) -- (\x,3, \y);
                \draw[grid line] (0,\x, \y) -- (3,\x, \y);
                \draw[grid line] (\y,\x, 0) -- (\y,\x, 3);
            }
        }

        \draw[zero line] (0,0,0) -- (0,0,2);
    
    \end{tikzpicture}
    \begin{tikzpicture}[baseline=-1.5cm]
        \draw[thick,->] (0,0) -- (1,0);
        \node[above] at (0.5,0) {(iii)};
    \end{tikzpicture}
    \begin{tikzpicture}[scale=1.0, x={(1cm,0cm)}, y={(0cm,1cm)}, z={(0.4cm,0.3cm)}]
        \tikzset{
            grid line/.style={thin, gray},
            zero line/.style={red, thick}
        }
    
        \foreach \x in {0,1,2} {
            \foreach \y in {0,1,2} {
                \draw[grid line] (\x,0, \y) -- (\x,3, \y);
                \draw[grid line] (0,\x, \y) -- (3,\x, \y);
                \draw[grid line] (\y,\x, 0) -- (\y,\x, 3);
            }
        }

        \draw[zero line] (0,0,0) -- (0,0,2);
        \foreach \x in {0,1,2} {
            \draw[zero line] (0,0,\x) -- (1,0,\x);
        }
    
    \end{tikzpicture}
    \begin{tikzpicture}[baseline=-1.5cm]
        \draw[thick,->] (0,0) -- (1,0);
        \node[above] at (0.5,0) {(iv)};
    \end{tikzpicture}
    \begin{tikzpicture}[scale=1.0, x={(1cm,0cm)}, y={(0cm,1cm)}, z={(0.4cm,0.3cm)}]
        \tikzset{
            grid line/.style={thin, gray},
            zero line/.style={red, thick}
        }
    
        \foreach \x in {0,1,2} {
            \foreach \y in {0,1,2} {
                \draw[grid line] (\x,0, \y) -- (\x,3, \y);
                \draw[grid line] (0,\x, \y) -- (3,\x, \y);
                \draw[grid line] (\y,\x, 0) -- (\y,\x, 3);
            }
        }

        \draw[zero line] (0,0,0) -- (0,0,2);
        \foreach \x in {0,1,2} {
            \draw[zero line] (0,0,\x) -- (2,0,\x);
        }
    
    \end{tikzpicture}
    \begin{tikzpicture}[baseline=-1.5cm]
        \draw[thick,->] (0,0) -- (1,0);
        \node[above] at (0.5,0) {(v)};
    \end{tikzpicture}
    \begin{tikzpicture}[scale=1.0, x={(1cm,0cm)}, y={(0cm,1cm)}, z={(0.4cm,0.3cm)}]
        \tikzset{
            grid line/.style={thin, gray},
            zero line/.style={red, thick}
        }
    
        \foreach \x in {0,1,2} {
            \foreach \y in {0,1,2} {
                \draw[grid line] (\x,0, \y) -- (\x,3, \y);
                \draw[grid line] (0,\x, \y) -- (3,\x, \y);
                \draw[grid line] (\y,\x, 0) -- (\y,\x, 3);
            }
        }

        \draw[zero line] (0,0,0) -- (0,0,2);
        \foreach \x in {0,1,2} {
            \draw[zero line] (0,0,\x) -- (2,0,\x);
        }
        \foreach \x in {0,1,2} {
            \foreach \y in {0,1,2} {
                \draw[zero line] (\x,0, \y) -- (\x,1, \y);
            }
        }
    
    \end{tikzpicture}
    \begin{tikzpicture}[baseline=-1.5cm]
        \draw[thick,->] (0,0) -- (1,0);
        \node[above] at (0.5,0) {(vi)};
    \end{tikzpicture}
    \begin{tikzpicture}[scale=1.0, x={(1cm,0cm)}, y={(0cm,1cm)}, z={(0.4cm,0.3cm)}]
        \tikzset{
            grid line/.style={thin, gray},
            zero line/.style={red, thick}
        }
    
        \foreach \x in {0,1,2} {
            \foreach \y in {0,1,2} {
                \draw[grid line] (\x,0, \y) -- (\x,3, \y);
                \draw[grid line] (0,\x, \y) -- (3,\x, \y);
                \draw[grid line] (\y,\x, 0) -- (\y,\x, 3);
            }
        }

        \draw[zero line] (0,0,0) -- (0,0,2);
        \foreach \x in {0,1,2} {
            \draw[zero line] (0,0,\x) -- (2,0,\x);
        }
        \foreach \x in {0,1,2} {
            \foreach \y in {0,1,2} {
                \draw[zero line] (\x,0, \y) -- (\x,2, \y);
            }
        }
    
    \end{tikzpicture}
    \begin{tikzpicture}
        \draw[white] (0,0) -- (1,0);
    \end{tikzpicture}
    \begin{tikzpicture}[scale=1.0, x={(1cm,0cm)}, y={(0cm,1cm)}, z={(0.4cm,0.3cm)}]
        \draw[thick,->] (0,0,0) -- (1,0,0);
        \draw[thick,->] (0,0,0) -- (0,1,0);
        \draw[thick,->] (0,0,0) -- (0,0,1);
        \node[above] at (1,0,0) {$1$};
        \node[above] at (0,1,0) {$0$};
        \node[above] at (0,0,1) {$2$};
    \end{tikzpicture}
    \caption{Illustration of the gauge fixing steps. The gauge fields on the red links are fixed to zero.}
    \label{fig:appendix_maxwell_allow_fix}
\end{figure}

The gauge fixing is done step by step. First we use Eq.~\eqref{eq:appendix_maxwell_allow_local} to fix some gauge fields into zero. We show this process in Fig.~\ref{fig:appendix_maxwell_allow_fix}. (i) We apply Eq.~\eqref{eq:appendix_maxwell_allow_local} on the site $(x_0,x_1,x_2)=(0,0,1)$ to fix $A_{(0,0,0);2}=0$. (ii) Then we go to the site $(0,0,2)$, $(0,0,3)$, ..., $(0,0,N_2-1)$ to fix $A_{(0,0,1);2}=0$, $A_{(0,0,2);2}=0$, ..., $A_{(0,0,N_2-2);2}=0$, respectively. Note that $A_{(0,0,N_2-1);2}$ cannot be fixed to zero, because it will finally carry the information of $\sum_{x_2=0}^{N_2-1} A_{(0,0,x_2);2}$, which is a gauge invariant quantity. (iii) We then apply Eq.~\eqref{eq:appendix_maxwell_allow_local} on the sites $(0,1,x_2)$, $\forall x_2\in\{0,1,\dots,N_2-1\}$, to fix $A_{(0,0,x_2);1}=0$. (iv) We go to the sites $(0,2,x_2)$, $(0,3,x_2)$, ..., $(0,N_1-1,x_2)$ to fix $A_{(0,1,x_2);1}=0$, $A_{(0,2,x_2);1}=0$, ..., $A_{(0,N_1-2,x_2);1}=0$, respectively. Similarly, note that $A_{(0,N_1-1,x_2);1}$ cannot be fixed to zero. (v) Apply on $(1,x_1,x_2)$, $\forall x_1\in\{0,1,\dots,N_1-1\}$, $\forall x_2\in\{0,1,\dots,N_2-1\}$, to fix $A_{(0,x_1,x_2);0}=0$. (vi) Finally, go to $(2,x_1,x_2)$, $(3,x_1,x_2)$, ..., $(N_0-1,x_1,x_2)$ to fix $A_{(1,x_1,x_2);0}=0$, $A_{(2,x_1,x_2);0}=0$, ..., $A_{(N_0-2,x_1,x_2);0}=0$, respectively. Note that the top layer $A_{(N_0-1,x_1,x_2);0}$ cannot be fixed to zero.

\begin{figure}[htp!]
    \centering
    \begin{tikzpicture}[scale=1.0, x={(1cm,0cm)}, y={(0cm,1cm)}, z={(-0.4cm,0.4cm)}]
        \tikzset{
            grid line/.style={thin, gray},
            zero line/.style={red, thick}
        }
    
        \foreach \x in {0,1,2} {
            \foreach \y in {0,1,2} {
                \draw[zero line] (\y,\x, 0) -- (\y,\x, 1);
            }
        }

        \foreach \x in {0,1,2} {
            \draw[grid line] (\x,0, 0) -- (\x,3, 0);
            \draw[grid line] (0,\x, 0) -- (3,\x, 0);
        }

        \draw[zero line] (0,0,0) -- (0,2,0);
        \foreach \x in {0,1,2} {
            \draw[zero line] (0,\x,0) -- (2,\x,0);
        }
        
        \node[above] at (1.5,3,0) {Bottom Layer};
    
    \end{tikzpicture}
    \begin{tikzpicture}
        \draw[white] (0,0) -- (1,0);
    \end{tikzpicture}
    \begin{tikzpicture}[scale=1.0, x={(1cm,0cm)}, y={(0cm,1cm)}, z={(-0.4cm,0.4cm)}]
        \tikzset{
            grid line/.style={thin, gray},
            zero line/.style={red, thick}
        }
    
        \foreach \x in {0,1,2} {
            \foreach \y in {0,1,2} {
                \draw[zero line] (\y,\x, 0) -- (\y,\x, 1);
            }
        }

        \foreach \x in {0,1,2} {
            \draw[grid line] (\x,0, 0) -- (\x,3, 0);
            \draw[grid line] (0,\x, 0) -- (3,\x, 0);
        }
        
        \node[above] at (1.5,3,0) {Middle Layer};
    
    \end{tikzpicture}
    \begin{tikzpicture}
        \draw[white] (0,0) -- (1,0);
    \end{tikzpicture}
    \begin{tikzpicture}[scale=1.0, x={(1cm,0cm)}, y={(0cm,1cm)}, z={(-0.4cm,0.4cm)}]
        \tikzset{
            grid line/.style={thin, gray},
            zero line/.style={red, thick}
        }
    
        \foreach \x in {0,1,2} {
            \foreach \y in {0,1,2} {
                \draw[grid line] (\y,\x, 0) -- (\y,\x, 1);
            }
        }

        \foreach \x in {0,1,2} {
            \draw[grid line] (\x,0, 0) -- (\x,3, 0);
            \draw[grid line] (0,\x, 0) -- (3,\x, 0);
        }
        
        \node[above] at (1.5,3,0) {Top Layer};
    
    \end{tikzpicture}
    \begin{tikzpicture}[scale=1.0, x={(1cm,0cm)}, y={(0cm,1cm)}, z={(-0.4cm,0.4cm)}]
        \draw[thick,->] (0,0,0) -- (1,0,0);
        \draw[thick,->] (0,0,0) -- (0,1,0);
        \draw[thick,->] (0,0,0) -- (0,0,1);
        \node[above] at (1,0,0) {$1$};
        \node[above] at (0,1,0) {$2$};
        \node[above] at (0,0,1) {$0$};
    \end{tikzpicture}
    \caption{View of the bottom layer, middle layers, and the top layer after first-step gauge fixing. The gauge fields on the red links are fixed to zero.}
    \label{fig:appendix_maxwell_allow_fix2}
\end{figure}
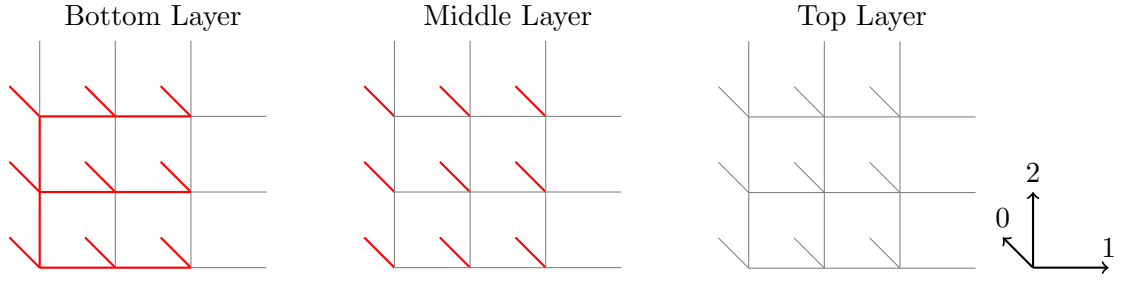

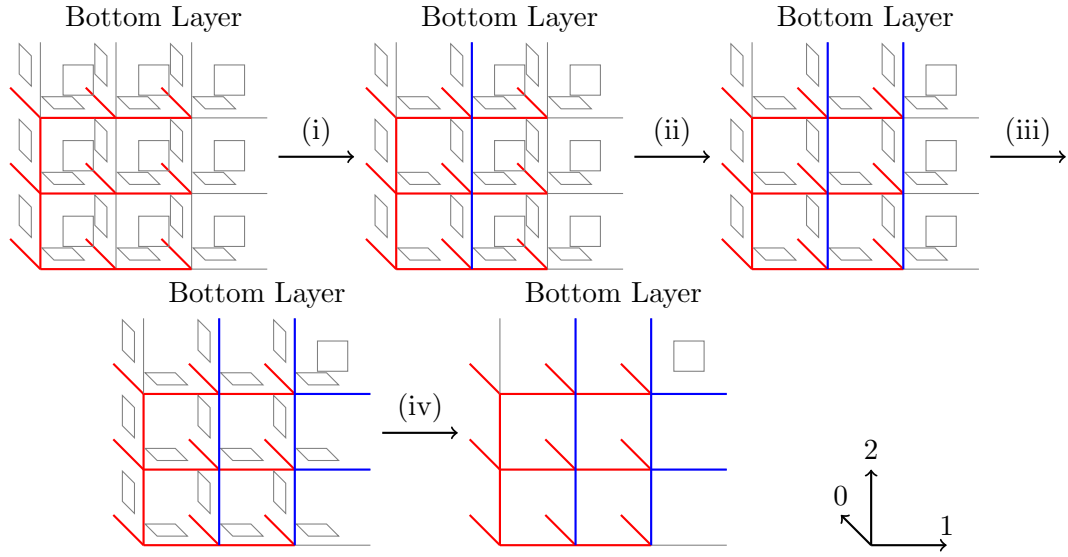
\begin{figure}[htp!]
    \centering
    \begin{tikzpicture}[scale=1.0, x={(1cm,0cm)}, y={(0cm,1cm)}, z={(-0.4cm,0.4cm)}]
        \tikzset{
            grid line/.style={thin, gray},
            zero line/.style={red, thick},
            absorb line/.style={blue, thick},
            small square/.style={thin, gray}
        }
    
        \foreach \x in {0,1,2} {
            \foreach \y in {0,1,2} {
                \draw[zero line] (\y,\x, 0) -- (\y,\x, 1);
            }
        }

        \foreach \x in {0,1,2} {
            \draw[grid line] (\x,0, 0) -- (\x,3, 0);
            \draw[grid line] (0,\x, 0) -- (3,\x, 0);
        }

        \draw[zero line] (0,0,0) -- (0,2,0);
        \foreach \x in {0,1,2} {
            \draw[zero line] (0,\x,0) -- (2,\x,0);
        }    

        \foreach \a in {0.3} {
            \foreach \b in {0.7} {
                \foreach \x in {0,1,2} {
                    \foreach \y in {0,1,2} {
                        \foreach \z in {0} {
                            \draw[small square] (\x+\a,\y+\a, \z) -- (\x+\b,\y+\a, \z) -- (\x+\b,\y+\b, \z) -- (\x+\a,\y+\b, \z) -- cycle;
            
                            \draw[small square] (\x,\y+\a, \z+\a) -- (\x,\y+\b, \z+\a) -- (\x,\y+\b, \z+\b) -- (\x,\y+\a, \z+\b) -- cycle;
            
                            \draw[small square] (\x+\a,\y, \z+\a) -- (\x+\b,\y, \z+\a) -- (\x+\b,\y, \z+\b) -- (\x+\a,\y, \z+\b) -- cycle;
                        }
                    }
                }
            }
        }
        
        \node[above] at (1.5,3,0) {Bottom Layer};
    
    \end{tikzpicture}
    \begin{tikzpicture}[baseline=-1.5cm]
        \draw[thick,->] (0,0) -- (1,0);
        \node[above] at (0.5,0) {(i)};
    \end{tikzpicture}
    \begin{tikzpicture}[scale=1.0, x={(1cm,0cm)}, y={(0cm,1cm)}, z={(-0.4cm,0.4cm)}]
        \tikzset{
            grid line/.style={thin, gray},
            zero line/.style={red, thick},
            absorb line/.style={blue, thick},
            small square/.style={thin, gray}
        }
    
        \foreach \x in {0,1,2} {
            \foreach \y in {0,1,2} {
                \draw[zero line] (\y,\x, 0) -- (\y,\x, 1);
            }
        }

        \foreach \x in {0,1,2} {
            \draw[grid line] (\x,0, 0) -- (\x,3, 0);
            \draw[grid line] (0,\x, 0) -- (3,\x, 0);
        }

        \draw[zero line] (0,0,0) -- (0,2,0);
        \foreach \x in {0,1,2} {
            \draw[zero line] (0,\x,0) -- (2,\x,0);
        }    

        \foreach \a in {0.3} {
            \foreach \b in {0.7} {
                \foreach \x in {0,1,2} {
                    \foreach \y in {0,1,2} {
                        \foreach \z in {0} {
                            \ifthenelse{\x=0}{}{
                            \draw[small square] (\x+\a,\y+\a, \z) -- (\x+\b,\y+\a, \z) -- (\x+\b,\y+\b, \z) -- (\x+\a,\y+\b, \z) -- cycle;
                            }
            
                            \draw[small square] (\x,\y+\a, \z+\a) -- (\x,\y+\b, \z+\a) -- (\x,\y+\b, \z+\b) -- (\x,\y+\a, \z+\b) -- cycle;
            
                            \draw[small square] (\x+\a,\y, \z+\a) -- (\x+\b,\y, \z+\a) -- (\x+\b,\y, \z+\b) -- (\x+\a,\y, \z+\b) -- cycle;
                        }
                    }
                }
            }
        }

        \draw[absorb line] (1,0,0) -- (1,3,0);
        
        \node[above] at (1.5,3,0) {Bottom Layer};
    
    \end{tikzpicture}
    \begin{tikzpicture}[baseline=-1.5cm]
        \draw[thick,->] (0,0) -- (1,0);
        \node[above] at (0.5,0) {(ii)};
    \end{tikzpicture}
    \begin{tikzpicture}[scale=1.0, x={(1cm,0cm)}, y={(0cm,1cm)}, z={(-0.4cm,0.4cm)}]
        \tikzset{
            grid line/.style={thin, gray},
            zero line/.style={red, thick},
            absorb line/.style={blue, thick},
            small square/.style={thin, gray}
        }
    
        \foreach \x in {0,1,2} {
            \foreach \y in {0,1,2} {
                \draw[zero line] (\y,\x, 0) -- (\y,\x, 1);
            }
        }

        \foreach \x in {0,1,2} {
            \draw[grid line] (\x,0, 0) -- (\x,3, 0);
            \draw[grid line] (0,\x, 0) -- (3,\x, 0);
        }

        \draw[zero line] (0,0,0) -- (0,2,0);
        \foreach \x in {0,1,2} {
            \draw[zero line] (0,\x,0) -- (2,\x,0);
        }    

        \foreach \a in {0.3} {
            \foreach \b in {0.7} {
                \foreach \x in {0,1,2} {
                    \foreach \y in {0,1,2} {
                        \foreach \z in {0} {
                            \ifthenelse{\x=2}{
                            \draw[small square] (\x+\a,\y+\a, \z) -- (\x+\b,\y+\a, \z) -- (\x+\b,\y+\b, \z) -- (\x+\a,\y+\b, \z) -- cycle;
                            }{}
            
                            \draw[small square] (\x,\y+\a, \z+\a) -- (\x,\y+\b, \z+\a) -- (\x,\y+\b, \z+\b) -- (\x,\y+\a, \z+\b) -- cycle;
            
                            \draw[small square] (\x+\a,\y, \z+\a) -- (\x+\b,\y, \z+\a) -- (\x+\b,\y, \z+\b) -- (\x+\a,\y, \z+\b) -- cycle;
                        }
                    }
                }
            }
        }
        
        \draw[absorb line] (1,0,0) -- (1,3,0);
        \draw[absorb line] (2,0,0) -- (2,3,0);
        
        \node[above] at (1.5,3,0) {Bottom Layer};
    
    \end{tikzpicture}
    \begin{tikzpicture}[baseline=-1.5cm]
        \draw[thick,->] (0,0) -- (1,0);
        \node[above] at (0.5,0) {(iii)};
    \end{tikzpicture}
    \begin{tikzpicture}[scale=1.0, x={(1cm,0cm)}, y={(0cm,1cm)}, z={(-0.4cm,0.4cm)}]
        \tikzset{
            grid line/.style={thin, gray},
            zero line/.style={red, thick},
            absorb line/.style={blue, thick},
            small square/.style={thin, gray}
        }
    
        \foreach \x in {0,1,2} {
            \foreach \y in {0,1,2} {
                \draw[zero line] (\y,\x, 0) -- (\y,\x, 1);
            }
        }

        \foreach \x in {0,1,2} {
            \draw[grid line] (\x,0, 0) -- (\x,3, 0);
            \draw[grid line] (0,\x, 0) -- (3,\x, 0);
        }

        \draw[zero line] (0,0,0) -- (0,2,0);
        \foreach \x in {0,1,2} {
            \draw[zero line] (0,\x,0) -- (2,\x,0);
        }    

        \foreach \a in {0.3} {
            \foreach \b in {0.7} {
                \foreach \x in {0,1,2} {
                    \foreach \y in {0,1,2} {
                        \foreach \z in {0} {
                            \ifthenelse{\x=2 \and \y=2}{
                            \draw[small square] (\x+\a,\y+\a, \z) -- (\x+\b,\y+\a, \z) -- (\x+\b,\y+\b, \z) -- (\x+\a,\y+\b, \z) -- cycle;
                            }{}
            
                            \draw[small square] (\x,\y+\a, \z+\a) -- (\x,\y+\b, \z+\a) -- (\x,\y+\b, \z+\b) -- (\x,\y+\a, \z+\b) -- cycle;
            
                            \draw[small square] (\x+\a,\y, \z+\a) -- (\x+\b,\y, \z+\a) -- (\x+\b,\y, \z+\b) -- (\x+\a,\y, \z+\b) -- cycle;
                        }
                    }
                }
            }
        }
        
        \draw[absorb line] (1,0,0) -- (1,3,0);
        \draw[absorb line] (2,0,0) -- (2,3,0);
        \draw[absorb line] (2,1,0) -- (3,1,0);
        \draw[absorb line] (2,2,0) -- (3,2,0);
        
        \node[above] at (1.5,3,0) {Bottom Layer};
    
    \end{tikzpicture}
    \begin{tikzpicture}[baseline=-1.5cm]
        \draw[thick,->] (0,0) -- (1,0);
        \node[above] at (0.5,0) {(iv)};
    \end{tikzpicture}
    \begin{tikzpicture}[scale=1.0, x={(1cm,0cm)}, y={(0cm,1cm)}, z={(-0.4cm,0.4cm)}]
        \tikzset{
            grid line/.style={thin, gray},
            zero line/.style={red, thick},
            absorb line/.style={blue, thick},
            small square/.style={thin, gray}
        }
    
        \foreach \x in {0,1,2} {
            \foreach \y in {0,1,2} {
                \draw[zero line] (\y,\x, 0) -- (\y,\x, 1);
            }
        }

        \foreach \x in {0,1,2} {
            \draw[grid line] (\x,0, 0) -- (\x,3, 0);
            \draw[grid line] (0,\x, 0) -- (3,\x, 0);
        }

        \draw[zero line] (0,0,0) -- (0,2,0);
        \foreach \x in {0,1,2} {
            \draw[zero line] (0,\x,0) -- (2,\x,0);
        }    

        \foreach \a in {0.3} {
            \foreach \b in {0.7} {
                \foreach \x in {0,1,2} {
                    \foreach \y in {0,1,2} {
                        \foreach \z in {0} {
                            \ifthenelse{\x=2 \and \y=2}{
                            \draw[small square] (\x+\a,\y+\a, \z) -- (\x+\b,\y+\a, \z) -- (\x+\b,\y+\b, \z) -- (\x+\a,\y+\b, \z) -- cycle;
                            }{}
                        }
                    }
                }
            }
        }
        
        \draw[absorb line] (1,0,0) -- (1,3,0);
        \draw[absorb line] (2,0,0) -- (2,3,0);
        \draw[absorb line] (2,1,0) -- (3,1,0);
        \draw[absorb line] (2,2,0) -- (3,2,0);
        
        \node[above] at (1.5,3,0) {Bottom Layer};
    
    \end{tikzpicture}
    \begin{tikzpicture}
        \draw[white] (0,0) -- (1,0);
    \end{tikzpicture}
    \begin{tikzpicture}[scale=1.0, x={(1cm,0cm)}, y={(0cm,1cm)}, z={(-0.4cm,0.4cm)}]
        \draw[thick,->] (0,0,0) -- (1,0,0);
        \draw[thick,->] (0,0,0) -- (0,1,0);
        \draw[thick,->] (0,0,0) -- (0,0,1);
        \node[above] at (1,0,0) {$1$};
        \node[above] at (0,1,0) {$2$};
        \node[above] at (0,0,1) {$0$};
    \end{tikzpicture}
    \caption{Illustration of the gauge fixing steps on the bottom layer. The gauge fields on the blue links absorb the integer degrees of freedom on nearby plaquettes and are lifted to $\mathbb{R}$. In the last step (iv), plaquettes are absorbed by the links on the upper middle layer.}
    \label{fig:appendix_maxwell_allow_fix_bottom}
\end{figure}

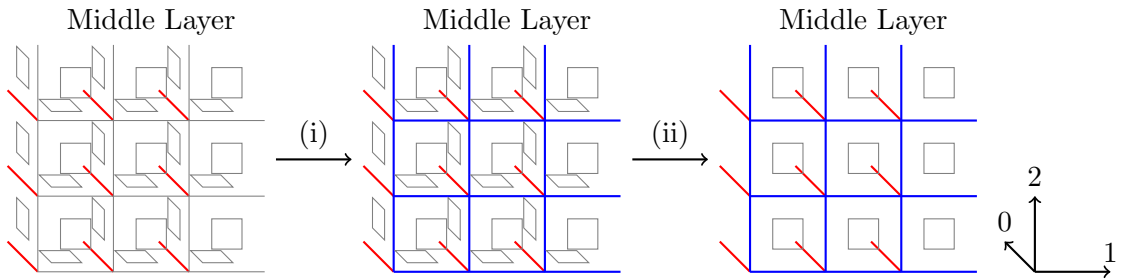
\begin{figure}[htp!]
    \centering
    \begin{tikzpicture}[scale=1.0, x={(1cm,0cm)}, y={(0cm,1cm)}, z={(-0.4cm,0.4cm)}]
        \tikzset{
            grid line/.style={thin, gray},
            zero line/.style={red, thick},
            absorb line/.style={blue, thick},
            small square/.style={thin, gray}
        }
    
        \foreach \x in {0,1,2} {
            \foreach \y in {0,1,2} {
                \draw[zero line] (\y,\x, 0) -- (\y,\x, 1);
            }
        }

        \foreach \x in {0,1,2} {
            \draw[grid line] (\x,0, 0) -- (\x,3, 0);
            \draw[grid line] (0,\x, 0) -- (3,\x, 0);
        } 

        \foreach \a in {0.3} {
            \foreach \b in {0.7} {
                \foreach \x in {0,1,2} {
                    \foreach \y in {0,1,2} {
                        \foreach \z in {0} {
                            \draw[small square] (\x+\a,\y+\a, \z) -- (\x+\b,\y+\a, \z) -- (\x+\b,\y+\b, \z) -- (\x+\a,\y+\b, \z) -- cycle;
            
                            \draw[small square] (\x,\y+\a, \z+\a) -- (\x,\y+\b, \z+\a) -- (\x,\y+\b, \z+\b) -- (\x,\y+\a, \z+\b) -- cycle;
            
                            \draw[small square] (\x+\a,\y, \z+\a) -- (\x+\b,\y, \z+\a) -- (\x+\b,\y, \z+\b) -- (\x+\a,\y, \z+\b) -- cycle;
                        }
                    }
                }
            }
        }
        
        \node[above] at (1.5,3,0) {Middle Layer};
    
    \end{tikzpicture}
    \begin{tikzpicture}[baseline=-1.5cm]
        \draw[thick,->] (0,0) -- (1,0);
        \node[above] at (0.5,0) {(i)};
    \end{tikzpicture}
    \begin{tikzpicture}[scale=1.0, x={(1cm,0cm)}, y={(0cm,1cm)}, z={(-0.4cm,0.4cm)}]
        \tikzset{
            grid line/.style={thin, gray},
            zero line/.style={red, thick},
            absorb line/.style={blue, thick},
            small square/.style={thin, gray}
        }
    
        \foreach \x in {0,1,2} {
            \foreach \y in {0,1,2} {
                \draw[zero line] (\y,\x, 0) -- (\y,\x, 1);
            }
        }

        \foreach \x in {0,1,2} {
            \draw[absorb line] (\x,0, 0) -- (\x,3, 0);
            \draw[absorb line] (0,\x, 0) -- (3,\x, 0);
        }

        \foreach \a in {0.3} {
            \foreach \b in {0.7} {
                \foreach \x in {0,1,2} {
                    \foreach \y in {0,1,2} {
                        \foreach \z in {0} {
                            \draw[small square] (\x+\a,\y+\a, \z) -- (\x+\b,\y+\a, \z) -- (\x+\b,\y+\b, \z) -- (\x+\a,\y+\b, \z) -- cycle;
            
                            \draw[small square] (\x,\y+\a, \z+\a) -- (\x,\y+\b, \z+\a) -- (\x,\y+\b, \z+\b) -- (\x,\y+\a, \z+\b) -- cycle;
            
                            \draw[small square] (\x+\a,\y, \z+\a) -- (\x+\b,\y, \z+\a) -- (\x+\b,\y, \z+\b) -- (\x+\a,\y, \z+\b) -- cycle;
                        }
                    }
                }
            }
        }
        
        \node[above] at (1.5,3,0) {Middle Layer};
    
    \end{tikzpicture}
    \begin{tikzpicture}[baseline=-1.5cm]
        \draw[thick,->] (0,0) -- (1,0);
        \node[above] at (0.5,0) {(ii)};
    \end{tikzpicture}
    \begin{tikzpicture}[scale=1.0, x={(1cm,0cm)}, y={(0cm,1cm)}, z={(-0.4cm,0.4cm)}]
        \tikzset{
            grid line/.style={thin, gray},
            zero line/.style={red, thick},
            absorb line/.style={blue, thick},
            small square/.style={thin, gray}
        }
    
        \foreach \x in {0,1,2} {
            \foreach \y in {0,1,2} {
                \draw[zero line] (\y,\x, 0) -- (\y,\x, 1);
            }
        }

        \foreach \x in {0,1,2} {
            \draw[absorb line] (\x,0, 0) -- (\x,3, 0);
            \draw[absorb line] (0,\x, 0) -- (3,\x, 0);
        }

        \foreach \a in {0.3} {
            \foreach \b in {0.7} {
                \foreach \x in {0,1,2} {
                    \foreach \y in {0,1,2} {
                        \foreach \z in {0} {
                            \draw[small square] (\x+\a,\y+\a, \z) -- (\x+\b,\y+\a, \z) -- (\x+\b,\y+\b, \z) -- (\x+\a,\y+\b, \z) -- cycle;
                        }
                    }
                }
            }
        }
        
        \node[above] at (1.5,3,0) {Middle Layer};
    
    \end{tikzpicture}
    \begin{tikzpicture}[scale=1.0, x={(1cm,0cm)}, y={(0cm,1cm)}, z={(-0.4cm,0.4cm)}]
        \draw[thick,->] (0,0,0) -- (1,0,0);
        \draw[thick,->] (0,0,0) -- (0,1,0);
        \draw[thick,->] (0,0,0) -- (0,0,1);
        \node[above] at (1,0,0) {$1$};
        \node[above] at (0,1,0) {$2$};
        \node[above] at (0,0,1) {$0$};
    \end{tikzpicture}
    \caption{Illustration of the gauge fixing steps on the middle layer. The gauge fields on the blue links absorb the integer degrees of freedom on nearby plaquettes and are lifted to $\mathbb{R}$. (i) Links absorb the plaquettes from the lower layer. (ii) Plaquettes are absorbed by the links from the upper layer.}
    \label{fig:appendix_maxwell_allow_fix_middle}
\end{figure}

\begin{figure}[htp!]
    \centering
    \begin{tikzpicture}[scale=1.0, x={(1cm,0cm)}, y={(0cm,1cm)}, z={(-0.4cm,0.4cm)}]
        \tikzset{
            grid line/.style={thin, gray},
            zero line/.style={red, thick},
            absorb line/.style={blue, thick},
            small square/.style={thin, gray}
        }
    
        \foreach \x in {0,1,2} {
            \foreach \y in {0,1,2} {
                \draw[grid line] (\y,\x, 0) -- (\y,\x, 1);
            }
        }

        \foreach \x in {0,1,2} {
            \draw[grid line] (\x,0, 0) -- (\x,3, 0);
            \draw[grid line] (0,\x, 0) -- (3,\x, 0);
        } 

        \foreach \a in {0.3} {
            \foreach \b in {0.7} {
                \foreach \x in {0,1,2} {
                    \foreach \y in {0,1,2} {
                        \foreach \z in {0} {
                            \draw[small square] (\x+\a,\y+\a, \z) -- (\x+\b,\y+\a, \z) -- (\x+\b,\y+\b, \z) -- (\x+\a,\y+\b, \z) -- cycle;
            
                            \draw[small square] (\x,\y+\a, \z+\a) -- (\x,\y+\b, \z+\a) -- (\x,\y+\b, \z+\b) -- (\x,\y+\a, \z+\b) -- cycle;
            
                            \draw[small square] (\x+\a,\y, \z+\a) -- (\x+\b,\y, \z+\a) -- (\x+\b,\y, \z+\b) -- (\x+\a,\y, \z+\b) -- cycle;
                        }
                    }
                }
            }
        }
        
        \node[above] at (1.5,3,0) {Top Layer};
    
    \end{tikzpicture}
    \begin{tikzpicture}[baseline=-1.5cm]
        \draw[thick,->] (0,0) -- (1,0);
        \node[above] at (0.5,0) {(i)};
    \end{tikzpicture}
    \begin{tikzpicture}[scale=1.0, x={(1cm,0cm)}, y={(0cm,1cm)}, z={(-0.4cm,0.4cm)}]
        \tikzset{
            grid line/.style={thin, gray},
            zero line/.style={red, thick},
            absorb line/.style={blue, thick},
            small square/.style={thin, gray}
        }
    
        \foreach \x in {0,1,2} {
            \foreach \y in {0,1,2} {
                \draw[grid line] (\y,\x, 0) -- (\y,\x, 1);
            }
        }

        \foreach \x in {0,1,2} {
            \draw[absorb line] (\x,0, 0) -- (\x,3, 0);
            \draw[absorb line] (0,\x, 0) -- (3,\x, 0);
        }

        \foreach \a in {0.3} {
            \foreach \b in {0.7} {
                \foreach \x in {0,1,2} {
                    \foreach \y in {0,1,2} {
                        \foreach \z in {0} {
                            \draw[small square] (\x+\a,\y+\a, \z) -- (\x+\b,\y+\a, \z) -- (\x+\b,\y+\b, \z) -- (\x+\a,\y+\b, \z) -- cycle;
            
                            \draw[small square] (\x,\y+\a, \z+\a) -- (\x,\y+\b, \z+\a) -- (\x,\y+\b, \z+\b) -- (\x,\y+\a, \z+\b) -- cycle;
            
                            \draw[small square] (\x+\a,\y, \z+\a) -- (\x+\b,\y, \z+\a) -- (\x+\b,\y, \z+\b) -- (\x+\a,\y, \z+\b) -- cycle;
                        }
                    }
                }
            }
        }
        
        \node[above] at (1.5,3,0) {Top Layer};
    
    \end{tikzpicture}
    \begin{tikzpicture}[baseline=-1.5cm]
        \draw[thick,->] (0,0) -- (1,0);
        \node[above] at (0.5,0) {(ii)};
    \end{tikzpicture}
    \begin{tikzpicture}[scale=1.0, x={(1cm,0cm)}, y={(0cm,1cm)}, z={(-0.4cm,0.4cm)}]
        \tikzset{
            grid line/.style={thin, gray},
            zero line/.style={red, thick},
            absorb line/.style={blue, thick},
            small square/.style={thin, gray}
        }
    
        \foreach \x in {0,1,2} {
            \foreach \y in {0,1,2} {
                \draw[grid line] (\y,\x, 0) -- (\y,\x, 1);
            }
        }

        \foreach \x in {0,1,2} {
            \draw[absorb line] (\x,0, 0) -- (\x,3, 0);
            \draw[absorb line] (0,\x, 0) -- (3,\x, 0);
        }

        \foreach \a in {0.3} {
            \foreach \b in {0.7} {
                \foreach \x in {0,1,2} {
                    \foreach \y in {0,1,2} {
                        \foreach \z in {0} {
                            \draw[small square] (\x+\a,\y+\a, \z) -- (\x+\b,\y+\a, \z) -- (\x+\b,\y+\b, \z) -- (\x+\a,\y+\b, \z) -- cycle;
            
                            \ifthenelse{\x=0 \and \y=0}{}{
                            \draw[small square] (\x,\y+\a, \z+\a) -- (\x,\y+\b, \z+\a) -- (\x,\y+\b, \z+\b) -- (\x,\y+\a, \z+\b) -- cycle;
                            }
            
                            \draw[small square] (\x+\a,\y, \z+\a) -- (\x+\b,\y, \z+\a) -- (\x+\b,\y, \z+\b) -- (\x+\a,\y, \z+\b) -- cycle;
                        }
                    }
                }
            }
        }

        \draw[absorb line] (0,1, 0) -- (0,1, 1);
        
        \node[above] at (1.5,3,0) {Top Layer};
    
    \end{tikzpicture}
    \begin{tikzpicture}[baseline=-1.5cm]
        \draw[thick,->] (0,0) -- (1,0);
        \node[above] at (0.5,0) {(iii)};
    \end{tikzpicture}
    \begin{tikzpicture}[scale=1.0, x={(1cm,0cm)}, y={(0cm,1cm)}, z={(-0.4cm,0.4cm)}]
        \draw[thick,->] (0,0,0) -- (1,0,0);
        \draw[thick,->] (0,0,0) -- (0,1,0);
        \draw[thick,->] (0,0,0) -- (0,0,1);
        \node[above] at (1,0,0) {$1$};
        \node[above] at (0,1,0) {$2$};
        \node[above] at (0,0,1) {$0$};
    \end{tikzpicture}
    \begin{tikzpicture}[scale=1.0, x={(1cm,0cm)}, y={(0cm,1cm)}, z={(-0.4cm,0.4cm)}]
        \tikzset{
            grid line/.style={thin, gray},
            zero line/.style={red, thick},
            absorb line/.style={blue, thick},
            small square/.style={thin, gray}
        }
    
        \foreach \x in {0,1,2} {
            \foreach \y in {0,1,2} {
                \draw[grid line] (\y,\x, 0) -- (\y,\x, 1);
            }
        }

        \foreach \x in {0,1,2} {
            \draw[absorb line] (\x,0, 0) -- (\x,3, 0);
            \draw[absorb line] (0,\x, 0) -- (3,\x, 0);
        }

        \foreach \a in {0.3} {
            \foreach \b in {0.7} {
                \foreach \x in {0,1,2} {
                    \foreach \y in {0,1,2} {
                        \foreach \z in {0} {
                            \draw[small square] (\x+\a,\y+\a, \z) -- (\x+\b,\y+\a, \z) -- (\x+\b,\y+\b, \z) -- (\x+\a,\y+\b, \z) -- cycle;
            
                            \ifthenelse{\x=0 \and \not\y=2}{}{
                            \draw[small square] (\x,\y+\a, \z+\a) -- (\x,\y+\b, \z+\a) -- (\x,\y+\b, \z+\b) -- (\x,\y+\a, \z+\b) -- cycle;
                            }
            
                            \draw[small square] (\x+\a,\y, \z+\a) -- (\x+\b,\y, \z+\a) -- (\x+\b,\y, \z+\b) -- (\x+\a,\y, \z+\b) -- cycle;
                        }
                    }
                }
            }
        }
        
        \draw[absorb line] (0,1, 0) -- (0,1, 1);
        \draw[absorb line] (0,2, 0) -- (0,2, 1);
        
        \node[above] at (1.5,3,0) {Top Layer};
    
    \end{tikzpicture}
    \begin{tikzpicture}[baseline=-1.5cm]
        \draw[thick,->] (0,0) -- (1,0);
        \node[above] at (0.5,0) {(iv)};
    \end{tikzpicture}
    \begin{tikzpicture}[scale=1.0, x={(1cm,0cm)}, y={(0cm,1cm)}, z={(-0.4cm,0.4cm)}]
        \tikzset{
            grid line/.style={thin, gray},
            zero line/.style={red, thick},
            absorb line/.style={blue, thick},
            small square/.style={thin, gray}
        }
    
        \foreach \x in {0,1,2} {
            \foreach \y in {0,1,2} {
                \draw[grid line] (\y,\x, 0) -- (\y,\x, 1);
            }
        }

        \foreach \x in {0,1,2} {
            \draw[absorb line] (\x,0, 0) -- (\x,3, 0);
            \draw[absorb line] (0,\x, 0) -- (3,\x, 0);
        }

        \foreach \a in {0.3} {
            \foreach \b in {0.7} {
                \foreach \x in {0,1,2} {
                    \foreach \y in {0,1,2} {
                        \foreach \z in {0} {
                            \draw[small square] (\x+\a,\y+\a, \z) -- (\x+\b,\y+\a, \z) -- (\x+\b,\y+\b, \z) -- (\x+\a,\y+\b, \z) -- cycle;
            
                            \ifthenelse{\x=0 \and \not\y=2}{}{
                            \draw[small square] (\x,\y+\a, \z+\a) -- (\x,\y+\b, \z+\a) -- (\x,\y+\b, \z+\b) -- (\x,\y+\a, \z+\b) -- cycle;
                            }
            
                            \ifthenelse{\x=0}{}{
                            \draw[small square] (\x+\a,\y, \z+\a) -- (\x+\b,\y, \z+\a) -- (\x+\b,\y, \z+\b) -- (\x+\a,\y, \z+\b) -- cycle;
                            }
                        }
                    }
                }
            }
        }
        
        \foreach \x in {0,1} {
            \foreach \y in {0,1,2} {
                \ifthenelse{\x=0 \and \y=0}{}{
                \draw[absorb line] (\x,\y, 0) -- (\x,\y, 1);
                }
            }
        }
        
        \node[above] at (1.5,3,0) {Top Layer};
    
    \end{tikzpicture}
    \begin{tikzpicture}[baseline=-1.5cm]
        \draw[thick,->] (0,0) -- (1,0);
        \node[above] at (0.5,0) {(v)};
    \end{tikzpicture}
    \begin{tikzpicture}[scale=1.0, x={(1cm,0cm)}, y={(0cm,1cm)}, z={(-0.4cm,0.4cm)}]
        \tikzset{
            grid line/.style={thin, gray},
            zero line/.style={red, thick},
            absorb line/.style={blue, thick},
            small square/.style={thin, gray}
        }
    
        \foreach \x in {0,1,2} {
            \foreach \y in {0,1,2} {
                \draw[grid line] (\y,\x, 0) -- (\y,\x, 1);
            }
        }

        \foreach \x in {0,1,2} {
            \draw[absorb line] (\x,0, 0) -- (\x,3, 0);
            \draw[absorb line] (0,\x, 0) -- (3,\x, 0);
        }

        \foreach \a in {0.3} {
            \foreach \b in {0.7} {
                \foreach \x in {0,1,2} {
                    \foreach \y in {0,1,2} {
                        \foreach \z in {0} {
                            \draw[small square] (\x+\a,\y+\a, \z) -- (\x+\b,\y+\a, \z) -- (\x+\b,\y+\b, \z) -- (\x+\a,\y+\b, \z) -- cycle;
            
                            \ifthenelse{\x=0 \and \not\y=2}{}{
                            \draw[small square] (\x,\y+\a, \z+\a) -- (\x,\y+\b, \z+\a) -- (\x,\y+\b, \z+\b) -- (\x,\y+\a, \z+\b) -- cycle;
                            }
            
                            \ifthenelse{\x=2}{
                            \draw[small square] (\x+\a,\y, \z+\a) -- (\x+\b,\y, \z+\a) -- (\x+\b,\y, \z+\b) -- (\x+\a,\y, \z+\b) -- cycle;
                            }{}
                        }
                    }
                }
            }
        }
        
        \foreach \x in {0,1,2} {
            \foreach \y in {0,1,2} {
                \ifthenelse{\x=0 \and \y=0}{}{
                \draw[absorb line] (\x,\y, 0) -- (\x,\y, 1);
                }
            }
        }
        
        \node[above] at (1.5,3,0) {Top Layer};
    
    \end{tikzpicture}
    \caption{Illustration of the gauge fixing steps on the top layer. The gauge fields on the blue links absorb the integer degrees of freedom on nearby plaquettes and are lifted to $\mathbb{R}$. (i) Links absorb the plaquettes from the lower layer. (ii-v) Links absorb the plaquettes nearby.}
    \label{fig:appendix_maxwell_allow_fix_top}
\end{figure}
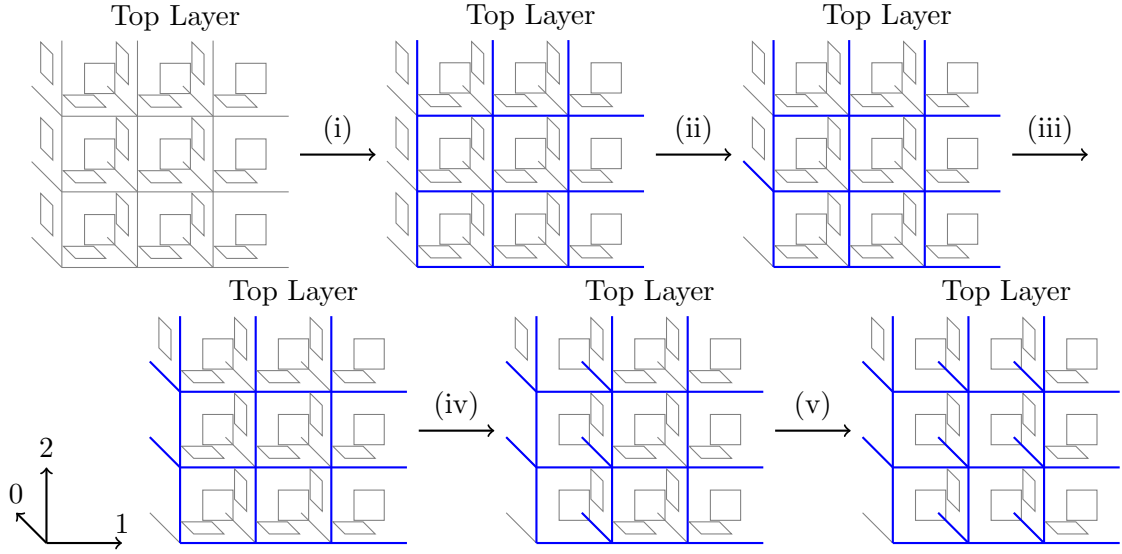

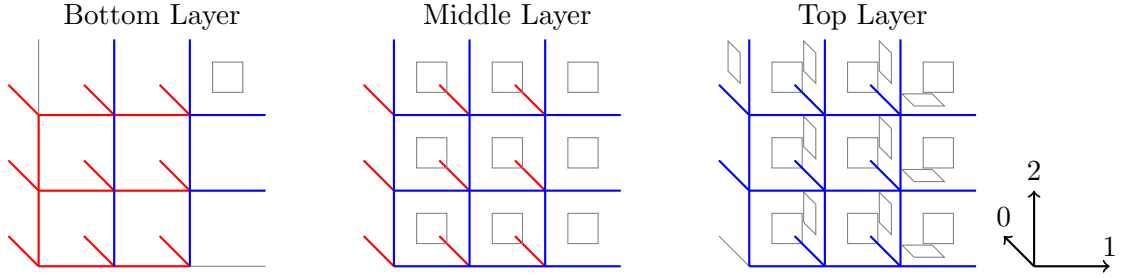
\begin{figure}[htp!]
    \centering
    \begin{tikzpicture}[scale=1.0, x={(1cm,0cm)}, y={(0cm,1cm)}, z={(-0.4cm,0.4cm)}]
        \tikzset{
            grid line/.style={thin, gray},
            zero line/.style={red, thick},
            absorb line/.style={blue, thick},
            small square/.style={thin, gray}
        }
    
        \foreach \x in {0,1,2} {
            \foreach \y in {0,1,2} {
                \draw[zero line] (\y,\x, 0) -- (\y,\x, 1);
            }
        }

        \foreach \x in {0,1,2} {
            \draw[grid line] (\x,0, 0) -- (\x,3, 0);
            \draw[grid line] (0,\x, 0) -- (3,\x, 0);
        }

        \draw[zero line] (0,0,0) -- (0,2,0);
        \foreach \x in {0,1,2} {
            \draw[zero line] (0,\x,0) -- (2,\x,0);
        }    

        \foreach \a in {0.3} {
            \foreach \b in {0.7} {
                \foreach \x in {0,1,2} {
                    \foreach \y in {0,1,2} {
                        \foreach \z in {0} {
                            \ifthenelse{\x=2 \and \y=2}{
                            \draw[small square] (\x+\a,\y+\a, \z) -- (\x+\b,\y+\a, \z) -- (\x+\b,\y+\b, \z) -- (\x+\a,\y+\b, \z) -- cycle;
                            }{}
                        }
                    }
                }
            }
        }
        
        \draw[absorb line] (1,0,0) -- (1,3,0);
        \draw[absorb line] (2,0,0) -- (2,3,0);
        \draw[absorb line] (2,1,0) -- (3,1,0);
        \draw[absorb line] (2,2,0) -- (3,2,0);
        
        \node[above] at (1.5,3,0) {Bottom Layer};
    
    \end{tikzpicture}
    \begin{tikzpicture}
        \draw[white] (0,0) -- (1,0);
    \end{tikzpicture}
    \begin{tikzpicture}[scale=1.0, x={(1cm,0cm)}, y={(0cm,1cm)}, z={(-0.4cm,0.4cm)}]
        \tikzset{
            grid line/.style={thin, gray},
            zero line/.style={red, thick},
            absorb line/.style={blue, thick},
            small square/.style={thin, gray}
        }
    
        \foreach \x in {0,1,2} {
            \foreach \y in {0,1,2} {
                \draw[zero line] (\y,\x, 0) -- (\y,\x, 1);
            }
        }

        \foreach \x in {0,1,2} {
            \draw[absorb line] (\x,0, 0) -- (\x,3, 0);
            \draw[absorb line] (0,\x, 0) -- (3,\x, 0);
        }

        \foreach \a in {0.3} {
            \foreach \b in {0.7} {
                \foreach \x in {0,1,2} {
                    \foreach \y in {0,1,2} {
                        \foreach \z in {0} {
                            \draw[small square] (\x+\a,\y+\a, \z) -- (\x+\b,\y+\a, \z) -- (\x+\b,\y+\b, \z) -- (\x+\a,\y+\b, \z) -- cycle;
                        }
                    }
                }
            }
        }
        
        \node[above] at (1.5,3,0) {Middle Layer};
    
    \end{tikzpicture}
    \begin{tikzpicture}
        \draw[white] (0,0) -- (1,0);
    \end{tikzpicture}
    \begin{tikzpicture}[scale=1.0, x={(1cm,0cm)}, y={(0cm,1cm)}, z={(-0.4cm,0.4cm)}]
        \tikzset{
            grid line/.style={thin, gray},
            zero line/.style={red, thick},
            absorb line/.style={blue, thick},
            small square/.style={thin, gray}
        }
    
        \foreach \x in {0,1,2} {
            \foreach \y in {0,1,2} {
                \draw[grid line] (\y,\x, 0) -- (\y,\x, 1);
            }
        }

        \foreach \x in {0,1,2} {
            \draw[absorb line] (\x,0, 0) -- (\x,3, 0);
            \draw[absorb line] (0,\x, 0) -- (3,\x, 0);
        }

        \foreach \a in {0.3} {
            \foreach \b in {0.7} {
                \foreach \x in {0,1,2} {
                    \foreach \y in {0,1,2} {
                        \foreach \z in {0} {
                            \draw[small square] (\x+\a,\y+\a, \z) -- (\x+\b,\y+\a, \z) -- (\x+\b,\y+\b, \z) -- (\x+\a,\y+\b, \z) -- cycle;
            
                            \ifthenelse{\x=0 \and \not\y=2}{}{
                            \draw[small square] (\x,\y+\a, \z+\a) -- (\x,\y+\b, \z+\a) -- (\x,\y+\b, \z+\b) -- (\x,\y+\a, \z+\b) -- cycle;
                            }
            
                            \ifthenelse{\x=2}{
                            \draw[small square] (\x+\a,\y, \z+\a) -- (\x+\b,\y, \z+\a) -- (\x+\b,\y, \z+\b) -- (\x+\a,\y, \z+\b) -- cycle;
                            }{}
                        }
                    }
                }
            }
        }
        
        \foreach \x in {0,1,2} {
            \foreach \y in {0,1,2} {
                \ifthenelse{\x=0 \and \y=0}{}{
                \draw[absorb line] (\x,\y, 0) -- (\x,\y, 1);
                }
            }
        }
        
        \node[above] at (1.5,3,0) {Top Layer};
    
    \end{tikzpicture}
    \begin{tikzpicture}[scale=1.0, x={(1cm,0cm)}, y={(0cm,1cm)}, z={(-0.4cm,0.4cm)}]
        \draw[thick,->] (0,0,0) -- (1,0,0);
        \draw[thick,->] (0,0,0) -- (0,1,0);
        \draw[thick,->] (0,0,0) -- (0,0,1);
        \node[above] at (1,0,0) {$1$};
        \node[above] at (0,1,0) {$2$};
        \node[above] at (0,0,1) {$0$};
    \end{tikzpicture}
    \caption{View of the bottom layer, middle layers, and the top layer after second-step gauge fixing. The gauge fields on the red links are fixed to zero, on the blue links are lifted to $\mathbb{R}$.}
    \label{fig:appendix_maxwell_allow_fix3}
\end{figure}

The second step of gauge fixing is to absorb some integer degrees of freedom and lift some gauge fields to $\mathbb{R}$. We apply Eqs.~\eqref{eq:appendix_maxwell_allow_A_1} and~\eqref{eq:appendix_maxwell_allow_A_2} to the links on the bottom layer, then on middle layers, and finally on the top layer. Figure~\ref{fig:appendix_maxwell_allow_fix_bottom} shows on the bottom layer how the integer degrees of freedom are absorbed and the gauge fields are lifted to $\mathbb{R}$. Some integer degrees of freedom on the bottom layer are then absorbed by the gauge fields on the upper middle layer, see (iv) in Fig.~\ref{fig:appendix_maxwell_allow_fix_bottom} and (i) in Fig.~\ref{fig:appendix_maxwell_allow_fix_middle}. Figure~\ref{fig:appendix_maxwell_allow_fix_middle} shows the process of second-step gauge fixing on the middle layer, and Figure~\ref{fig:appendix_maxwell_allow_fix_top} shows the top layer.

Figure~\ref{fig:appendix_maxwell_allow_fix3} summarize the results after fixing all gauge degrees of freedom. Each surviving gauge field variable or integer plaquette actually encodes some true physical information that is gauge invariant. The bottom layer has $N_1 N_2 - 1$ number of $\mathbb{R}$-ranged gauge fields (blue), each including the information of the flux in a space-like plaquette. It also has two ($\frac{2\pi}{a}$)-ranged gauge fields (grey), each including the information of the flux through a space-like non-contractible cycle. Moreover, it has one remaining integer degree of freedom (plaquette), including the information of the total flux through the bottom plane, which is a torus in the space. By the Dirac quantization condition we know that this total flux is (integer)$\times 2\pi$~\cite{Dirac:1931quantize}, which agrees with the remaining integer degree of freedom.

The middle layers each has $2 N_1 N_2$ number of $\mathbb{R}$-ranged gauge fields, which include the information of the fluxes in time-like plaquettes. It also has $N_1 N_2$ number of integer degrees of freedom, each including the information of the total flux through a unit cell, i.e. the number of instantons in a unit cell. As we will see later, in an instanton-suppressed theory, these integer degrees of freedom in the middle layers will disappear.

In the top layer, there are $3 N_1 N_2 - 1$ number of $\mathbb{R}$-ranged gauge fields including the information of the fluxes in time-like plaquettes. There is one ($\frac{2\pi}{d\tau}$)-ranged gauge field including the information of the flux through a time-like non-contractible cycle. There are $2 N_1 N_2 + 1$ number of integer degrees of freedom, two of them including the information of the total flux through time-like planes, and the other $2 N_1 N_2 - 1$ of them including information of instantons in unit cells. Note that similar to the fact that monopoles have to come in pairs, the total number of instantons will sum to zero. 

After identifying all the physical degrees of freedoms, we are ready to build up the Hilbert space with the set of basis which are the gauge field configurations in a middle layer. The integer degrees of freedom are regarded as sectors that need to be sum over in the partition function, not entering the trace over the Hilbert space. From the analysis above of middle-layer degrees of freedom, a basis vector is labeled by $\mathbb{R}$-ranged gauge fields on every links. The bottom layer needs to borrow $|\mathbb{Z}|^2\times|\mathbb{R}|^{N_1 N_2-1}\times|\mathbb{Z}|^{N_1 N_2-1}$ to match the dimension of variables on the middle layer. The $|\mathbb{Z}|^2$ comes from lifting the two ($\frac{2\pi}{a}$)-ranged gauge fields (grey), the $|\mathbb{R}|^{N_1 N_2-1}$ comes from the gauge fields fixed to be zero (red), and the $|\mathbb{Z}|^{N_1 N_2-1}$ comes from the missing plaquettes. The top layer has extra degrees of freedom which will become operator constraints on the Hilbert space. There are $N_1 N_2 + 1$ extra $\mathbb{Z}$ plaquettes, $N_1 N_2 - 1$ extra $\mathbb{R}$ gauge fields, and one extra ($\frac{2\pi}{a}$)-ranged gauge field. By summing or integrating out those extra degrees of freedom, $|\mathbb{Z}|^{N_1 N_2+1}\times|\mathbb{R}|^{N_1 N_2-1}$ will be returned, which cancels with the borrowed dimensions and just leaves us an unimportant finite volume factor in the partition function.

Now with the Hilbert space ready, we can define the transfer matrix between a pair of neighboring middle layers. Assuming the lower layer has configuration $|A\rangle$ and the upper layer has configuration $|A'\rangle$, we can read out the transfer matrix element from the action in Eq.~\eqref{eq:appendix_maxwell_allow_S}:
\begin{align}
    \langle A'| \hat{T} |A\rangle = \sum_n \exp\left[-\frac{\beta_0 a^2}{2 d\tau}\sum_{\text{links}}(A' - A)^2 - \frac{\beta d\tau}{2 a^2} \sum_{\text{plaq.}} (a\,\Box A + 2\pi n)^2\right],
\label{eq:appendix_maxwell_allow_T_element}
\end{align}
where $(\Box A)_{x;1,2}\equiv A_{x;1} + A_{x+\hat{1};2} - A_{x+\hat{2};1} - A_{x;2}$.

We do a Fourier transform to get the formulation of $\hat{T}\equiv e^{-\hat{H}d\tau}$, where $\hat{H}$ is the Hamiltonian we want to derive. Assume 
\begin{align}
    \hat{T}(\hat{p}, \hat{A}) = f(\hat{p}) \sum_n \exp\left[- \frac{\beta d\tau}{2 a^2} \sum_{\text{plaq.}} (a\,\Box \hat{A} + 2\pi n)^2\right],
\label{eq:appendix_maxwell_allow_T}
\end{align}
where $\hat{p}$ is the canonical conjugate momentum operator of $\hat{A}$. $f$ is a function we want to solve. Inserting the complete momentum basis $\mathbb{I}=\int_{-\infty}^\infty \frac{dp}{2\pi} |p\rangle\langle p|$, we have
\begin{align}
    \langle A'| f(\hat{p}) |A\rangle &= \int_{-\infty}^\infty \frac{dp}{2\pi} e^{i p (A'-A)} f(p) = \exp\left[-\frac{\beta_0 a^2}{2 d\tau}\sum_{\text{links}}(A' - A)^2\right].
\end{align}
We can solve the function $f(p)$ by the Fourier transform 
\begin{align}
    f(p) \propto \exp\left[-\frac{d\tau}{2 \beta_0 a^2}\sum_{\text{links}}p^2\right]. 
\label{eq:appendix_maxwell_allow_f}
\end{align}
The other term in Eq.~\eqref{eq:appendix_maxwell_allow_T} can be approximated by the reversed Villain approximation, which is good when $d\tau\to 0$:
\begin{align}
    \sum_n \exp\left[- \frac{\beta d\tau}{2 a^2} \sum_{\text{plaq.}} (a\,\Box \hat{A} + 2\pi n)^2\right] \approx \exp\left[ \frac{\beta d\tau}{a^2} \sum_{\text{plaq.}} \cos(a\,\Box \hat{A})\right].
\end{align}
Therefore, up to $O(d\tau^2)$, 
\begin{align}
    \hat{T}(\hat{p}, \hat{A}) \propto \exp\left[-\frac{d\tau}{2 \beta_0 a^2}\sum_{\text{links}}\hat{p}^2 + \frac{\beta d\tau}{a^2} \sum_{\text{plaq.}} \cos(a\,\Box \hat{A}) \right], 
\label{eq:appendix_maxwell_allow_T_propto}
\end{align}
and we get the familiar compact lattice Maxwell Hamiltonian (with instantons allowed):
\begin{align}
    \hat{H} = \frac{1}{2 \beta_0 a^2}\sum_{\text{links}}\hat{p}^2 - \frac{\beta}{a^2} \sum_{\text{plaq.}} \cos(a\,\Box \hat{A}).
\end{align}

However, this is not the end of the derivation. So far we allow $\mathbb{R}$-ranged gauge fields, and we do not have Gauss' law yet. We need to look at the top layer to figure out the constraints on the Hilbert space. In Figure~\ref{fig:appendix_maxwell_allow_fix3}, we see that the top layer has $N_1 N_2 + 1$ extra time-like plaquettes, $N_1 N_2 - 1$ non-zero $\mathbb{R}$-ranged $A_0$, and one non-zero ($\frac{2\pi}{d\tau}$)-ranged $A_0$. With these non-zero variables, the transfer matrix element in Eq.~\eqref{eq:appendix_maxwell_allow_T_element} becomes
\begin{align}
    \langle A'| \hat{T} |A\rangle = \sum_{n,m,A_0} \exp &\left[ - \frac{\beta_0 a^2}{2 d\tau}\sum_{\substack{x\in\text{sites} \\ i\in\{1,2\}}}(A_{x;i}' - A_{x;i} + \frac{d\tau}{a}(A_{x;0} - A_{x+\hat{i};0})+\frac{2\pi}{a}m_{x;0,i})^2 \right. \nonumber \\
    &\quad\left. - \frac{\beta d\tau}{2 a^2} \sum_{\text{plaq.}} (a\,\Box A + 2\pi n)^2\right],
\end{align}
where we rename the integer degrees of freedom on time-like plaquettes as $m$, not to confuse with space-like $n$. This modification affects the Fourier transform in Eq.~\eqref{eq:appendix_maxwell_allow_f}:
\begin{align}
    f(p) \propto \sum_{m,A_0} \exp\left[\sum_{\substack{x\in\text{sites} \\ i\in\{1,2\}}} -\frac{d\tau}{2 \beta_0 a^2}p_{x;i}^2 + i p_{x;i} \left(\frac{d\tau}{a}(A_{x;0} - A_{x+\hat{i};0})+\frac{2\pi}{a}m_{x;0,i}\right)\right] ,
\end{align}
and also Eq.~\eqref{eq:appendix_maxwell_allow_T_propto}:
\begin{align}
    \hat{T}(\hat{p}, \hat{A}) \propto&\ \exp\left[-\frac{d\tau}{2 \beta_0 a^2}\sum_{\text{links}}\hat{p}^2 + \frac{\beta d\tau}{a^2} \sum_{\text{plaq.}} \cos(a\,\Box \hat{A}) \right] 
    \nonumber \\
    &\times \sum_{m,A_0} \exp\left[\sum_{\substack{x\in\text{sites} \\ i\in\{1,2\}}} i \hat{p}_{x;i} \left(\frac{d\tau}{a}(A_{x;0} - A_{x+\hat{i};0})+\frac{2\pi}{a}m_{x;0,i}\right)\right].
\end{align}
Here $A_0$'s and $m$'s are not operators. They are variables to be summed over in the partition function.

Note that the extra term is equivalent to:
\begin{align}
    &\exp\left[\sum_{\substack{x\in\text{sites} \\ i\in\{1,2\}}} i \hat{p}_{x;i} \left(\frac{d\tau}{a}(A_{x;0} - A_{x+\hat{i};0})+\frac{2\pi}{a}m_{x;0,i}\right)\right] \nonumber \\
    =&\ \prod_{x\in\text{sites}} \exp\left[i\frac{d\tau}{a}A_{x;0}(\hat{p}_{x;1} + \hat{p}_{x;2} - \hat{p}_{x-\hat{1};1} - \hat{p}_{x-\hat{2};2})\right] \prod_{\text{links}}\exp\left[i\frac{2\pi}{a}m \hat{p}\right].
\end{align}
The sum over ($m$, $A_0$) can be explicitly written as
\begin{align}
    \sum_{m,A_0} \equiv \int_0^{\frac{2\pi}{d\tau}} dA_{x=(0,0);0} \prod_{x\neq(0,0)}\int_{-\infty}^{\infty} dA_{x;0} \prod_{l \in \left\{\substack{\text{links where time-like} \\ \text{plaquettes are present}}\right\}}\sum_{m_l=-\infty}^\infty .
\end{align}
Applying the sum to the extra term, we can see that it does not vanish if and only if
\begin{align}
\label{eq:appendix_maxwell_allow_constraint1}
    \int_0^{\frac{2\pi}{d\tau}} dA_{x;0} \exp\left[i\frac{d\tau}{a}A_{x;0}(\hat{p}_{x;1} + \hat{p}_{x;2} - \hat{p}_{x-\hat{1};1} - \hat{p}_{x-\hat{2};2})\right] &\neq 0, \quad \text{for } x=(0,0), \\
\label{eq:appendix_maxwell_allow_constraint2}
    \int_{-\infty}^{\infty} dA_{x;0} \exp\left[i\frac{d\tau}{a}A_{x;0}(\hat{p}_{x;1} + \hat{p}_{x;2} - \hat{p}_{x-\hat{1};1} - \hat{p}_{x-\hat{2};2})\right] &\neq 0, \quad \forall x\neq(0,0), \\
\label{eq:appendix_maxwell_allow_constraint3}
    \sum_{m_l=-\infty}^\infty \exp\left[i\frac{2\pi}{a} m_l\,\hat{p}_l\right] &\neq 0, \quad \forall l \in \left\{\substack{\text{links where time-like} \\ \text{plaquettes are present}}\right\}.
\end{align}
Equation~\eqref{eq:appendix_maxwell_allow_constraint2} implies
\begin{align}
\label{eq:appendix_maxwell_allow_constraint22}
    \hat{p}_{x;1} + \hat{p}_{x;2} - \hat{p}_{x-\hat{1};1} - \hat{p}_{x-\hat{2};2} &= 0, \quad \forall x\neq(0,0), 
\end{align}
which is the Gauss' law. Because $\sum_{x\in\text{sites}}(\hat{p}_{x;1} + \hat{p}_{x;2} - \hat{p}_{x-\hat{1};1} - \hat{p}_{x-\hat{2};2})=0$, the Gauss' law in Eq.~\eqref{eq:appendix_maxwell_allow_constraint22} holds for $x=(0,0)$ as well. This also means Eq.~\eqref{eq:appendix_maxwell_allow_constraint1} is automatically satisfied.

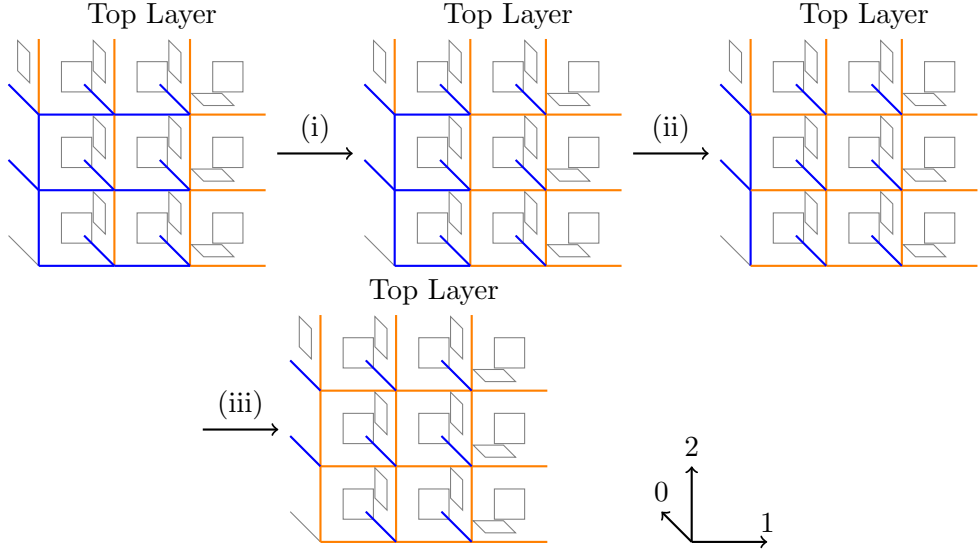
\begin{figure}[htp!]
    \centering
    \begin{tikzpicture}
        \draw[white] (0,0) -- (1,0);
    \end{tikzpicture}
    \begin{tikzpicture}[scale=1.0, x={(1cm,0cm)}, y={(0cm,1cm)}, z={(-0.4cm,0.4cm)}]
        \tikzset{
            grid line/.style={thin, gray},
            zero line/.style={red, thick},
            absorb line/.style={blue, thick},
            int line/.style={orange, thick},
            small square/.style={thin, gray}
        }
    
        \foreach \x in {0,1,2} {
            \foreach \y in {0,1,2} {
                \draw[grid line] (\y,\x, 0) -- (\y,\x, 1);
            }
        }

        \foreach \a in {0.3} {
            \foreach \b in {0.7} {
                \foreach \x in {0,1,2} {
                    \foreach \y in {0,1,2} {
                        \foreach \z in {0} {
                            \draw[small square] (\x+\a,\y+\a, \z) -- (\x+\b,\y+\a, \z) -- (\x+\b,\y+\b, \z) -- (\x+\a,\y+\b, \z) -- cycle;
            
                            \ifthenelse{\x=0 \and \not\y=2}{
                            \draw[absorb line] (\x,\y, \z) -- (\x,\y+1, \z);
                            }{
                            \draw[small square] (\x,\y+\a, \z+\a) -- (\x,\y+\b, \z+\a) -- (\x,\y+\b, \z+\b) -- (\x,\y+\a, \z+\b) -- cycle;
                            \draw[int line] (\x,\y, \z) -- (\x,\y+1, \z);
                            }
            
                            \ifthenelse{\x=2}{
                            \draw[small square] (\x+\a,\y, \z+\a) -- (\x+\b,\y, \z+\a) -- (\x+\b,\y, \z+\b) -- (\x+\a,\y, \z+\b) -- cycle;
                            \draw[int line] (\x,\y, \z) -- (\x+1,\y, \z);
                            }{
                            \draw[absorb line] (\x,\y, \z) -- (\x+1,\y, \z);
                            }
                        }
                    }
                }
            }
        }
        
        \foreach \x in {0,1,2} {
            \foreach \y in {0,1,2} {
                \ifthenelse{\x=0 \and \y=0}{}{
                \draw[absorb line] (\x,\y, 0) -- (\x,\y, 1);
                }
            }
        }
        
        \node[above] at (1.5,3,0) {Top Layer};
    
    \end{tikzpicture}
    \begin{tikzpicture}[baseline=-1.5cm]
        \draw[thick,->] (0,0) -- (1,0);
        \node[above] at (0.5,0) {(i)};
    \end{tikzpicture}
    \begin{tikzpicture}[scale=1.0, x={(1cm,0cm)}, y={(0cm,1cm)}, z={(-0.4cm,0.4cm)}]
        \tikzset{
            grid line/.style={thin, gray},
            zero line/.style={red, thick},
            absorb line/.style={blue, thick},
            int line/.style={orange, thick},
            small square/.style={thin, gray}
        }
    
        \foreach \x in {0,1,2} {
            \foreach \y in {0,1,2} {
                \draw[grid line] (\y,\x, 0) -- (\y,\x, 1);
            }
        }

        \foreach \a in {0.3} {
            \foreach \b in {0.7} {
                \foreach \x in {0,1,2} {
                    \foreach \y in {0,1,2} {
                        \foreach \z in {0} {
                            \draw[small square] (\x+\a,\y+\a, \z) -- (\x+\b,\y+\a, \z) -- (\x+\b,\y+\b, \z) -- (\x+\a,\y+\b, \z) -- cycle;
            
                            \ifthenelse{\x=0 \and \not\y=2}{
                            \draw[absorb line] (\x,\y, \z) -- (\x,\y+1, \z);
                            }{
                            \draw[small square] (\x,\y+\a, \z+\a) -- (\x,\y+\b, \z+\a) -- (\x,\y+\b, \z+\b) -- (\x,\y+\a, \z+\b) -- cycle;
                            \draw[int line] (\x,\y, \z) -- (\x,\y+1, \z);
                            }
            
                            \ifthenelse{\x=2}{
                            \draw[small square] (\x+\a,\y, \z+\a) -- (\x+\b,\y, \z+\a) -- (\x+\b,\y, \z+\b) -- (\x+\a,\y, \z+\b) -- cycle;
                            }{}

                            \ifthenelse{\not\x=0}{
                            \draw[int line] (\x,\y, \z) -- (\x+1,\y, \z);
                            }{
                            \draw[absorb line] (\x,\y, \z) -- (\x+1,\y, \z);
                            } 
                        }
                    }
                }
            }
        }
        
        \foreach \x in {0,1,2} {
            \foreach \y in {0,1,2} {
                \ifthenelse{\x=0 \and \y=0}{}{
                \draw[absorb line] (\x,\y, 0) -- (\x,\y, 1);
                }
            }
        }
        
        \node[above] at (1.5,3,0) {Top Layer};
    
    \end{tikzpicture}
    \begin{tikzpicture}[baseline=-1.5cm]
        \draw[thick,->] (0,0) -- (1,0);
        \node[above] at (0.5,0) {(ii)};
    \end{tikzpicture}
    \begin{tikzpicture}[scale=1.0, x={(1cm,0cm)}, y={(0cm,1cm)}, z={(-0.4cm,0.4cm)}]
        \tikzset{
            grid line/.style={thin, gray},
            zero line/.style={red, thick},
            absorb line/.style={blue, thick},
            int line/.style={orange, thick},
            small square/.style={thin, gray}
        }
    
        \foreach \x in {0,1,2} {
            \foreach \y in {0,1,2} {
                \draw[grid line] (\y,\x, 0) -- (\y,\x, 1);
            }
        }

        \foreach \a in {0.3} {
            \foreach \b in {0.7} {
                \foreach \x in {0,1,2} {
                    \foreach \y in {0,1,2} {
                        \foreach \z in {0} {
                            \draw[small square] (\x+\a,\y+\a, \z) -- (\x+\b,\y+\a, \z) -- (\x+\b,\y+\b, \z) -- (\x+\a,\y+\b, \z) -- cycle;
            
                            \ifthenelse{\x=0 \and \not\y=2}{
                            \draw[absorb line] (\x,\y, \z) -- (\x,\y+1, \z);
                            }{
                            \draw[small square] (\x,\y+\a, \z+\a) -- (\x,\y+\b, \z+\a) -- (\x,\y+\b, \z+\b) -- (\x,\y+\a, \z+\b) -- cycle;
                            \draw[int line] (\x,\y, \z) -- (\x,\y+1, \z);
                            }
            
                            \ifthenelse{\x=2}{
                            \draw[small square] (\x+\a,\y, \z+\a) -- (\x+\b,\y, \z+\a) -- (\x+\b,\y, \z+\b) -- (\x+\a,\y, \z+\b) -- cycle;
                            }{}

                            \draw[int line] (\x,\y, \z) -- (\x+1,\y, \z);
                        }
                    }
                }
            }
        }
        
        \foreach \x in {0,1,2} {
            \foreach \y in {0,1,2} {
                \ifthenelse{\x=0 \and \y=0}{}{
                \draw[absorb line] (\x,\y, 0) -- (\x,\y, 1);
                }
            }
        }
        
        \node[above] at (1.5,3,0) {Top Layer};
    
    \end{tikzpicture}
    \begin{tikzpicture}
        \draw[white] (0,0) -- (1,0);
    \end{tikzpicture}
    \begin{tikzpicture}[baseline=-1.5cm]
        \draw[thick,->] (0,0) -- (1,0);
        \node[above] at (0.5,0) {(iii)};
    \end{tikzpicture}
    \begin{tikzpicture}[scale=1.0, x={(1cm,0cm)}, y={(0cm,1cm)}, z={(-0.4cm,0.4cm)}]
        \tikzset{
            grid line/.style={thin, gray},
            zero line/.style={red, thick},
            absorb line/.style={blue, thick},
            int line/.style={orange, thick},
            small square/.style={thin, gray}
        }
    
        \foreach \x in {0,1,2} {
            \foreach \y in {0,1,2} {
                \draw[grid line] (\y,\x, 0) -- (\y,\x, 1);
            }
        }

        \foreach \x in {0,1,2} {
            \draw[int line] (\x,0, 0) -- (\x,3, 0);
            \draw[int line] (0,\x, 0) -- (3,\x, 0);
        }

        \foreach \a in {0.3} {
            \foreach \b in {0.7} {
                \foreach \x in {0,1,2} {
                    \foreach \y in {0,1,2} {
                        \foreach \z in {0} {
                            \draw[small square] (\x+\a,\y+\a, \z) -- (\x+\b,\y+\a, \z) -- (\x+\b,\y+\b, \z) -- (\x+\a,\y+\b, \z) -- cycle;
            
                            \ifthenelse{\x=0 \and \not\y=2}{}{
                            \draw[small square] (\x,\y+\a, \z+\a) -- (\x,\y+\b, \z+\a) -- (\x,\y+\b, \z+\b) -- (\x,\y+\a, \z+\b) -- cycle;
                            }
            
                            \ifthenelse{\x=2}{
                            \draw[small square] (\x+\a,\y, \z+\a) -- (\x+\b,\y, \z+\a) -- (\x+\b,\y, \z+\b) -- (\x+\a,\y, \z+\b) -- cycle;
                            }{}
                        }
                    }
                }
            }
        }
        
        \foreach \x in {0,1,2} {
            \foreach \y in {0,1,2} {
                \ifthenelse{\x=0 \and \y=0}{}{
                \draw[absorb line] (\x,\y, 0) -- (\x,\y, 1);
                }
            }
        }
        
        \node[above] at (1.5,3,0) {Top Layer};
    
    \end{tikzpicture}
    \begin{tikzpicture}
        \draw[white] (0,0) -- (1,0);
    \end{tikzpicture}
    \begin{tikzpicture}[scale=1.0, x={(1cm,0cm)}, y={(0cm,1cm)}, z={(-0.4cm,0.4cm)}]
        \draw[thick,->] (0,0,0) -- (1,0,0);
        \draw[thick,->] (0,0,0) -- (0,1,0);
        \draw[thick,->] (0,0,0) -- (0,0,1);
        \node[above] at (1,0,0) {$1$};
        \node[above] at (0,1,0) {$2$};
        \node[above] at (0,0,1) {$0$};
    \end{tikzpicture}
    \caption{Illustration of the Hilbert space constraints on the top layer. The $\hat{p}$ operator, canonical conjugate momentum to the gauge field, on the orange links are constrained to take value in $a\mathbb{Z}$. Initially the orange links are under the time-like plaquettes. Equipped with the Gauss' law constraints, i.e. the sum of four $\hat{p}$'s sharing a common site is zero, all space-like links are colored orange step by step, which means all $\hat{p}\in a\mathbb{Z}$.}
    \label{fig:appendix_maxwell_allow_p}
\end{figure}

Equation~\eqref{eq:appendix_maxwell_allow_constraint3} implies
\begin{align}
\label{eq:appendix_maxwell_allow_constraint32}
    \exp\left[i\frac{2\pi}{a} \hat{p}_l\right] &= 1, \quad \hat{p}_l\in a\mathbb{Z}, \quad \forall l \in \left\{\substack{\text{links where time-like} \\ \text{plaquettes are present}}\right\}.
\end{align}
The links $l$ are under the extra time-like plaquettes in the top layer, which are shown as the orange links in the first pannel in Fig.~\ref{fig:appendix_maxwell_allow_p}.

Combining Eq.~\eqref{eq:appendix_maxwell_allow_constraint32} with Eq.~\eqref{eq:appendix_maxwell_allow_constraint22}, we can propagate the set of links where $\hat{p}\in a\mathbb{Z}$. For a given site, when three of the four links that share this site have $\hat{p}\in a\mathbb{Z}$ on them, then the last one should also have $\hat{p}\in a\mathbb{Z}$ by Eq.~\eqref{eq:appendix_maxwell_allow_constraint22}. In Figure.~\ref{fig:appendix_maxwell_allow_p}, we demonstrate this process and show that the constraints imply that all $\hat{p}\in a\mathbb{Z}$, i.e. all the gauge fields are defined with a period $\frac{2\pi}{a}$.

With the Gauss' law constraints and the momentum constraints derived, we complete the whole story for the pure lattice Maxwell theory with instantons allowed.

\subsection{Pure Maxwell Lattice Theory Without Instantons}
\label{subsec:appendix_Maxwell_Instanton_Suppressed}

As reviewed in Sec.~\ref{subsec:appendix_instanton}, the instantons in the 2+1D U(1) lattice gauge theory create a gap of the theory. We can completely suppress the instantons to get back to a gapless theroy, with behaves more like the pure Maxwell theory in the continuum. 

\begin{figure}[htp!]
    \centering
    \begin{tikzpicture}[scale=1.0, x={(1cm,0cm)}, y={(0cm,1cm)}, z={(-0.4cm,0.4cm)}]
        \tikzset{
            grid line/.style={thin, gray},
            zero line/.style={red, thick},
            absorb line/.style={blue, thick},
            small square/.style={thin, gray},
            small 0 square/.style={thin, orange},
            small 1 square/.style={thin, cyan},
            small 2 square/.style={thin, green}
        }
    
        \foreach \x in {0,1,2} {
            \foreach \y in {0,1,2} {
                \draw[zero line] (\y,\x, 0) -- (\y,\x, 1);
            }
        }

        \foreach \x in {0,1,2} {
            \draw[grid line] (\x,0, 0) -- (\x,3, 0);
            \draw[grid line] (0,\x, 0) -- (3,\x, 0);
        }

        \draw[zero line] (0,0,0) -- (0,2,0);
        \foreach \x in {0,1,2} {
            \draw[zero line] (0,\x,0) -- (2,\x,0);
        }    

        \foreach \a in {0.3} {
            \foreach \b in {0.7} {
                \foreach \x in {0,1,2} {
                    \foreach \y in {0,1,2} {
                        \foreach \z in {0} {
                            \ifthenelse{\x=2 \and \y=2}{
                            \draw[small 0 square] (\x+\a,\y+\a, \z) -- (\x+\b,\y+\a, \z) -- (\x+\b,\y+\b, \z) -- (\x+\a,\y+\b, \z) -- cycle;
                            }{}
                        }
                    }
                }
            }
        }
        
        \draw[absorb line] (1,0,0) -- (1,3,0);
        \draw[absorb line] (2,0,0) -- (2,3,0);
        \draw[absorb line] (2,1,0) -- (3,1,0);
        \draw[absorb line] (2,2,0) -- (3,2,0);
        
        \node[above] at (1.5,3,0) {Bottom Layer};
    
    \end{tikzpicture}
    \begin{tikzpicture}
        \draw[white] (0,0) -- (1,0);
    \end{tikzpicture}
    \begin{tikzpicture}[scale=1.0, x={(1cm,0cm)}, y={(0cm,1cm)}, z={(-0.4cm,0.4cm)}]
        \tikzset{
            grid line/.style={thin, gray},
            zero line/.style={red, thick},
            absorb line/.style={blue, thick},
            small square/.style={thin, gray},
            small 0 square/.style={thin, orange},
            small 1 square/.style={thin, cyan},
            small 2 square/.style={thin, green}
        }
    
        \foreach \x in {0,1,2} {
            \foreach \y in {0,1,2} {
                \draw[zero line] (\y,\x, 0) -- (\y,\x, 1);
            }
        }

        \foreach \x in {0,1,2} {
            \draw[absorb line] (\x,0, 0) -- (\x,3, 0);
            \draw[absorb line] (0,\x, 0) -- (3,\x, 0);
        }

        \foreach \a in {0.3} {
            \foreach \b in {0.7} {
                \foreach \x in {0,1,2} {
                    \foreach \y in {0,1,2} {
                        \foreach \z in {0} {
                            \ifthenelse{\x=2 \and \y=2}{
                            \draw[small 0 square] (\x+\a,\y+\a, \z) -- (\x+\b,\y+\a, \z) -- (\x+\b,\y+\b, \z) -- (\x+\a,\y+\b, \z) -- cycle;
                            }{}
                        }
                    }
                }
            }
        }
        
        \node[above] at (1.5,3,0) {Middle Layer};
    
    \end{tikzpicture}
    \begin{tikzpicture}
        \draw[white] (0,0) -- (1,0);
    \end{tikzpicture}
    \begin{tikzpicture}[scale=1.0, x={(1cm,0cm)}, y={(0cm,1cm)}, z={(-0.4cm,0.4cm)}]
        \tikzset{
            grid line/.style={thin, gray},
            zero line/.style={red, thick},
            absorb line/.style={blue, thick},
            small square/.style={thin, gray},
            small 0 square/.style={thin, orange},
            small 1 square/.style={thin, cyan},
            small 2 square/.style={thin, green}
        }
    
        \foreach \x in {0,1,2} {
            \foreach \y in {0,1,2} {
                \draw[grid line] (\y,\x, 0) -- (\y,\x, 1);
            }
        }

        \foreach \x in {0,1,2} {
            \draw[absorb line] (\x,0, 0) -- (\x,3, 0);
            \draw[absorb line] (0,\x, 0) -- (3,\x, 0);
        }

        \foreach \a in {0.3} {
            \foreach \b in {0.7} {
                \foreach \x in {0,1,2} {
                    \foreach \y in {0,1,2} {
                        \foreach \z in {0} {
                            \ifthenelse{\x=2 \and \y=2}{
                            \draw[small 0 square] (\x+\a,\y+\a, \z) -- (\x+\b,\y+\a, \z) -- (\x+\b,\y+\b, \z) -- (\x+\a,\y+\b, \z) -- cycle;
                            }{}
            
                            \ifthenelse{\y=2}{
                                \ifthenelse{\x=0}{
                                \draw[small 1 square] (\x,\y+\a, \z+\a) -- (\x,\y+\b, \z+\a) -- (\x,\y+\b, \z+\b) -- (\x,\y+\a, \z+\b) -- cycle;
                                }{
                                \draw[small 1 square] (\x,\y+\a, \z+\a) -- (\x,\y+\b, \z+\a) -- (\x,\y+\b, \z+\b) -- (\x,\y+\a, \z+\b) -- cycle;
                                }
                            }{}
            
                            \ifthenelse{\x=2}{
                                \ifthenelse{\y=0}{
                                \draw[small 2 square] (\x+\a,\y, \z+\a) -- (\x+\b,\y, \z+\a) -- (\x+\b,\y, \z+\b) -- (\x+\a,\y, \z+\b) -- cycle;
                                }{
                                \draw[small 2 square] (\x+\a,\y, \z+\a) -- (\x+\b,\y, \z+\a) -- (\x+\b,\y, \z+\b) -- (\x+\a,\y, \z+\b) -- cycle;
                                }
                            }{}
                        }
                    }
                }
            }
        }
        
        \foreach \x in {0,1,2} {
            \foreach \y in {0,1,2} {
                \ifthenelse{\x=0 \and \y=0}{}{
                \draw[absorb line] (\x,\y, 0) -- (\x,\y, 1);
                }
            }
        }
        
        \node[above] at (1.5,3,0) {Top Layer};
    
    \end{tikzpicture}
    \begin{tikzpicture}[scale=1.0, x={(1cm,0cm)}, y={(0cm,1cm)}, z={(-0.4cm,0.4cm)}]
        \draw[thick,->] (0,0,0) -- (1,0,0);
        \draw[thick,->] (0,0,0) -- (0,1,0);
        \draw[thick,->] (0,0,0) -- (0,0,1);
        \node[above] at (1,0,0) {$1$};
        \node[above] at (0,1,0) {$2$};
        \node[above] at (0,0,1) {$0$};
    \end{tikzpicture}
    \caption{View of the bottom layer, middle layers, and the top layer after gauge fixing. Instantons are suppressed. The gauge fields on the red links are fixed to zero, on the blue links are lifted to $\mathbb{R}$. The plaquettes with three different colors, orange, cyan, and green, represent the three independent integer degrees of freedom. The plaquettes with the same color are not free degrees of freedom: they are enforced to have the same integer value.}
    \label{fig:appendix_maxwell_suppress_fix}
\end{figure}
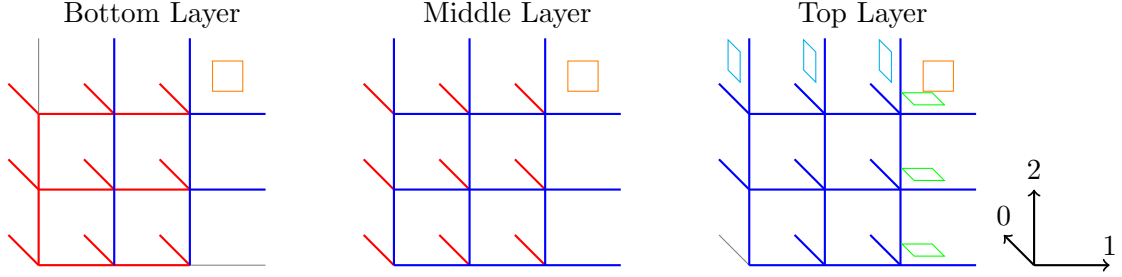

We can repeat most of the derivations shown in the previous section, where we study the instanton-allowed version of the theory. The route differs after we fix all gauge degrees of freedom. Recall that in Fig.~\ref{fig:appendix_maxwell_allow_fix3}, most of the plaquettes are the integer degrees of freedom including information about the instantons in unit cells. For an instanton-suppressed theory, these integer degrees of freedom are eliminated. Therefore, we have a modified version for the physical degrees of freedom in each layers, as shown in Fig.~\ref{fig:appendix_maxwell_suppress_fix}. The gauge field degrees of freedom are in the same situation as the instanton-allowed case. However, there are only three remaining independent integer degrees of freedom, each includes the information of the total flux through a plane. Because there is no instanton, all planes with the same orientation have the same total flux. Therefore, to demonstrate the dependence on their values, we use three colors, orange, cyan, and green, to color the plaquettes in Fig.~\ref{fig:appendix_maxwell_suppress_fix}. The plaquettes with the same color have the same integer value and represent one integer degree of freedom.

Because the gauge field degrees of freedom are the same, the Hilbert space basis vectors are also the same as in the previous section: We label each basis vector with a configuration of $\mathbb{R}$-ranged gauge fields on every links. We can define the transfer matrix between a pair of neighboring middle layers. Assuming the lower layer has configuration $|A\rangle$ and the upper layer has configuration $|A'\rangle$, we can read out the transfer matrix element from the action in Eq.~\eqref{eq:appendix_maxwell_allow_S}:
\begin{align}
    &\langle A'| \hat{T} |A\rangle = \nonumber \\
    &\exp\left[-\frac{\beta_0 a^2}{2 d\tau}\sum_{\text{links}}(A' - A)^2 - \frac{\beta d\tau}{2 a^2} \left(\sum_{x\neq(N_1-1,N_2-1)} (a\,\Box A_{x;1,2})^2 + (a\,\Box A_{(N_1-1,N_2-1);1,2} + 2\pi n)^2\right)\right],
\label{eq:appendix_maxwell_suppress_T_element}
\end{align}
where $(\Box A)_{x;1,2}\equiv A_{x;1} + A_{x+\hat{1};2} - A_{x+\hat{2};1} - A_{x;2}$. $n$ is the integer degree of freedom on space-like plaquettes (orange in Fig.~\ref{fig:appendix_maxwell_suppress_fix}), which is fixed for the transfer matrix, and should be summed over in the partition function as denoting different total-flux sectors.

Performing a similar Fourier transform as in the previous section, we obtain the Hamiltonian for the total-flux-$n$ sector:
\begin{align}
    \hat{H}_n = \frac{1}{2 \beta_0 a^2}\sum_{\text{links}}\hat{p}^2 + \frac{\beta}{2 a^2} \left(\sum_{x\neq(N_1-1,N_2-1)} (a\,\Box A_{x;1,2})^2 + (a\,\Box A_{(N_1-1,N_2-1);1,2} + 2\pi n)^2\right).
\label{eq:appendix_maxwell_suppress_H}
\end{align}
The $n=0$ sector looks like a non-compact Maxwell theory. However, this theory is indeed compact after including all $n$ sectors and the constraints we will discuss soon. It is the compact Maxwell theory with instantons suppressed. As we reviewed in Sec.~\ref{subsec:appendix_instanton}, this theory is gappless with massless photons. The $n\neq 0$ sector can be considered as fluctuations around a classical configuration with a background magnetic field, which generates a total flux $2\pi n$.  

For the constraints on the Hilbert space, we look at the top layer in Fig.~\ref{fig:appendix_maxwell_suppress_fix}. Similar to the previous section, we get an extra phase factor to the transfer matrix in the top layer:
\begin{align}
    &\exp\left[\sum_{\substack{x\in\text{sites} \\ i\in\{1,2\}}} i \hat{p}_{x;i} \left(\frac{d\tau}{a}(A_{x;0} - A_{x+\hat{i};0})+\frac{2\pi}{a}m_{x;0,i}\right)\right] \nonumber \\
    =&\ \prod_{x\in\text{sites}} \exp\left[i\frac{d\tau}{a}A_{x;0}(\hat{p}_{x;1} + \hat{p}_{x;2} - \hat{p}_{x-\hat{1};1} - \hat{p}_{x-\hat{2};2})\right] \nonumber \\
    &\quad \times \exp\left[i\frac{2\pi}{a}m_1 \sum_{x_1=0}^{N_1-1}\hat{p}_{(x_1,N_2-1);2}\right] \times \exp\left[i\frac{2\pi}{a}m_2 \sum_{x_2=0}^{N_2-1}\hat{p}_{(N_1-1,x_2);1}\right],
\end{align}
where we denote the integer degree of freedom on the $1$-oriented plaquettes (cyan in Fig.~\ref{fig:appendix_maxwell_suppress_fix}) as $m_1$, and the integer degree of freedom on the $2$-oriented plaquettes (green in Fig.~\ref{fig:appendix_maxwell_suppress_fix}) as $m_2$. In the partition function, we sum over the extra degrees of freedom by
\begin{align}
    \sum_{m,A_0} \equiv \int_0^{\frac{2\pi}{d\tau}} dA_{x=(0,0);0} \prod_{x\neq(0,0)}\int_{-\infty}^{\infty} dA_{x;0} \sum_{m_1=-\infty}^\infty \sum_{m_2=-\infty}^\infty .
\end{align}
The gauge field part implies the Gauss' law $(\hat{p}_{x;1} + \hat{p}_{x;2} - \hat{p}_{x-\hat{1};1} - \hat{p}_{x-\hat{2};2})=0$ for all sites $x$ using the same argument as in the previous section.

The sum over $m_1$ sets a constraint on the loop operator $e^{2\pi i\hat{L}_1} = 1$, where
\begin{align}
    \hat{L}_1 = \begin{tikzpicture}[baseline=-0.5ex]
    \tikzset{mid arrow/.style={
        decoration={markings, mark=at position 0.75 with {\arrow{>}}},
        postaction={decorate},
        thick
    }}
    \foreach \x in {0,1,2} {
        \draw[mid arrow] (\x,-0.5) -- (\x,0.5);
    }
    \node[left] at (0,0) {$\cdots$};
    \node[right] at (3,0) {$\cdots$};
    \node[right] at (2,0) {$\frac{1}{a}\hat{p}_2$};
    \end{tikzpicture}  = \frac{1}{a} \sum_{x_1=0}^{N_1-1} \hat{p}_{(x_1,N_2-1);2} 
    ,
\end{align}
and the sum over $m_2$ sets a constraint on the loop operator $e^{2\pi i\hat{L}_2} = 1$, where
\begin{align}
    \hat{L}_2 = \begin{tikzpicture}[baseline=-0.5ex]
    \tikzset{mid arrow/.style={
        decoration={markings, mark=at position 0.75 with {\arrow{>}}},
        postaction={decorate},
        thick
    }}
    \foreach \y in {-1.5,-0.5,0.5} {
        \draw[mid arrow] (-0.5,\y) -- (0.5,\y);
    }
    \node[below] at (0,-1.5) {$\vdots$};
    \node[above] at (0,1.5) {$\vdots$};
    \node[above] at (0,0.5) {$\frac{1}{a}\hat{p}_1$};
    \end{tikzpicture}  
    = \frac{1}{a} \sum_{x_2=0}^{N_2-1} \hat{p}_{(N_1-1,x_2);1} 
    .
\end{align}
We see the structure that is similar to the loop operators in Eqs.~\eqref{eq:constraint_L1} and~\eqref{eq:constraint_L2} in the main text. These loop operators here can also be deformed by adding or subtracting the Gauss' law, showing their properties as 1-form symmetry operators.

We complete the story for the pure compact lattice Maxwell theory with instantons suppressed. The Hamiltonian is shown in Eq.~\eqref{eq:appendix_maxwell_suppress_H}. The Hilbert space has $\mathbb{R}$-ranged gauge fields on every links, but is subject to the Gauss' law constraints and two 1-form constraints shown above.

\subsection{Maxwell-Chern-Simons Lattice Theory Without Instantons}
\label{subsec:appendix_Maxwell_Chern_Simons}

With the preparations from the earlier sections, we can now start our derivation for the lattice Maxwell-Chern-Simons Hamiltonian. As reviewed in Sec.~\ref{subsec:appendix_instanton}, we need to suppress the instantons in this theory, which makes its Hilbert space structure very similar to the pure lattice Maxwell theory with instantons suppressed. We have seen that in the previous section.

The starting point of the lattice Maxwell-Chern-Simons theory is the lattice action in Eq.~\eqref{eq:S_total}, which has gauge transformations written in Eqs.~\eqref{eq:local_transformation} -~\eqref{eq:A_2_transform}. The $\varphi$ degrees of freedom are introduced to impose the instanton-suppression condition: Summing over $\varphi$ in the partition function enforces the total flux through any unit cell is zero. Their existence is equivalent to our earlier elimination of instanton-related integer degrees of freedom in the previous section, and thus does not affect our discussion here.

Therefore, we can perform the exact same gauge fixing process as in the previous section. After gauge fixing, we are left with the physical degrees of freedom shown in Fig.~\ref{fig:appendix_maxwell_suppress_fix}. We also have the same Hilbert space basis vectors, labeled by configurations of $\mathbb{R}$-ranged gauge fields on every links.

To get the transfer matrix elements, we evaluate the lattice action in Eq.~\eqref{eq:S_total} on two neighboring middle layers in Fig.~\ref{fig:appendix_maxwell_suppress_fix}. The upper layer has a gauge field configuration $|A'\rangle$ and the lower layer has $|A\rangle$. Because the time-like gauge fields and the time-like integer plaquettes are gauge fixed to zero, the result is largely simplified:
\begin{align}
    \langle A'| \hat{T} |A\rangle = &\ \exp\left[-\frac{\beta_0 a^2}{2 d\tau}\sum_{\text{links}}(A' - A)^2\right.  \nonumber \\
    &\quad\quad - \frac{\beta d\tau}{2 a^2} \left(\sum_{x\neq(N_1-1,N_2-1)} (a\,\Box A_{x;1,2})^2 + (a\,\Box A_{(N_1-1,N_2-1);1,2} + 2\pi n)^2\right) \nonumber \\
    &\quad\quad \left. + \frac{i k a^2}{4\pi} \sum_{x\in\text{sites}} \left( - A_{x-\hat{1};1} (A_{x;2}'-A_{x;2}) + A_{x-\hat{2};2} (A_{x;1}'-A_{x;1})\right)\right],
\label{eq:appendix_maxwell_chern_T_element}
\end{align}
where $(\Box A)_{x;1,2}\equiv A_{x;1} + A_{x+\hat{1};2} - A_{x+\hat{2};1} - A_{x;2}$. $n$ is the integer degree of freedom on space-like plaquettes (orange in Fig.~\ref{fig:appendix_maxwell_suppress_fix}), which denotes the flux sector and is fixed for the transfer matrix.

We apply the following general Fourier transform result:
\begin{align}
    \int_{-\infty}^{\infty}d(A'-A)e^{-ip(A'-A)}e^{-\frac{c}{2}(A'-A+d)^2+ib(A'-A)} \propto e^{-\frac{1}{2c}(p-b)^2}e^{i(p-b)d}.
\end{align}
We repeat the process in the earlier sections and derive the Hamiltonian for the total-flux-$n$ sector:
\begin{align}
    \hat{H}_n &= \frac{1}{2 \beta_0 a^2}\sum_{x\in\text{sites}}\left[(\hat{p}_{x;1}-\frac{ka^2}{4\pi}\hat{A}_{x-\hat{2};2})^2 + (\hat{p}_{x;2}+\frac{ka^2}{4\pi}\hat{A}_{x-\hat{1};1})^2\right] \nonumber \\
    &\quad + \frac{\beta}{2 a^2} \left(\sum_{x\neq(N_1-1,N_2-1)} (a\,\Box \hat{A}_{x;1,2})^2 + (a\,\Box \hat{A}_{(N_1-1,N_2-1);1,2} + 2\pi n)^2\right).
\label{eq:appendix_maxwell_chern_H_n}
\end{align}

For the constraints on the Hilbert space, we look at the top layer in Fig.~\ref{fig:appendix_maxwell_suppress_fix}. Due to the Chern-Simons action in Eq.~\eqref{eq:S_CS_full_diagram}, the extra phase factor to the transfer matrix in the top layer is much more complicated:
\begin{align}
    \exp & \left\{\sum_{\substack{x\in\text{sites} \\ i\in\{1,2\}}} i \left(\hat{p}_{x;i} - \epsilon_{ij}\frac{ka^2}{4\pi}\hat{A}_{x-\hat{j};j}\right) \left(\frac{d\tau}{a}(A_{x;0} - A_{x+\hat{i};0})+\frac{2\pi}{a}m_{x;0,i}\right)\right. \nonumber \\
    &\quad + \frac{ika^2}{4\pi} \sum_{x\in\text{sites}} \left[ \frac{d\tau}{a} \left( A_{x;0}\,\Box \hat{A}_{x;1,2}' - \hat{A}_{x-\hat{1};1}(A_{x;0} - A_{x+\hat{2};0}) + \hat{A}_{x-\hat{2};2}(A_{x;0} - A_{x+\hat{1};0}) \right) \right. \nonumber \\
    &\qquad\qquad\qquad\quad + \frac{2\pi d\tau}{a^2} (A_{x=(N_1-1,N_2-1);0} + A_{x=(0,0);0})\,n \nonumber \\
    &\qquad\qquad\qquad\quad\left.\left. + \frac{2\pi}{a} \left( - \hat{A}_{x-\hat{1};1}m_{x;0,2} - \hat{A}_{x+\hat{2};1}' m_{x;0,2} + \hat{A}_{x-\hat{2};2}m_{x;0,1} + \hat{A}_{x+\hat{1};2}' m_{x;0,1} \right) \right]\right\},
\label{eq:appendix_maxwell_chern_phase}
\end{align}
where we denote the integer degree of freedom on the time-like plaquettes as $m$. Here $A_0$'s and $m$'s are not operators. They are variables to be summed over in the partition function. $\hat{A}'$ means this operator is placed between $\langle A'|$ and $\hat{T}$. Note that some terms in Eq.~\eqref{eq:appendix_maxwell_chern_phase} can cancel with each other. We simplify it into
\begin{align}
    \exp & \left\{\sum_{\substack{x\in\text{sites} \\ i\in\{1,2\}}} i \hat{p}_{x;i} \left(\frac{d\tau}{a}(A_{x;0} - A_{x+\hat{i};0})+\frac{2\pi}{a}m_{x;0,i}\right)\right. \nonumber \\
    &\quad + \frac{ika^2}{4\pi} \sum_{x\in\text{sites}} \left[ \frac{d\tau}{a} A_{x;0}\,\Box \hat{A}_{x;1,2}'  + \frac{2\pi d\tau}{a^2} (A_{x=(N_1-1,N_2-1);0} + A_{x=(0,0);0})\,n \right.  \nonumber \\
    &\qquad\qquad\qquad\quad\left.\left. + \frac{2\pi}{a} \left( - \hat{A}_{x+\hat{2};1}' m_{x;0,2} + \hat{A}_{x+\hat{1};2}' m_{x;0,1} \right) \right]\right\}.
\label{eq:appendix_maxwell_chern_phase2}
\end{align}
Further organizing the terms, we get
\begin{align}
    &\prod_{x\in\text{sites}} \exp\left[i\frac{d\tau}{a}A_{x;0} \left(\hat{p}_{x;1} + \hat{p}_{x;2} - \hat{p}_{x-\hat{1};1} - \hat{p}_{x-\hat{2};2} + \frac{ka^2}{4\pi}\Box \hat{A}_{x;1,2}' + \frac{ka}{2} (\delta_{x=(N_1-1,N_2-1)}+\delta_{x=(0,0)}) n \right)\right] \nonumber \\
    &\times \exp\left[i\frac{2\pi}{a}m_1 \left(\sum_{x_1=0}^{N_1-1}\hat{p}_{(x_1,N_2-1);2} - \frac{ka^2}{4\pi} \hat{A}_{(x_1,N_2);1}' \right)\right]  \nonumber \\
    &\times \exp\left[i\frac{2\pi}{a}m_2 \left(\sum_{x_2=0}^{N_2-1}\hat{p}_{(N_1-1,x_2);1} + \frac{ka^2}{4\pi} \hat{A}_{(N_1,x_2);2}' \right) \right],
\label{eq:appendix_maxwell_chern_phase3}
\end{align}
where we denote the integer degree of freedom on the $1$-oriented plaquettes (cyan in Fig.~\ref{fig:appendix_maxwell_suppress_fix}) as $m_1$, and the integer degree of freedom on the $2$-oriented plaquettes (green in Fig.~\ref{fig:appendix_maxwell_suppress_fix}) as $m_2$.

In the partition function, we sum over the extra degrees of freedom by
\begin{align}
    \sum_{m,A_0} \equiv \int_0^{\frac{2\pi}{d\tau}} dA_{x=(0,0);0} \prod_{x\neq(0,0)}\int_{-\infty}^{\infty} dA_{x;0} \sum_{m_1=-\infty}^\infty \sum_{m_2=-\infty}^\infty .
\end{align}
The gauge field part implies 
\begin{align}
\label{eq:appendix_maxwell_chern_constraint11}
    \hat{p}_{x;1} + \hat{p}_{x;2} - \hat{p}_{x-\hat{1};1} - \hat{p}_{x-\hat{2};2} + \frac{ka^2}{4\pi}\Box \hat{A}_{x;1,2}' + \frac{ka}{2} n = 0, &\quad \text{for}\ x=(N_1-1,N_2-1), \\
\label{eq:appendix_maxwell_chern_constraint12}
    \hat{p}_{x;1} + \hat{p}_{x;2} - \hat{p}_{x-\hat{1};1} - \hat{p}_{x-\hat{2};2} + \frac{ka^2}{4\pi}\Box \hat{A}_{x;1,2}' = 0, &\quad \text{for}\ x\neq(N_1-1,N_2-1) \text{ or } (0,0),  
\end{align}
and for $x=(0,0)$
\begin{align}
\label{eq:appendix_maxwell_chern_constraint13}
    \int_0^{\frac{2\pi}{d\tau}} dA_{x;0} \exp\left[i\frac{d\tau}{a}A_{x;0}\left(\hat{p}_{x;1} + \hat{p}_{x;2} - \hat{p}_{x-\hat{1};1} - \hat{p}_{x-\hat{2};2} + \frac{ka^2}{4\pi}\Box \hat{A}_{x;1,2}' + \frac{ka}{2} n\right)\right] \neq 0.
\end{align}
Note that $\sum_{x\in\text{sites}} (\hat{p}_{x;1} + \hat{p}_{x;2} - \hat{p}_{x-\hat{1};1} - \hat{p}_{x-\hat{2};2} + \frac{ka^2}{4\pi}\Box \hat{A}_{x;1,2}') = 0$. Therefore, Eqs.~\eqref{eq:appendix_maxwell_chern_constraint11} and~\eqref{eq:appendix_maxwell_chern_constraint12} implies that $(\hat{p}_{x;1} + \hat{p}_{x;2} - \hat{p}_{x-\hat{1};1} - \hat{p}_{x-\hat{2};2} + \frac{ka^2}{4\pi}\Box \hat{A}_{x;1,2}') = \frac{ka}{2}n$ for $x=(0,0)$. Inserting it into Eq.~\eqref{eq:appendix_maxwell_chern_constraint13}, we get a constraint
\begin{align}
\label{eq:appendix_maxwell_chern_constraint14}
    &\int_0^{\frac{2\pi}{d\tau}} dA_{x;0} \exp\left[i\frac{d\tau}{a}A_{x;0}\left(\frac{ka}{2} n + \frac{ka}{2} n\right)\right] = \int_0^{\frac{2\pi}{d\tau}} dA_{x;0} \exp\left(i d\tau A_{x;0}\, k\, n\right) \neq 0.
\end{align}
Because we assume an even integer $k\neq 0$, this constraint implies that the integer $n$ has to be zero. i.e. The Maxwell-Chern-Simons theory only allows the zero-total-flux sector.

With knowing $n=0$, Eqs.~\eqref{eq:appendix_maxwell_chern_constraint11} and~\eqref{eq:appendix_maxwell_chern_constraint12} can be summarized into
\begin{align}
    \hat{p}_{x;1} + \hat{p}_{x;2} - \hat{p}_{x-\hat{1};1} - \hat{p}_{x-\hat{2};2} + \frac{ka^2}{4\pi}\Box \hat{A}_{x;1,2}' = 0, \quad \text{for all sites}\ x,
\end{align}
as constraints on physical states $\langle A'|$. We can equally derive the hermitian conjugate version with $\hat{A}'\to\hat{A}$, which is exactly the Gauss' law in Eq.~\eqref{eq:Gauss_law} in the main text. 

We can also just keep the zero-total-flux sector in Eq.~\eqref{eq:appendix_maxwell_chern_H_n}. The resulting Hamiltonian is
\begin{align}
    \hat{H} &= \frac{1}{2 \beta_0 a^2}\sum_{x\in\text{sites}}\left[(\hat{p}_{x;1}-\frac{ka^2}{4\pi}\hat{A}_{x-\hat{2};2})^2 + (\hat{p}_{x;2}+\frac{ka^2}{4\pi}\hat{A}_{x-\hat{1};1})^2\right] \nonumber \\
    &\quad + \frac{\beta}{2 a^2} \sum_{x\in\text{sites}} (a\,\Box \hat{A}_{x;1,2})^2.
\label{eq:appendix_maxwell_chern_H}
\end{align}
Setting $\beta_0=\beta=\frac{1}{e^2}$, we derive exactly the Hamiltonian in Eq.~\eqref{eq:Hamiltonian} in the main text.

Now we look at the $m_1$ part in Eq.~\eqref{eq:appendix_maxwell_chern_phase3}. For the part to be non-vanishing under the sum $\sum_{m_1=-\infty}^{\infty}$, we have the constraint $e^{2\pi i\hat{L}_1} = 1$, where
\begin{align}
    \hat{L}_1 &= \begin{tikzpicture}[baseline=-0.5ex]
    \tikzset{mid arrow/.style={
        decoration={markings, mark=at position 0.75 with {\arrow{>}}},
        postaction={decorate},
        thick
    }}
    \foreach \x in {0,1,2} {
        \draw[mid arrow] (\x,-1) -- (\x,0);
        \draw[mid arrow] (\x + 1, 0) -- (\x,0);
    }
    \node[left] at (0,0) {$\cdots$};
    \node[right] at (3,0) {$\cdots$};
    \node[above] at (1.5,0) {$-\frac{k a}{4\pi}\hat{A}_1$};
    \node[right] at (2,-0.5) {$\frac{1}{a}\hat{p}_2$};
    \end{tikzpicture}  \nonumber \\
    &= \sum_{x_1=0}^{N_1-1} \left(\frac{1}{a} \hat{p}_{(x_1,N_2-1);2} - \frac{k a}{4\pi} \hat{A}_{(x_1,N_2);1}'\right)
    .
\end{align}
For the $m_2$ part in Eq.~\eqref{eq:appendix_maxwell_chern_phase3}, similarly, we have the constraint $e^{2\pi i\hat{L}_2} = 1$, where
\begin{align}
    \hat{L}_2 &= \begin{tikzpicture}[baseline=-0.5ex]
    \tikzset{mid arrow/.style={
        decoration={markings, mark=at position 0.75 with {\arrow{>}}},
        postaction={decorate},
        thick
    }}
    \foreach \y in {-1.5,-0.5,0.5} {
        \draw[mid arrow] (-1,\y) -- (0,\y);
        \draw[mid arrow] (0, \y) -- (0,\y + 1);
    }
    \node[below] at (0,-1.5) {$\vdots$};
    \node[above] at (0,1.5) {$\vdots$};
    \node[right] at (0,0) {$\frac{k a}{4\pi}\hat{A}_2$};
    \node[above] at (-0.5,0.5) {$\frac{1}{a}\hat{p}_1$};
    \end{tikzpicture}  \nonumber \\
    &= \sum_{x_2=0}^{N_2-1} \left(\frac{1}{a} \hat{p}_{(N_1-1,x_2);1} + \frac{k a}{4\pi} \hat{A}_{(N_1,x_2);2}'\right)
    .
\end{align}
Note that the Gauss' law allows us to shift $\hat{L}_1$ and $\hat{L}_2$ to arbitrary locations. Moreover, they can be arbitrarily deformed as long as being topologically equivalent. i.e. They are 1-form symmetry operators. We can equally derive their hermitian conjugate versions with $\hat{A}'\to\hat{A}$, which are exactly the loop operator constraints in Eqs.~\eqref{eq:constraint_L1} and~\eqref{eq:constraint_L2} in the main text. (We are allowed to add two global phases to the constraints by twisting the boundary conditions of the gauge bundle.)

\subsection{Details of Analytical Solution}
\label{subsec:appendix_analytical_solution}

In this subsection, we add some details about the analytical solution mentioned in Sec.~\ref{sec:solution}. In Eq.~\eqref{eq:Hamiltonian_fourier_transformed}, we saw that in the momentum sector, $q=(q_1, q_2)$, the Fourier transformed Hamiltonian was
\begin{align}
    \hat{H}_q = &\quad \frac{e^2}{2 a^2} \left[
    \left(
    \hat{p}_{q;1} - \frac{k a^2}{4\pi} e^{i q_2} \hat{A}_{q;2}
    \right)\left(
    \hat{p}_{-q;1} - \frac{k a^2}{4\pi} e^{-i q_2} \hat{A}_{-q;2}
    \right)\right. \nonumber \\
    &\qquad \qquad +\left.\left(
    \hat{p}_{q;2} + \frac{k a^2}{4\pi} e^{i q_1} \hat{A}_{q;1}
    \right)\left(
    \hat{p}_{-q;2} + \frac{k a^2}{4\pi} e^{-i q_1} \hat{A}_{-q;1}
    \right)
    \right] \nonumber \\
    &+ \frac{1}{2 e^2} \left[
    \left(1-e^{-i q_2}\right) \hat{A}_{q;1}
    - \left(1-e^{-i q_1}\right) \hat{A}_{q;2}
    \right] \times \nonumber \\
    &\quad \qquad \left[
    \left(1-e^{i q_2}\right) \hat{A}_{-q;1}
    - \left(1-e^{i q_1}\right) \hat{A}_{-q;2}
    \right].
    \label{eq:Hamiltonian_sector_q}
\end{align}

We define the following change of variables (the operator notation $\hat{\cdot}$ is omitted for simplicity):
\begin{align}
    X &= \frac{(1-e^{-i q_2})\hat{A}_{q;1} + (1-e^{i q_2})\hat{A}_{-q;1} - (1-e^{-i q_1})\hat{A}_{q;2} - (1-e^{i q_1})\hat{A}_{-q;2}}{2\sqrt{(1-\cos q_1)+(1-\cos q_2)}} \\
    \Tilde{X} &= \frac{(1-e^{i q_1})\hat{A}_{q;1} + (1-e^{-i q_1})\hat{A}_{-q;1} + (1-e^{i q_2})\hat{A}_{q;2} + (1-e^{-i q_2})\hat{A}_{-q;2}}{2\sqrt{(1-\cos q_1)+(1-\cos q_2)}} \\
    Y &= \frac{(1-e^{-i q_2})\hat{A}_{q;1} - (1-e^{i q_2})\hat{A}_{-q;1} - (1-e^{-i q_1})\hat{A}_{q;2} + (1-e^{i q_1})\hat{A}_{-q;2}}{2 i\sqrt{(1-\cos q_1)+(1-\cos q_2)}} \\
    \Tilde{Y} &= \frac{(1-e^{i q_1})\hat{A}_{q;1} - (1-e^{-i q_1})\hat{A}_{-q;1} + (1-e^{i q_2})\hat{A}_{q;2} - (1-e^{-i q_2})\hat{A}_{-q;2}}{2 i\sqrt{(1-\cos q_1)+(1-\cos q_2)}} \\
    P_X &= \frac{(1-e^{-i q_2})\hat{p}_{q;1} + (1-e^{i q_2})\hat{p}_{-q;1} - (1-e^{-i q_1})\hat{p}_{q;2} - (1-e^{i q_1})\hat{p}_{-q;2}}{2\sqrt{(1-\cos q_1)+(1-\cos q_2)}} \\
    \Tilde{P}_X &= \frac{(1-e^{i q_1})\hat{p}_{q;1} + (1-e^{-i q_1})\hat{p}_{-q;1} + (1-e^{i q_2})\hat{p}_{q;2} + (1-e^{-i q_2})\hat{p}_{-q;2}}{2\sqrt{(1-\cos q_1)+(1-\cos q_2)}} \\
    P_Y &= \frac{(1-e^{-i q_2})\hat{p}_{q;1} - (1-e^{i q_2})\hat{p}_{-q;1} - (1-e^{-i q_1})\hat{p}_{q;2} + (1-e^{i q_1})\hat{p}_{-q;2}}{2 i\sqrt{(1-\cos q_1)+(1-\cos q_2)}} \\
    \Tilde{P}_Y &= \frac{(1-e^{i q_1})\hat{p}_{q;1} - (1-e^{-i q_1})\hat{p}_{-q;1} + (1-e^{i q_2})\hat{p}_{q;2} - (1-e^{-i q_2})\hat{p}_{-q;2}}{2 i\sqrt{(1-\cos q_1)+(1-\cos q_2)}}. 
\end{align}
Given the commutation relation in Eq.~\eqref{eq:A_p_commutator_fourier_transformed}, one can check that $[X, P_X]=[\Tilde{X}, \Tilde{P}_X]=[Y, P_Y]=[\Tilde{Y}, \Tilde{P}_Y]=i$, and all other commutators are zero.

The Gauss' law Eq.~\eqref{eq:Gauss_law} after the Fourier transform becomes
\begin{align}
    \left\{(1-e^{i q_1})\hat{p}_{q;1}+(1-e^{i q_2})\hat{p}_{q;2}+\frac{k a^2}{4\pi}\left[(1-e^{-i q_2})\hat{A}_{q;1}-(1-e^{-i q_1})\hat{A}_{q;2}\right]\right\}|\psi\rangle = 0 ,
    \label{eq:Gauss_law_fourier_transformed}
\end{align}
for any physical state $|\psi\rangle$ in the Hilbert space, and any $q \in \{0,\frac{2\pi}{N_1},\dots,\frac{2\pi}{N_1}(N_1-1)\} \times \{0,\frac{2\pi}{N_2},\dots,\frac{2\pi}{N_2}(N_2-1)\}$. Note that Eq.~\eqref{eq:Gauss_law_fourier_transformed} is trivially satisfied for the zero mode $q=(0,0)$. The zero mode is restricted by the two 1-form constraints Eq.~\eqref{eq:constraint_L1} and Eq.~\eqref{eq:constraint_L2}. We will analyze the zero mode separately later. For non-zero modes, the two 1-form constraints do not introduce further restrictions.

Note that Eq.~\eqref{eq:Gauss_law_fourier_transformed} also works for $-q$, which says
\begin{align}
    \left\{(1-e^{-i q_1})\hat{p}_{-q;1}+(1-e^{-i q_2})\hat{p}_{-q;2}+\frac{k a^2}{4\pi}\left[(1-e^{i q_2})\hat{A}_{-q;1}-(1-e^{i q_1})\hat{A}_{-q;2}\right]\right\}|\psi\rangle = 0.
    \label{eq:Gauss_law_fourier_transformed_neg_q}
\end{align}

Linearly combining the Gauss' law constraints for $q$ and $-q$, (Eq.~\eqref{eq:Gauss_law_fourier_transformed} $+$ Eq.~\eqref{eq:Gauss_law_fourier_transformed_neg_q}) $\implies$
\begin{align}
    \label{eq:Gauss_law_PX_X}
    \left(\Tilde{P}_X +\frac{k a^2}{4\pi} X\right)|\psi\rangle = 0,
\end{align}
and (Eq.~\eqref{eq:Gauss_law_fourier_transformed} $-$ Eq.~\eqref{eq:Gauss_law_fourier_transformed_neg_q}) $\implies$
\begin{align}
    \label{eq:Gauss_law_PY_Y}
    \left(\Tilde{P}_Y +\frac{k a^2}{4\pi} Y\right)|\psi\rangle = 0,
\end{align}

We then do another change of variables:
\begin{align}
    Q_X &= P_X + \frac{k a^2}{4\pi} \Tilde{X} \\
    \Tilde{Q}_X &= \Tilde{P}_X + \frac{k a^2}{4\pi} X \\
    Q_Y &= P_Y + \frac{k a^2}{4\pi} \Tilde{Y} \\
    \Tilde{Q}_Y &= \Tilde{P}_Y + \frac{k a^2}{4\pi} Y.
\end{align}
One can check that $[X, Q_X]=[\Tilde{X}, \Tilde{Q}_X]=[Y, Q_Y]=[\Tilde{Y}, \Tilde{Q}_Y]=i$, and all other commutators are zero. With the new variables, the Gauss' law constraints Eq.~\eqref{eq:Gauss_law_PX_X} and Eq.~\eqref{eq:Gauss_law_PY_Y} is further simplified:
\begin{align}
    \label{eq:Gauss_law_QX}
    \Tilde{Q}_X|\psi\rangle &= 0, \\
    \label{eq:Gauss_law_QY}
    \Tilde{Q}_Y|\psi\rangle &= 0. 
\end{align}

The inverse change of variables is:
\begin{align}
    \hat{A}_{q;1} &= \frac{(1-e^{i q_2})(X+i Y) + (1-e^{-i q_1})(\Tilde{X}+i\Tilde{Y})}{2\sqrt{(1-\cos q_1)+(1-\cos q_2)}} \\
    \hat{A}_{-q;1} &= \frac{(1-e^{-i q_2})(X-i Y) + (1-e^{i q_1})(\Tilde{X}-i\Tilde{Y})}{2\sqrt{(1-\cos q_1)+(1-\cos q_2)}} \\
    \hat{A}_{q;2} &= \frac{-(1-e^{i q_1})(X+i Y) + (1-e^{-i q_2})(\Tilde{X}+i\Tilde{Y})}{2\sqrt{(1-\cos q_1)+(1-\cos q_2)}} \\
    \hat{A}_{-q;2} &= \frac{-(1-e^{-i q_1})(X-i Y) + (1-e^{i q_2})(\Tilde{X}-i\Tilde{Y})}{2\sqrt{(1-\cos q_1)+(1-\cos q_2)}} \\
    \hat{p}_{q;1} &= \frac{(1-e^{i q_2})(P_X+i P_Y) + (1-e^{-i q_1})(\Tilde{P}_X+i\Tilde{P}_Y)}{2\sqrt{(1-\cos q_1)+(1-\cos q_2)}} \\
    \hat{p}_{-q;1} &= \frac{(1-e^{-i q_2})(P_X-i P_Y) + (1-e^{i q_1})(\Tilde{P}_X-i\Tilde{P}_Y)}{2\sqrt{(1-\cos q_1)+(1-\cos q_2)}} \\
    \hat{p}_{q;2} &= \frac{-(1-e^{i q_1})(P_X+i P_Y) + (1-e^{-i q_2})(\Tilde{P}_X+i\Tilde{P}_Y)}{2\sqrt{(1-\cos q_1)+(1-\cos q_2)}} \\
    \hat{p}_{-q;2} &= \frac{-(1-e^{-i q_1})(P_X-i P_Y) + (1-e^{i q_2})(\Tilde{P}_X-i\Tilde{P}_Y)}{2\sqrt{(1-\cos q_1)+(1-\cos q_2)}} \\
    P_X &= Q_X - \frac{k a^2}{4\pi} \Tilde{X} \\
    \Tilde{P}_X &= \Tilde{Q}_X - \frac{k a^2}{4\pi} X \\
    P_Y &= Q_Y - \frac{k a^2}{4\pi} \Tilde{Y} \\
    \Tilde{P}_Y &= \Tilde{Q}_Y - \frac{k a^2}{4\pi} Y.
\end{align}

Inserting everything into the Fourier transformed Hamiltonian in Eq.~\eqref{eq:Hamiltonian_sector_q}, after some algebra we get
\begin{align}
    \hat{H}_q &= \frac{a^2}{4 e^2} \left\{\frac{1}{a^2} \left[2(1-\cos q_1) + 2(1-\cos q_2)\right] + (\frac{k e^2}{4\pi})^2\left[2+2\cos(q_1 + q_2)\right]\right\}(X^2+Y^2) \nonumber \\
    &\quad + \frac{e^2}{4 a^2}\left\{Q_X^2 + Q_Y^2 + \Tilde{Q}_X^2 + \Tilde{Q}_Y^2 - \frac{k a^2}{4\pi}\left[2(1-\cos q_1) + 2(1-\cos q_2)\right](X\Tilde{Q}_X + Y\Tilde{Q}_Y)\right. \nonumber \\
    &\qquad\qquad \left. + \frac{k a^2}{4\pi}2\sin(q_1+q_2)(Y\Tilde{Q}_X - X\Tilde{Q}_Y)\right\}.
\end{align}
Applying the Gauss' law constraints Eq.~\eqref{eq:Gauss_law_QX} and Eq.~\eqref{eq:Gauss_law_QY}, we can set $\Tilde{Q}_X=\Tilde{Q}_Y=0$. We define an angular frequency, which also appeared in Eq.~\eqref{eq:solution_omega} in the main text:
\begin{align}
    \label{eq:appendix_solution_omega}
    \omega^2 = \frac{1}{a^2} \left[2(1-\cos q_1) + 2(1-\cos q_2)\right] + (\frac{k e^2}{4\pi})^2\left[2+2\cos(q_1 + q_2)\right].
\end{align}
The Hamiltonian in sector $q$ finally looks simple and familiar:
\begin{align}
    \hat{H}_q = \frac{e^2}{4 a^2}(Q_X^2 + Q_Y^2) + \frac{a^2}{4 e^2} \omega^2 (X^2 + Y^2).
\end{align}

Note that we actually have combined the $q$ and $-q$ sectors together in the analysis above. On both sectors, the total Hamiltonian is 
\begin{align}
    \hat{H}_q + \hat{H}_{-q} = \frac{e^2}{2 a^2}(Q_X^2 + Q_Y^2) + \frac{a^2}{2 e^2} \omega^2 (X^2 + Y^2).
\end{align}
Simply rescaling the momentum and coordinate, we see that the $q$ and $-q$ sectors become two decoupled simple harmonic oscillators, both having the same angular frequency $\omega$ described in Eq.~\eqref{eq:appendix_solution_omega}.

In the special case when $q=(0,\pi)$, $(\pi,0)$, or $(\pi,\pi)$, i.e. when $q$ and $-q$ are identical, one can proceed a similar analysis with half of the degrees of freedom (note that all $Y$-related variables become zero). The result is that the sector is equivalent to one simple harmonic oscillator with an angular frequency given by Eq.~\eqref{eq:appendix_solution_omega} as well (without the $Y$, $Q_Y$ degrees of freedom). Since the angular frequency for these special momentum values is given by the same equation, the band structure is smooth and has no singularity on these special momentum values.

The zero mode $q=(0,0)$, however, need to be treated separately.

First, for $q=(0,0)$, the Hamiltonian in Eq.~\eqref{eq:Hamiltonian_sector_q} does not have the magnetic potential term, and the electric kinetic terms are quadratic:
\begin{align}
    \hat{H}_{q=(0,0)} = \frac{e^2}{2 a^2} \left[
    \left(
    \hat{p}_{q=(0,0);1} - \frac{k a^2}{4\pi} \hat{A}_{q=(0,0);2}
    \right)^2 + \left(
    \hat{p}_{q=(0,0);2} + \frac{k a^2}{4\pi} \hat{A}_{q=(0,0);1}
    \right)^2 \right].
    \label{eq:Hamiltonian_sector_zero_mode}
\end{align}

Second, the Gauss' law constraint acts trivially on the zero mode. However, the two 1-form constraints Eq.~\eqref{eq:constraint_L1} and Eq.~\eqref{eq:constraint_L2} act non-trivially on the zero mode:
\begin{align}
    \label{eq:constraint_L1_zero_mode}
    \exp\left[i 2\pi \sqrt{\frac{N_1}{N_2}} \left( \frac{1}{a} \hat{p}_{q=(0,0);2} - \frac{k a}{4\pi} \hat{A}_{q=(0,0);1}\right)\right]|\psi\rangle &= e^{i\theta_1}|\psi\rangle, \\
    \label{eq:constraint_L2_zero_mode}
    \exp\left[i 2\pi \sqrt{\frac{N_2}{N_1}} \left( \frac{1}{a} \hat{p}_{q=(0,0);1} + \frac{k a}{4\pi} \hat{A}_{q=(0,0);2}\right)\right]|\psi\rangle &= e^{i\theta_2}|\psi\rangle, 
\end{align}
for any physical state $|\psi\rangle$ in the Hilbert space. Here $N_1$, $N_2$ are the number of lattice sites in the $1$ and $2$ directions, respectively. $\theta_1$ and $\theta_2$ are two constant global phases.

Third, the change of variables defined earlier does not work for the zero mode. We need to define a separate change of variables for the zero mode purpose (again, the operator notation $\hat{\cdot}$ is omitted for simplicity):
\begin{align}
    X_0 &= \frac{\hat{A}_{q=(0,0);1} + \frac{4\pi}{k a^2} \hat{p}_{q=(0,0);2}}{\sqrt{2}} ,\\
    Y_0 &= \frac{\hat{A}_{q=(0,0);1} - \frac{4\pi}{k a^2} \hat{p}_{q=(0,0);2}}{\sqrt{2}} ,\\
    P_{X_0} &= \frac{\hat{p}_{q=(0,0);1} - \frac{k a^2}{4\pi} \hat{A}_{q=(0,0);2}}{\sqrt{2}} ,\\
    P_{Y_0} &= \frac{\hat{p}_{q=(0,0);1} + \frac{k a^2}{4\pi} \hat{A}_{q=(0,0);2}}{\sqrt{2}}.
\end{align}
One can check that $[X_0, P_{X_0}]=[Y_0, P_{Y_0}]=i$ and all other commutators are zero.

Inserting the change of variables, the zero mode Hamiltonian in Eq.~\eqref{eq:Hamiltonian_sector_zero_mode} becomes
\begin{align}
    \label{eq:Hamiltonian_sector_zero_mode_simplified}
    \hat{H}_{q=(0,0)} &= \frac{e^2}{2 a^2} \cdot 2 \left[P_{X_0}^2 + \left(\frac{k a^2}{4\pi}\right)^2 X_0^2\right] \nonumber \\
    &= \frac{1}{2} \left(\frac{2 e^2}{a^2}\right) P_{X_0}^2 + \frac{1}{2} \left(\frac{a^2}{2 e^2}\right) \left(\frac{k e^2}{2 \pi}\right)^2 X_0^2,
\end{align}
which can be recognized as a simple harmonic oscillator with effective mass $\frac{a^2}{2 e^2}$ and the angular frequency
\begin{align}
    \omega_0 = \frac{k e^2}{2\pi}.
\end{align}
This is the same result as inserting $q_1=q_2=0$ into the band dispersion Eq.~\eqref{eq:appendix_solution_omega}. Therefore, the band is smooth and has no singularity at $q=(0,0)$ as well.

What makes the zero mode special, is that the constraints do not kill the degrees of freedom like in Eq.~\eqref{eq:Gauss_law_QX} and Eq.~\eqref{eq:Gauss_law_QY} for the non-zero modes. Instead, the constraints for the zero mode, Eq.~\eqref{eq:constraint_L1_zero_mode} and Eq.~\eqref{eq:constraint_L2_zero_mode}, after change of variables, say
\begin{align}
    \exp\left(i 2\pi \sqrt{\frac{N_1}{N_2}} \frac{k a}{4\pi} \sqrt{2} Y_0\right)|\psi\rangle &= e^{i\theta_1}|\psi\rangle, \\
    \exp\left(i 2\pi \sqrt{\frac{N_2}{N_1}} \frac{1}{a} \sqrt{2} P_{Y_0}\right)|\psi\rangle &= e^{i\theta_2}|\psi\rangle,
\end{align}
for any physical state $|\psi\rangle$ in the Hilbert space.

To make things cleaner, define a constant $b=\sqrt{\frac{N_2}{N_1}}\frac{4\pi}{k a}\frac{1}{\sqrt{2}}$. We rewrite the constraints above and get 
\begin{align}
    e^{i 2\pi Y_0 / b}|\psi\rangle &= e^{i\theta_1}|\psi\rangle, \\
    e^{i k b P_{Y_0} }|\psi\rangle &= e^{i\theta_2}|\psi\rangle,
\end{align}
which implies the spectrum of the operator $Y_0 \in b \mathbb{Z} + \frac{b \theta_1}{2\pi}$, and the spectrum of the operator $P_{Y_0} \in \frac{2\pi}{k b} \mathbb{Z} + \frac{\theta_2}{k b}$.
The latter further implies that, in the $Y_0$ basis, the wave function has to be periodic with the period $kb$, i.e. $\psi(Y_0 + k b) = e^{i\theta_2}\psi(Y_0)$.

\begin{figure}[htp!]
    \centering
    \begin{tikzpicture}
        \draw (0,0) circle (2cm);
    
        \foreach \x in {-90,-45,...,180} {
            \filldraw[black] (\x:2cm) circle (2pt);
        }
    
        \node at (225:1.7cm) {$\ddots$};
    
        \node at (157.5:2.2cm) {b};
        \node at (67.5:2.2cm) {b};
        \node at (112.5:2.2cm) {b};
    \end{tikzpicture}
    \caption{Illustration of the zero mode degree of freedom which generates the degenerate Hilbert space. It is equivalent to a particle on a ring of perimeter $k b$. There are $k$ equally spacing sites on the ring, having distance $b$ between the neighboring sites.}
    \label{fig:ring}
\end{figure}
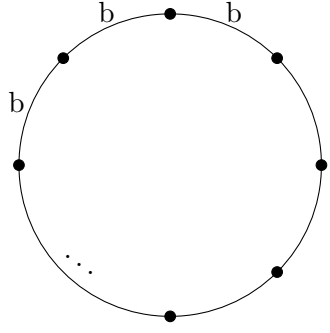

These constraints are very familiar to us. It is equivalent to a particle on a ring of perimeter $k b$, and the location of the particle can be chosen on $k$ different sites, with the neighboring sites having distance $b$ between them. See Fig.~\ref{fig:ring} for an illustration.

Because $Y_0$ does not appear in the Hamiltonian in Eq.~\eqref{eq:Hamiltonian_sector_zero_mode_simplified}, it is a decoupled degree of freedom with zero Hamiltonian. Therefore, its Hilbert space is a degenerate subspace, and is direct product to the other Hilbert spaces of the modes on the band structure.

Because the number of sites on the ring has to be an integer, we naturally get the quantization of $k$ (otherwise the discreteness $Y_0\in b\mathbb{Z} + \frac{b \theta_1}{2\pi}$ and the periodicity $\psi(Y_0 + k b) = e^{i\theta_2}\psi(Y_0)$ are incompatible). The number of independent states in the degenerate Hilbert subspace equals to the number of sites on the ring. Therefore, we also see the $k$-fold degeneracy of states from this analysis.

\bibliographystyle{jhep}

\bibliography {papers}

\end{document}